\documentclass[proof]{WileyASNA-v1}
\articletype{Article Type}%

\received{}
\revised{}
\accepted{6}

\usepackage{tikz}

\newcommand*\annotatedFigureBoxCustom[8]{\draw[#5,thick,rounded corners] (#1) rectangle (#2);\node at (#4) [fill=#6,thick,shape=circle,draw=#7,inner sep=2pt,font=\sffamily,text=#8] {\textbf{#3}};}
\newcommand*\annotatedFigureBox[4]{\annotatedFigureBoxCustom{#1}{#2}{#3}{#4}{white}{white}{white}{black}} 

\newenvironment {annotatedFigure}[1]{\centering\begin{tikzpicture}
    \node[anchor=south west,inner sep=0] (image) at (0,0) { #1};\begin{scope}[x={(image.south east)},y={(image.north west)}]}{\end{scope}\end{tikzpicture}}

\newcommand{\kepler}{\textit{Kepler}}

\newcommand{\Dnu}{\Delta\nu}

\newcommand{\numax}{\nu_{\mathrm{max}}}
\newcommand{\Pgran}{P_{\mathrm{gran}}}
\newcommand{\agran}{\alpha_{\mathrm{gran}}}
\newcommand{\taugran}{\tau_{\mathrm{gran}}}

\newcommand{\Pact}{P_{\mathrm{act}}}
\newcommand{\aact}{\alpha_{\mathrm{act}}}
\newcommand{\tauact}{\tau_{\mathrm{act}}}

\newcommand{\taueff}{\tau_{\mathrm{eff}}}

\newcommand{\Teff}{T_{\mathrm{eff}}}

\newcommand{\Henv}{H_{\mathrm{env}}}

\newcommand{\Amax}{\mathcal{A}_{\mathrm{max}}}
\newcommand{\Alim}{\mathcal{A}_{\mathrm{lim}}}

\begin{document}

\title{A new method for extracting seismic indices and granulation parameters: results for more than 20,000 CoRoT and $\kepler$ red giants.}

\author[1]{R. de Assis Peralta*}
\author[1]{R. Samadi}
\author[1]{E. Michel}

\authormark{\textsc{R. de Assis Peralta et al}}

\address[1]{\orgname{LESIA, Observatoire de Paris, Universit\'e PSL, CNRS, Sorbonne Universit\'es, Univ. Paris Diderot, Sorbonne Paris Cit\'e}, \orgaddress{\state{Meudon}, \country{France}}}

\corres{*\email{raphael.peralta@obspm.fr}}


\abstract{We have developed a new automated method intended to perform the simultaneous -- and thus more consistent -- measurement of both the seismic indices 
characterizing the oscillations  and the parameters characterizing the granulation signature of red-giant stars.
This method, called MLEUP, takes advantage of the Maximum Likelihood Estimate (MLE) algorithm combined with the parametrized 
representation of red giant pulsation spectra following the Universal Pattern (UP).
Its performances have been tested on Monte Carlo simulations for observation conditions representative of CoRoT and $\kepler$ data.
These simulations allowed us to determine, calibrate and propose correction for the biases on the parameter estimates as well as on the error estimates produced with MLEUP. 
Finally, we applied MLEUP to  CoRoT and $\kepler$ data.
In total, MLEUP yields seismic indices for 20,122 red giant stars and granulation parameters for 17,109 of them. 
These data have been made available in the Stellar Seismic Indices database (\url{http://ssi.lesia.obspm.fr/}).}



\maketitle




\section{Introduction} \label{Intro}

Red giants are evolved low mass stars which left the main sequence after burning all the hydrogen in their core.
In the Hertzsprung--Russell (HR) diagram, they are located in a narrow temperature range: $3500~\leqslant~\Teff~\leqslant~5600$ K
and in a broader luminosity range: $0.5~\leqslant~\log(L/\mathrm{L_\odot})~\leqslant~3.5$ \citep[see e.g.][]{2009A&A...503L..21M}.
Among red giants, one distinguishes stars ascending the red giant branch (RGB) and those in the red clump.
The first ones have not yet started helium burning in their core.
Consequently, while their core is contracting, their radius increases and they become brighter and brighter.
The second ones have started helium burning in their core.
Therefore, they are in a relatively stable state with a nearly constant radius and luminosity.

\cite{2009Natur.459..398D} presented the first study of several hundreds of red giants observed with CoRoT, 
showing that they exhibit non-radial oscillations with common patterns.
Indeed, red giants are solar-type pulsators, 
exhibiting oscillation modes intrinsically stable and stochastically excited by turbulent convection in the upper parts of their convective envelope. 
Stellar oscillations appear in the power spectrum of the light curve as an excess power, resulting from a balance between mode driving and damping.
\cite{2011A&A...525L...9M, 2013A&A...559A.137M} showed that the frequency distribution of modes follow a universal pattern (hereafter UP), 
valid for a large range of evolutionary stages, from main-sequence to AGB stars.
Thus, as for all solar-like pulsators, the pressure modes (p-modes) in red giants can be characterised to first order by three seismic indices 
\citep[e.g.][]{2013ARA&A..51..353C}: 
the frequency of the maximum power $\numax$, its height $\Henv$ and an equidistant frequency spacing $\Dnu$, called the mean large separation.
Both seismic indices, $\Dnu$ and $\numax$ are directly related to stellar physical properties 
such as the mean density $ <\rho>$, the surface gravity $g$ and the effective temperature $\Teff$
\citep[see e.g.][]{1986ApJ...306L..37U, 1991ApJ...368..599B, 2011A&A...530A.142B, 1995A&A...293...87K}.
Therefore, it is possible to deduce the \textit{seismic} mass and radius \citep{2010A&A...522A...1K} to a very good precision and therefore, 
study stellar populations \citep[see e.g.][]{2009A&A...503L..21M}.

Turbulent convection at the stellar photosphere induces an other signal observable in the light curve: granulation.
At the stellar surface, granulation appears under the form of irregular cellular patterns evolving with time.
In the power spectrum of the light curve, the signature of stellar granulation is located at low frequency. 
It can be characterised by two parameters \citep[e.g.][]{2014A&A...570A..41K}:
the characteristic amplitude $\sigma_\mathrm{g}^2$, which is the RMS (Root Mean Square) brightness fluctuation ($\sigma_{\rm g}^2$, thus corresponds to the total integrated energy of the granulation),
and the effective timescale $\taueff$, or \textit{e-folding time}, which measures the temporal coherence of the granulation in the time domain  \citep[e.g.][]{2011ApJ...741..119M,2014A&A...570A..41K}.
Thus, the granulation parameters carry information about stellar convection \citep[see e.g.][and references therein]{2013A26A...559A..39S}.

Various methods to automatically extract the stellar seismic indices and/or granulation parameters from light-curves exist.
Seismic indices extraction methods have been reported in \cite{2011MNRAS.415.3539V, 2011A&A...525A.131H} 
and granulation extraction methods in \cite{2011ApJ...741..119M}.

Regarding seismic indices extraction, pipelines usually deduce $\numax$ from the centroid of a Gaussian fit to the smoothed power spectrum,
except in the automated Bayesian method of \cite[][hereafter CAN]{2010A&A...522A...1K, 2014A&A...570A..41K}.
For $\Dnu$, there are three main approaches sharing same mathematical basis: the autocorrelation of the time series, the autocorrelation of the power spectrum
and the power spectrum of the power spectrum.
In addition, the autocorrelation method of \cite[][hereafter COR]{2011A&A...525L...9M} 
refines the $\Dnu$ estimate in a second step by maximizing the correlation between the raw spectrum and the UP.

The granulation parameters are extracted by fitting one or two Harvey-like functions \citep{1985ESASP.235..199H} on the smoothed spectrum.
For most methods, the least-squares algorithm is used, 
except for \cite{2010A&A...511A..46M}'s method (hereafter A2Z), which uses the MLE, 
coupled with the Levenberg-Marquardt (LM) algorithm \citep{numerical_recipes} for the optimization.
The CAN method fits the raw spectrum using a Bayesian Monte Carlo Markov Chain (MCMC) algorithm.

These studies show that it is difficult to simultaneously and consistently extract all these parameters at the same time. 
The problem comes mainly from the fact that smoothing the power spectrum is generally used to measure $\numax$ and the granulation parameters.
However, smoothing alters both the width and the height of the pulsation pics and the granulation profile.
On the other hand, the unsmoothed power spectrum follows a $\chi^2$ statistics with two degrees of freedom 
\citep{1984PhDT........34W} which requires for the optimisation the use of the Maximum Likelihood Estimator \citep[][hereafter MLE]{1994A&A...289..649T}
or the Bayesian approach \citep[e.g.][]{2010A&A...522A...1K}.


By observing several tens of thousands of red giants,
CoRoT \citep{2009IAUS..253...71B, 2006ESASP1306...33B} and $\kepler$ \citep{2010AAS...21510101B, 2010PASP..122..131G}
have enabled a breakthrough in our understanding of and way of studying these stars.
In this context, the Stellar Seismic Indices (SSI\footnote{\label{SSI_footnote}SSI database website: \url{http://ssi.lesia.obspm.fr/}}) 
database intends to provide the scientific community with a homogeneous set of parameters characterizing solar-type pulsators observed by CoRoT and $\kepler$.
For this purpose, we analyse this large set of stars, 
extracting simultaneously and consistently the seismic indices ($\Dnu$, $\numax$ and $\Henv$) 
together with the granulation parameters ($\taueff$ and $\sigma_\mathrm{g}^2$).
Furthermore, we want to characterise the error bars on the measurements provided. 
 Hence, we developed a new automated method, called MLEUP \citep{2017EPJWC.16001012D}.
The final estimates of seismic indices and granulation parameters (as well as their respective uncertainties) 
are obtained via the MLE adjustment of the unsmoothed Fourier 
spectrum with a parametric model including components for the pulsations, the granulation, the activity and the 
intrumental white noise. 

In Sect. \ref{Desciption_pipeline}, we describe our method. 
In Sect. \ref{simulations}, we assess its performances and limitations using simulated light-curves.
Accordingly, we have developed a light curve simulator designed to be as representative of red giants as possible. 
We use simulated light-curves to quantify the bias on the MLEUP results for seismic indices and granulation parameter 
values. We also quantify the biases between the real dispersion of the results and the formal errors provided by the MLE adjustment.
In Sect. \ref{application}, 
we apply our method to almost all stars observed by $\kepler$ and CoRoT and discuss the results obtained,
while taking into account correction of the biases and the uncertainty estimates. 
Finally, Sect. \ref{conclusion} is devoted to the discussion and conclusion.

%
%
%

\section{Description of the MLEUP method} \label{Desciption_pipeline}

MLEUP is an automated method based on the UP and designed to extract simultaneously the seismic indices ($\Dnu$, $\numax$ and $\Henv$) 
and the granulation parameters ($\taueff$ and $\sigma_\mathrm{g}^2$) of red-giant stars (see Sect. \ref{Intro}).  
In this section, we first describe the global model used in this method. 
Then, we describe the different steps used to adjust this model to the spectra.

\subsection{The stellar background and oscillations model} \label{BG+osc_model}

The global theoretical model we consider (Eq. (\ref{equation_BG+UP})) is composed of two main contributions: 
the stellar background model and the oscillation model.

\subsubsection{The background model} \label{background_model}

\cite{1985ESASP.235..199H} approximated the solar background signal as the sum of pseudo-Lorentzian functions: 
\begin{equation}
 P(\nu) = \sum_{i}^{N} \frac{4 \sigma_{\mathrm{g},i}^2 \tau_i}{1+(2\pi\nu\tau_i)^{\alpha_i}} \, ,
\end{equation}
with $P(\nu)$, the total power of the signal at frequency $\nu$; $N$, the number of background components;  
$\sigma_\mathrm{g,i}$, the characteristic amplitude of a given component; $\tau$, the characteristic timescale; and $\alpha_i$, 
the slope characterizing the frequency decay of pseudo-Lorentzian functions. 
For a Lorentzian function, as used by \cite{1985ESASP.235..199H}, $\alpha_i=2$, which corresponds to an exponential decay function in the temporal domain.
Each Lorentzian is associated with a different background component, such as the granulation or the activity.

Later, it was noticed that to better model solar granulation, 
it was necessary not only to change the value of the exponent $\alpha_{\rm gran}=2$ to $\alpha_{\rm gran}=4$ \citep[e.g.][]{Andersen1998}
or higher \citep[$\alpha_i=6.2$,][]{2012MNRAS.421.3170K}, but also to use two components instead of one \citep[e.g.][]{2005A&A...443L..11V}.
Then, with the arrival of the quasi-continuous long time series of CoRoT, 
granulation has also been observed in other solar type oscillators \citep{2008Sci...322..558M}.
In a few cases with the $\kepler$ data, it has even been possible to observe the double component of the granulation in the sub-giant phase \citep{2013ApJ...767...34K}
as well as in the red giant phase \citep{2010A&A...522A...1K}.
However, the origin of this double component characterizing granulation is not well understood. \\

In the framework of the SSI database, we want to be able to analyse, in a homogeneous way, as many of the stars observed by CoRoT and $\kepler$ as possible.
However, these data are disparate and for most of them, 
the signal-to-noise level and/or the resolution does not allow us to reliably fit two granulation components.
Hence, we only use one granulation component, modelled by a Lorentzian-like function (with $\alpha$ as a free parameter).
Nonetheless, this induces a bias in the measurements that we will quantify and correct a posteriori using simulated light-curves 
(see Sect. \ref{simulations}).

In addition to the granulation, we consider a second Lorentzian function ($\alpha=2$), 
in order to take into account the signal located at very low frequencies, 
usually accepted to be the signature of the stellar activity 
\citep[][]{1985ESASP.235..199H, 2010Sci...329.1032G, 2014A&A...562A.124M} and a possible residual of instrumental effects.
Finally, the background model is completed with a constant component to fit the white noise, 
corresponding to the photon and instrumental noise.

We obtain the following background model
\begin{equation}
BG(\nu) = W + \sum_{i=1}^{N=2} \frac {P_i} {1+(2 \pi \tau_i \nu)^{\alpha_i}} \, , 
   \label{equation_BG}
\end{equation}
with $N$, the number of background components; $W$, a constant for the white noise; 
$P_i$, the power of the Lorentzian-like profiles ($\Pact$, $\Pgran$) at a frequency of zero; 
$\tau_i$, the characteristic timescales ($\tauact$,  $\taugran$) and $\alpha_i$, the slope ($\aact=2$; $\agran$).\\

We note that $\Pgran$, $\taugran$ and $\agran$ are highly dependent on the type or number of the functions used.
However, they can be related in a simple way to the intrinsic parameters of the granulation, $\sigma_{\rm g}$ and $\taueff$. Indeed,  $\sigma_{\rm g}$ is given by the following relation \citep{2013ApJ...767...34K}

\begin{equation}
\sigma_\mathrm{g}^2 = \sum_{i}^{N_\mathrm{g}} \frac {1} {2} \frac {P_i} {\tau_i \alpha_i \sin(\frac {\pi} {\alpha_i})} \, ,
   \label{sigma}
\end{equation}
with $N_\mathrm{g}$, the number of granulation components. 
%
The ``e-folding time'', $\taueff$, is measured using the  autocorrelation function (hereafter ACF) of the granulation component \footnote{We compute $\taueff$ numerically. 
First, we computed the inverse Fourier transform ($\mathrm{FFT}^{-1}$) of the granulation component (one pseudo-Lorentzian) to get the corresponding ACF: 
$ACF=\mathrm{FFT}^{-1} [\Pgran/(1 + (2 \pi \taugran \nu)^{\agran})]$.
Then, we normalise the ACF by the first bin. $\taueff$ corresponds to where the  normalised ACF is equal to $1/e$.}.
Thus, unlike $\taugran$, $\taueff$ does not explicitly dependent on the number of granulation components, nor the function used.
If one uses a Lorentzian function to fit the granulation ($\alpha = 2$), $\taueff = \taugran$, 
while for a pseudo-Lorentzian, $\taueff \neq \taugran$. 



\subsubsection{The oscillation model} \label{oscillations_model}

In previous models \citep{2011ApJ...741..119M}, the oscillation spectrum is modeled using a Gaussian component and $\numax$ is determined by fitting this envelope on the raw or smoothed spectrum.
However, we know that the oscillations spectrum follows a discrete pattern which can be parametrized by the so-called {\it Universal Pattern} (UP) \citep{2011A&A...525L...9M}. 
Thus, in order to improve the seismic indices estimates, we replace the Gaussian profile by the UP. \\

To develop a synthetic and parametric pattern, we first generate a frequency comb following the asymptotic relation \citep{1980ApJS...43..469T, 2011A&A...525L...9M}. 
\begin{equation}
\nu_{n,\ell} = n + \frac{\ell}{2} + \varepsilon(\Dnu) - d_{0\ell}(\Dnu) + \frac{\gamma}{2}   \left(n - \frac{\numax}{\Dnu}\right)^{2} \Dnu \, , 
   \label{equation_freq_comb}
\end{equation}
where $n$ and $\ell$ are respectively the radial order and the degree of a given mode. 
$\varepsilon$ is an offset; $d_ {0\ell}$, the small separation; and $\gamma$, the curvature.

For each $n$, we use the first four angular degrees ($\ell=0,1,2,3$).
$\Dnu$ and $\numax$ are input parameters while $\varepsilon$, $d_{0\ell}$ and $\gamma$ are deduced from scaling relations depending on $\Dnu$. 
$\varepsilon$ and $d_{0\ell}$ follow a relation taken from \cite{2011A&A...525L...9M}, and $\gamma$ from \cite{2013A&A...550A.126M}.

Then, we describe each individual mode (Eq. (\ref{equation_freq_comb})) with a Lorentzian shape modulated by its individual visibility: 
\begin{equation}
L_{n,\ell} (\nu) = \frac {V_{\ell}^2} {1 + \lbrack 2 (\nu-\nu_{n,\ell}) / \Gamma \rbrack ^2 }  \, , 
   \label{equation_Lorentzian}
\end{equation}
where the linewidth $\Gamma$ is taken from \cite{2012sf2a.conf..173B} and the visibility $V_{\ell}$, 
the height of the individual modes of degree $\ell$, from \cite{2012A&A...537A..30M}.

Last of all, we multiply the Lorentzian profiles $L_{n,\ell}$ by a Gaussian envelope $G_\mathrm{env}(\nu)$, centred at $\numax$: 
\begin{equation}
G_\mathrm{env} (\nu) = \Henv \exp \left[\frac {-(\nu-\numax)^2} {\delta \nu_{\mathrm{env}}^2 / 4 \ln 2 } \right]  \, , 
   \label{equation_Gaussian}
\end{equation}
where the full width at half maximum ($\delta \nu_{\mathrm{env}}$) is taken as a function of $\numax$ following the scaling relation proposed by \cite{2012A&A...537A..30M}. The height of the Gaussian envelope ($\Henv$) is a fitted parameter.\\

\begin{figure}
\centering
\includegraphics[scale=0.35]{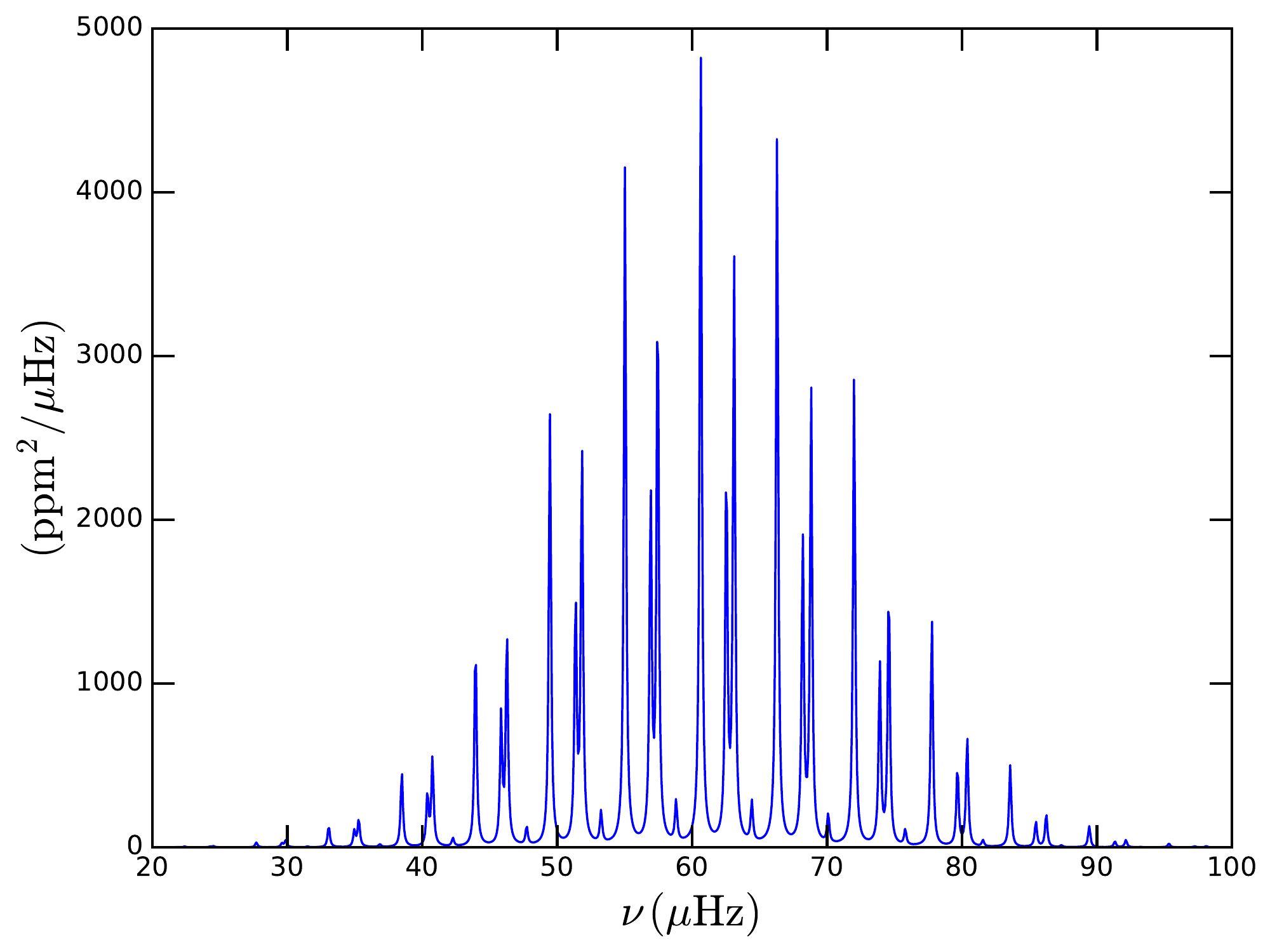}
 \caption{The \textit{Universal Pattern} (UP) is the red giant parametric oscillations pattern, used as model to fit the observed one.}
 \label{UP}
\end{figure}

Thus we finally get the UP, an oscillation pattern parametrized by three input parameters $\numax$, $\Dnu$ and $\Henv$ (see Fig. \ref{UP}):
\begin{equation}
UP (\nu) = G_\mathrm{env}(\nu) \times \sum_{n=1}^{n_{\mathrm{env}}} \sum_{\ell=0}^{3} L_{n,\ell} (\nu)  \, ,
   \label{equation_UP}
\end{equation}
with $n_{\mathrm{env}}$, the total number of radial orders considered. It is deduced from the scaling relation taken in \cite{2012A&A...537A..30M}.

\subsubsection{Global model} \label{global_model}

The global model has to consider a damping factor $\eta(\nu)$ which takes into account the distortion of the spectrum due to the integration time 
\citep{1993DSSN....6...19M, 2014A&A...570A..41K, 2014MNRAS.445..946C}.
This factor is particularly important for $\kepler$ long cadence data because of a low Nyquist frequency ($\nu_\mathrm{Nyq} \sim 287~\mu$Hz).
When the integration time is equal to the sampling time, the damping factor is expressed as:
\begin{equation}
\eta(\nu) = \mathrm{sinc} \left( \frac{\pi \nu}{2\nu_{\mathrm{Nyq}}} \right)  \, . 
   \label{damping_factor}
\end{equation}
This factor does not affect the white noise component.\\

The global model used to fit spectra is:
\begin{equation}
P (\nu) = W + \eta^2(\nu) \left[ \sum_{i=1}^{N} \frac {P_i} {1+(2 \pi \tau_i \nu)^{\alpha_i}} + UP(\nu) \right] \, ,
   \label{equation_BG+UP}
\end{equation}
with $N=2$ if the activity is taken into account in the fit, and $N=1$ otherwise.

\subsection{Detailed explanations of the algorithm} \label{Modelling}

\begin{figure*}
\includegraphics[width=160mm]{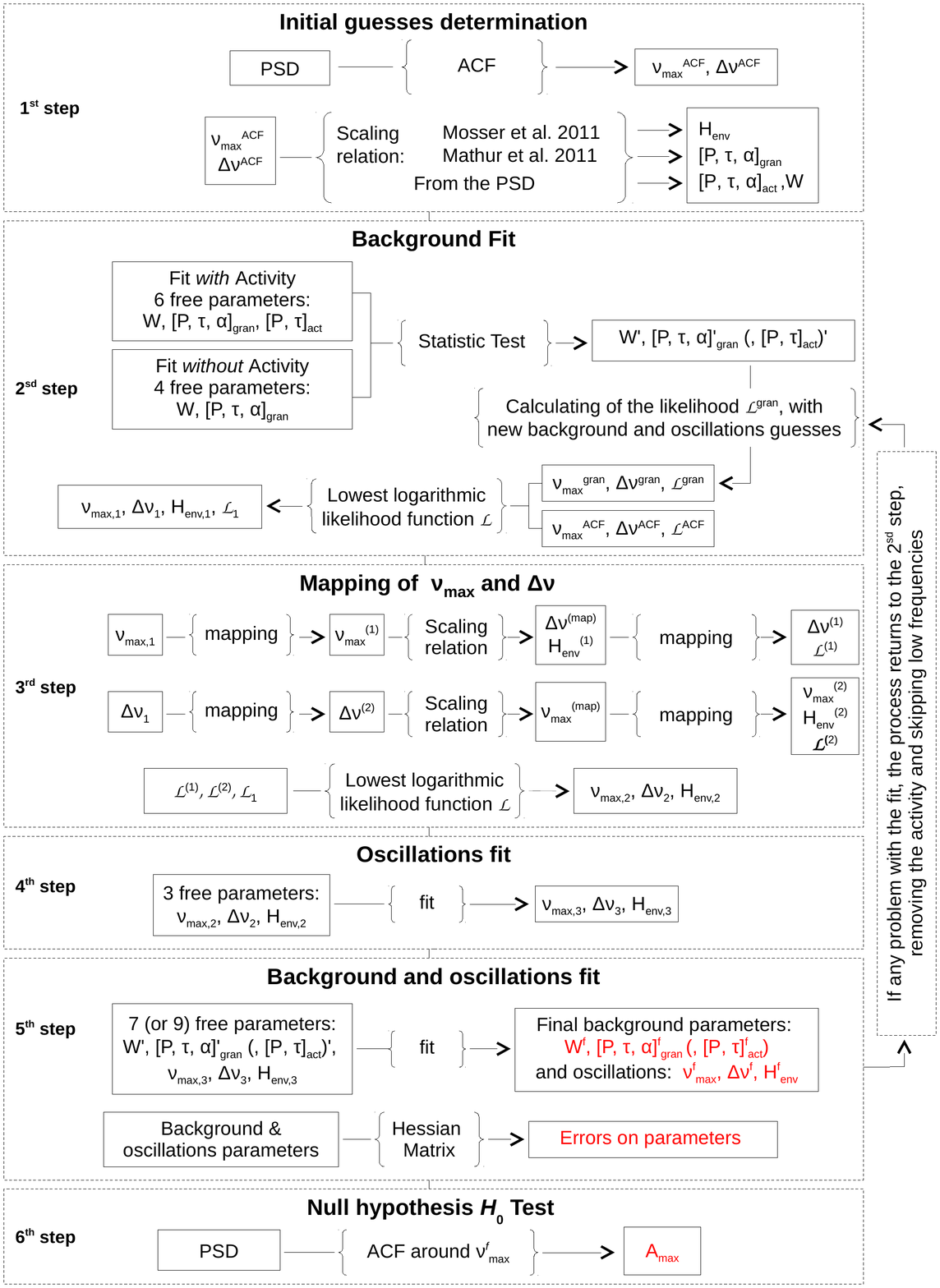}
 \caption{Diagram describing the different steps of the MLEUP method. In red, the outputs of MLEUP. \label{method}}   
\end{figure*}

To adjust our model to the power spectrum, we use the Maximum-Likelihood Estimator (MLE) algorithm \citep{1994A&A...289..649T}
coupled with the Levenberg-Marquardt (LM) algorithm \citep{numerical_recipes}  for the optimization.
Since we do not smooth the spectrum in order to preserve all the information (see Sect. \ref{Intro}),
we could not use the least-squares method which is suited for a Gaussian statistics only, 
whereas the raw spectrum follows a $\chi^2$ statistics with two degrees of freedom  \citep{1984PhDT........34W}.
An alternative approach to the MLE would have been to use the Bayesian/Monte-Carlo-Markov-Chain (MCMC) method 
to adjust our model to the spectrum (as for the CAN method). 
However, it has been considered too time consuming and therefore not acceptable for analyzing all CoRoT and $\kepler$ data.
Nonetheless, since the MLE/LM method is sensitive to guesses, 
we perform several steps before fitting all parameters together in order to improve the robustness of the final fit.

Figure \ref{method} summarizes the method and in the following section we describe each step.

\subsubsection{1\textsuperscript{st} step: Determining initial guesses} \label{step_1}

The first step consists in determining guesses for $\Dnu$ and $\numax$. 
For this purpose, we follow the COR method \citep{2009A&A...508..877M} which is based on the ACF method.
%
It deduces the large separation by calculating the autocorrelation of the time series, more precisely, 
the inverse Fourier transform of the filtered PSD.

The Power Spectral Density (PSD) is computed from the light curve with the Fast Lomb-Scargle periodogram algorithm 
developed by Leroy (2012)\footnote{The code with the Python interface developed by R\'eza Samadi can be found at: \url{https://pypi.python.org/pypi/pynfftls/1.2}}.
In the case of CoRoT, due to signatures of the low-Earth orbit \citep{2009A&A...506..411A}, data are contaminated by the 1 c.d$^{-1}$  frequency ($11.57~\mu$Hz) and its harmonics, 
as well as by the orbital frequency ($161.71~\mu$Hz). Thus, since the ACF is sensitive to regularities, 
we replace \textbf{(only for this step)} these frequencies by noise following the statistics of the PSD ($\chi^2$ with two degrees of freedom).

As was done in \cite{2009A&A...508..877M},  we filter the PSD using a cosine function centred at the frequency $\nu_\mathrm{c}$ with a width $\delta\nu_\mathrm{c}$. Here, 
$\nu_\mathrm{c}$ takes values from $3~\mu$Hz to $110~\mu$Hz for CoRoT (to avoid orbital frequencies above) and from $1~\mu$Hz to the Nyquist frequency for $\kepler$, 
following a geometric progression with a ratio $g=2^{0.1}$, which allows a filter overlap.
$\delta\nu_c$ is equal to $3\Delta\nu_\mathrm{c}$, with $\Delta\nu_\mathrm{c}$ proportional to $\nu_\mathrm{c}$ following the scaling law established in \cite{2012A&A...537A..30M}.

Afterwards, we compute the inverse Fourier transform of the PSD multiplied by the filter. 
For each filter, we obtain the envelope autocorrelation function (EACF) which reaches a maximum equal to $\Amax$ at time $\tau_{\Delta\nu}$ = 2/$\Dnu$.
The filter with the highest envelope $\Amax$ gives an estimate of $\Dnu^{\mathrm{ACF}}$ as well as $\numax^{\mathrm{ACF}}$, which is 
equal to the frequency position $\nu_{\mathrm{c}}$ of the corresponding filter. 
In order to improve these results, the process is repeated around $\numax^{\mathrm{ACF}}$ at $\pm \numax^{\mathrm{ACF}}/3 $ 
following an arithmetic progression with a ratio equal to the resolution of the spectrum.
We get our final  estimate of $\Dnu^{\mathrm{ACF}}$ and $\numax^{\mathrm{ACF}}$.

With $\numax^{\mathrm{ACF}}$, we deduce guesses for the following parameters: 
$\Henv$ from the scaling law in \cite{2012A&A...537A..30M} and $\taugran$ \citep{2011ApJ...741..119M}. 
For $\Pgran$, we compute the median of the spectrum between $\numax^{\mathrm{ACF}}/25$ and $\numax^{\mathrm{ACF}}/10$.
We estimate the white noise component $W$ by computing the median of the last points of the spectrum over a width of $3 \Dnu^{\mathrm{ACF}}$.
$\agran$ is initialized to 2.
Finally, as initial guesses, we set $\tauact = 10 \taugran$ and $\Pact$ equal to the first bin of the power of the spectrum. Meanwhile $\aact$ is fixed to 2.

\subsubsection{2\textsuperscript{sd} step: Background fit} \label{step_2}

The aim of this step is to obtain a better estimate of the background parameters. Thus, only its parameters will be free during the fit.
Since the spectrum is not smoothed in order to preserve all its informations, the Maximum-Likelihood Estimator (MLE) algorithm is used.
The MLE consists in maximising the likelihood function $L$, which is the probability to obtain the observed power spectrum $S(\nu)$
with a spectrum model $M(\nu,\lambda)$ given by Eq.~\ref{equation_BG+UP} and a set of free parameters $\lambda$ \citep[e.g.][]{1990ApJ...364..699A, 1994A&A...289..649T}. 
Knowing that the raw power spectrum of solar oscillations follows a $\chi^2$ probability distribution with two degrees of freedom (Woodard 1984), 
the likelihood function $L$ is defined as:
\begin{equation}
 L = \prod_{i=1}^{N} \frac{1}{M(\nu_i, \lambda)} e^{- \frac{S(\nu_i)}{M(\nu_i, \lambda)} }  \, ,
   \label{densite_proba}
\end{equation}
where $N$ is the number of independent frequencies $\nu_i$ of the power spectrum $S(\nu)$.\\

In practice, one will minimize the logarithmic likelihood function $\mathcal{L}$ instead of maximising $L$. $\mathcal{L}$ is calculated as:
\begin{equation}
 \mathcal{L} = - \ln L = - \sum_{i=1}^{N} \frac{S(\nu_i)}{M(\nu_i, \lambda)} + \ln M(\nu_i, \lambda)  \, .
   \label{densite_proba}
\end{equation}

Consequently, the position of the minimum of $\mathcal{L}$ in the $\lambda$-space gives the most likely value of $\lambda$ \citep{1998A&AS..132..107A}, 
i.e. the set of optimal parameters.
The formal error bars are then deduced by taking the diagonal elements of the inverse of the Hessian matrix $h$ \citep{numerical_recipes}:
\begin{equation}
 h_{ij} = \frac{\partial^2 \mathcal{L}}{\partial \lambda_i \partial \lambda_j}  \, .
   \label{matrice_hessienne}
\end{equation}

The minimization of $\mathcal{L}$ is done using a modified version of Powell's method \citep{Powell1964}.\\

The signal at very low frequency depends on the intensity of the stellar activity, instrumental effects and on the resolution of the spectrum.
Also, in some cases, it is important to take into account this component in order to improve the robustness and the quality of the fit.
Thus, we perform two fits with two different models. 
One {\it with} the activity component ($N=2$ in Eq. (\ref{equation_BG+UP})), using six free parameters ($W$, $\Pgran$, $\taugran$, $\agran$, $\Pact$ and $\tauact$)
and one {\it without} the activity component ($N=1$ in Eq. (\ref{equation_BG+UP})), using only four parameters ($W$, $\Pgran$, $\taugran$, $\agran$). 
In both cases, we consider all the spectrum. 
Then, to distinguish the best fit, we proceed to a statistical test computing the logarithmic likelihood ratio $\Lambda$ 
following \cite{wilks1938} \citep[see also][]{1998A&AS..132..107A,2012MNRAS.421.3170K}:
\begin{equation}
\ln \Lambda = \mathcal{L}(\lambda_{p+q}) - \mathcal{L}(\lambda_p) \, ,
   \label{log_ratio}
\end{equation}
with $\mathcal{L}$, the logarithmic likelihood function given by the set of parameters $\lambda$;
the index represents the number of free parameters considered for a given fit: 
$p$, in the case without the activity component  and $p+q$, in the case with the activity component (here $p=4$ and $q=2$).

Next, we compare the value of $-2 \ln \Lambda$ to the confidence level (CL) which follows a $\chi^2$ distribution with $q$ degrees of freedom, fixed at a given probability $P$.
Here, $q=2$ and we adopt $P=99\%$. Consequently, the confidence level is equal to $\mathrm{CL}=9.21$.\\
Thus, we can get three possible scenarios:

\begin{enumerate}
  \item[a.] $-2 \ln \Lambda \geq +\mathrm{CL}$: The fit with the activity is more significant than without, given the probability $P$. 
  So, we keep the activity component in the model.
  \item[b.] $-2 \ln \Lambda \leq -\mathrm{CL}$: The activity component is not significant. So, we remove the activity from the model.
  \item[c.] $-\mathrm{CL} < -2 \ln \Lambda < +\mathrm{CL}$: This third case is less clear-cut. 
  So, we remove the activity and we ignore frequencies below $\numax/20$.
\end{enumerate}

From new granulation parameters, we deduce new guesses for the seismic indices: $\numax^{\mathrm{gran}}$ \citep{2011ApJ...741..119M} and $\Dnu^{\mathrm{gran}}$ \citep{2012A&A...537A..30M}.
We compare these new guesses with the ones given by the ACF ($\numax^{\mathrm{ACF}}$ and $\Dnu^{\mathrm{ACF}}$),
according to their logarithmic likelihood function (hereafter LLF).
The pair giving the lowest LLF (denoted $\mathcal{L}_1$) is kept as seismic guesses ($\nu_{\mathrm{max,1}}$ and $\Dnu_{1}$) for the next step.  $H_{\mathrm{env,1}}$ is deduced from the scaling relation given in \cite{2012A&A...537A..30M}.

\subsubsection{3\textsuperscript{rd} step: mapping of $\numax$ and $\Dnu$ } \label{step_3}

\begin{figure}
\centering
 \includegraphics[scale=0.37, trim=0cm 0cm 1.5cm 1cm, clip=True]{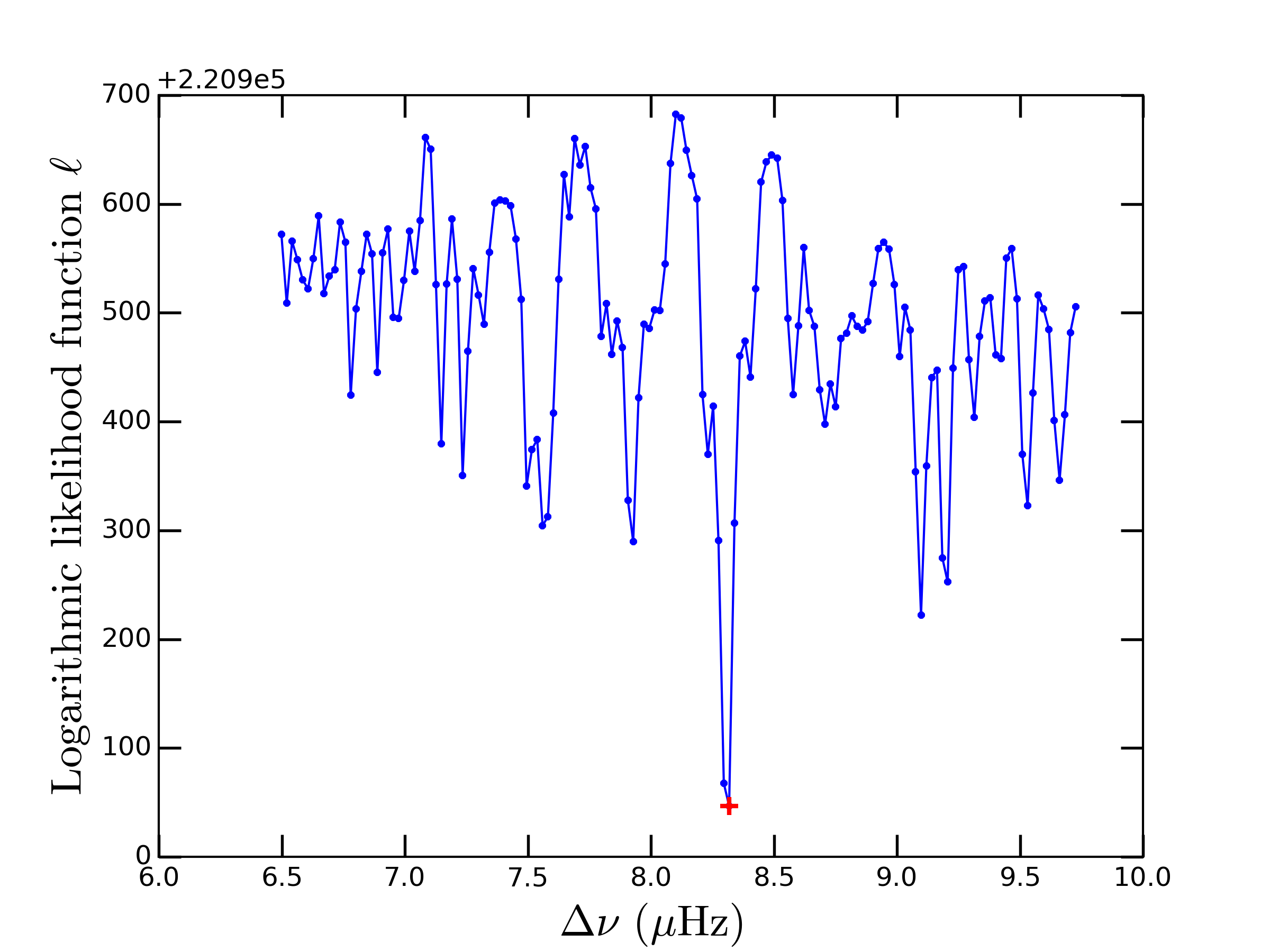}
 \caption{Topology of the logarithmic likelihood function $\mathcal{L}$ as a function of $\Dnu$ over the interval $\Dnu   \pm 20\%$ for KIC 5527304.
 The red cross indicates the lowest value of $\mathcal{L}$ corresponding to the best match between the UP and the observed oscillations spectrum.}
   \label{Dnu_mapping}
\end{figure}

In order to improve $\numax$ and $\Dnu$ estimates, we perform the mapping of the parameters.  
The mapping consists in computing the LLF for several values of a given parameter around a guess. 
We do not use in this step a simple minimization algorithm in order to avoid as much as possible local minima
due to the complexity of the likelihood topology, 
especially for $\Dnu$, which exhibits several valleys (cf. Fig. \ref{Dnu_mapping}). 
Indeed, when $\Dnu$ changes, three others parameters change proportionally $\varepsilon$, $d_{0\ell}$ and $\gamma$ (cf. Sect. \ref{oscillations_model}), thereby
modifying   the oscillation component. 
 The value giving the lowest likelihood corresponds to the combination where the synthetic  modes best match  the observations.

Since $\numax$ and $\Dnu$ are strongly dependent on each other, the best would be to do the mapping of both indices simultaneously. 
However, this is very time consuming. Therefore, we obtain two separate mappings.
The quality of the mapping, and therefore of the deduced results, depend on previous guesses of the background and oscillations.
So, it is more efficient to start finding the mapping in some cases by one parameter rather than by the other. 
Thus, we proceed with two optimization strategies:

\begin{enumerate}
  \item[(1)] We obtain the mapping of $\numax$ around $\nu_{\mathrm{max,1}} \pm 25\%$. 
  The lowest LLF gives us $\numax^{(1)}$ with which we deduce new guesses: 
  $\Dnu^{\mathrm{map}}$ and $\Henv^{(1)}$ following the scaling law from \cite{2012A&A...537A..30M}.
  Then, we obtain the mapping of $\Dnu$ around  $\Dnu^{\mathrm{map}} \pm 20\%$. 
  The value of $\Dnu$ with the lowest LLF gives $\Dnu^{(1)}$.
  
  \item[(2)] The strategy is reversed: First, the mapping of $\Dnu$ is performed, 
  giving $\Dnu^{(2)}$ from which $\numax^{\mathrm{map}}$ and $\Henv^{\mathrm{map}}$ are deduced following the scaling laws from \cite{2012A&A...537A..30M}.
  Then, we obtain the mapping of $\numax$ around $\numax^{\mathrm{map}}$ to obtain $\numax^{(2)}$. We then deduce $\Henv^{(2)}$ from the scaling relation taken from \cite{2012A&A...537A..30M}.
\end{enumerate}

At last, we compare final  LLF values given by the strategy (1), (2), as well as the one obtained in the second step ($\mathcal{L}_1$). 
Results with the lowest LLF are kept, giving new seismic indices estimates: $\nu_{\mathrm{max,2}}$, $\Dnu_2$ and $H_\mathrm{env,2}$.

\subsubsection{4\textsuperscript{th} step: Oscillations fit} \label{step_4}

The aim of this step is to optimize the three seismic indices determined in the previous step using the minimization algorithm
within  $\nu_{\mathrm{max,2}} \pm 6 \Dnu_2$.
At the end, we get new seismic guesses ($\nu_\mathrm{max,3}$, $\Dnu_{3}$ and $H_\mathrm{env,3}$).

\subsubsection{5\textsuperscript{th} step: Background and oscillations global fit} \label{step_5}

\begin{figure}
\centering
\includegraphics[scale=0.38, trim=0cm 0cm 1.5cm 1cm, clip=True]{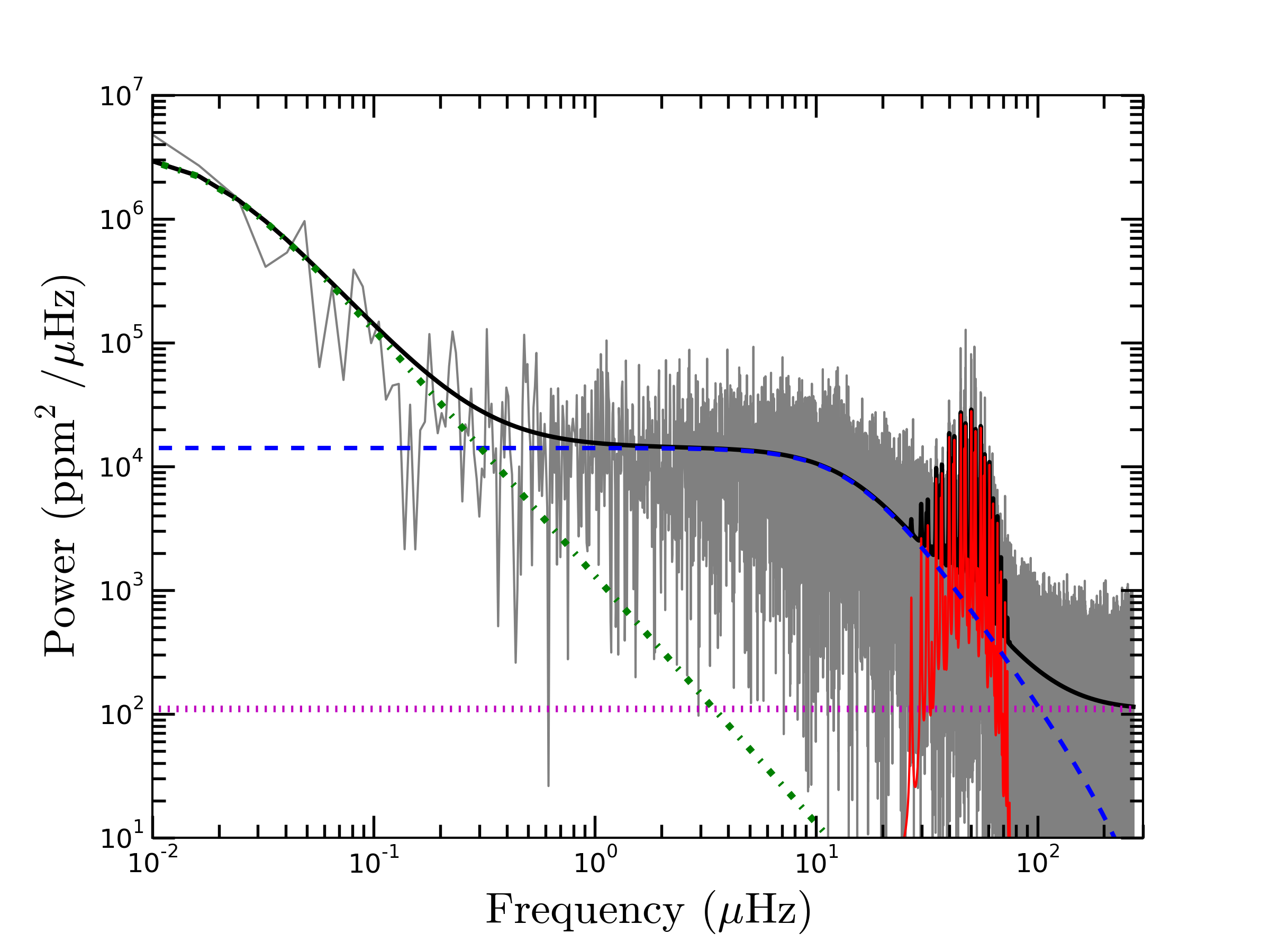}
 \caption{Results of the simultaneous adjustment of the background and oscillations at the 5\textsuperscript{th} step of MLEUP (black line). 
 In grey, the raw PSD of KIC 2850913. 
 The dash-dot green line and the dashed blue one correspond respectively to the activity and granulation component.
 The dotted magenta line represents the white noise component and the solid red one is the Universal Pattern.}
   \label{fit_final}
\end{figure}

For this last fit, the background and oscillations are fitted simultaneously (cf. Fig. \ref{fit_final}). 
Thus, all parameters of the model (Eq. (\ref{equation_BG+UP})) are free, except $\aact$ which is fixed to 2 (if we have kept the activity component).
We obtain final estimates of seismic indices: $\numax^\mathrm{f}$, $\Dnu^\mathrm{f}$ and $\Henv^\mathrm{f}$ and background parameters, 
including the granulation: $\Pgran^\mathrm{f}$, $\taugran^\mathrm{f}$, $\agran^\mathrm{f}$; 
the white noise $W^\mathrm{f}$ and, depending of the case, the activity: $\Pact^\mathrm{f}$, $\tauact^\mathrm{f}$.
Then, we compute internal errors from each parameter by inverting  the Hessian matrix \citep{numerical_recipes}.

If the fit did not converge while the activity was kept, the process returns to the second step, 
removing the activity component and skipping low frequencies ($\nu < \numax^{\mathrm{ACF}}/20$).

\subsubsection{6\textsuperscript{th} step: null hypothesis $H_0$ test} \label{step_7}
 
In this additional step, we use the ACF to produce an independent rejection criterion. 
We recompute the ACF around $\numax^\mathrm{f}$ at $\pm \numax^\mathrm{f}/3$, 
following an arithmetic progression with a  common difference  equal to the resolution of the spectrum and we keep the highest $\Amax$ value.
Following \cite{2009A&A...508..877M},  we apply to this  $\Amax$ value the null hypothesis $ H_0$ test. 
We keep the results  when $\Amax$  is above the  threshold $\Alim = 8$ which corresponds to a probability of $P= 1 \%$.

\section{Biases and error calibration using synthetic light-curves} \label{simulations}

In this section, our goal is to test the MLEUP algorithm on synthetic light-curves in order to calibrate 
(and be able to correct a posteriori) possible biases in the estimates of seismic indices and 
granulation parameters as well as in the associated error estimates. 

This approach is motivated by two main reasons.
The first one is that we anticipate possible biases, since the model used in the MLEUP 
algorithm (see Sect. \ref{Desciption_pipeline}) is different, simpler in fact, 
than the one suggested by 
our present understanding of solar-like pulsations and granulation. For the reasons developed in Sect. \ref{Desciption_pipeline}, our MLEUP-model 
only includes  pressure dominated modes and only one component for granulation. 
Synthetic light-curves including mixed modes and a two components description of the granulation will allow us to investigate the possible 
impact of such differences on the estimates of the various seismic indices and granulation parameters.

The second reason is that we want to test the strategy adopted for the MLEUP algorithm 
(see Sect. \ref{Modelling}) and in particular to which extent it allows us to reach the best solution, 
avoiding as much as possible solutions associated with secondary minima in the optimization process.

This approach relies on the assumption that the present understanding of solar-like pulsation 
and granulation patterns is advanced enough to allow us to produce  synthetic light-curves which are realistic 
and representative enough of the data obtained with CoRoT and $\kepler$.


We thus developed a solar-like light-curve simulator (SLS) to produce such synthetic light-curves 
representative of CoRoT and $\kepler$ data\footnote{The code is available at: \url{https://psls.lesia.obspm.fr}}. 
We generated sets of light-curves, 
for several representative durations, stellar magnitudes and evolutionary stages.
The detailed description of these synthetic light-curves and of the
statistical study of their MLEUP analysis are given in  \ref{full_simulations}.
Here we stress the main specificities of the input model considered for the SLS and we describe 
the corrections applied to the results obtained with  
the MLEUP algorithm when analysing CoRoT and $\kepler$ data in Sect. \ref{application}.

\subsection{The input model for synthetic light-curves} \label{simulations_parameters}

\begin{figure}
 \centering
\includegraphics[scale=0.35]{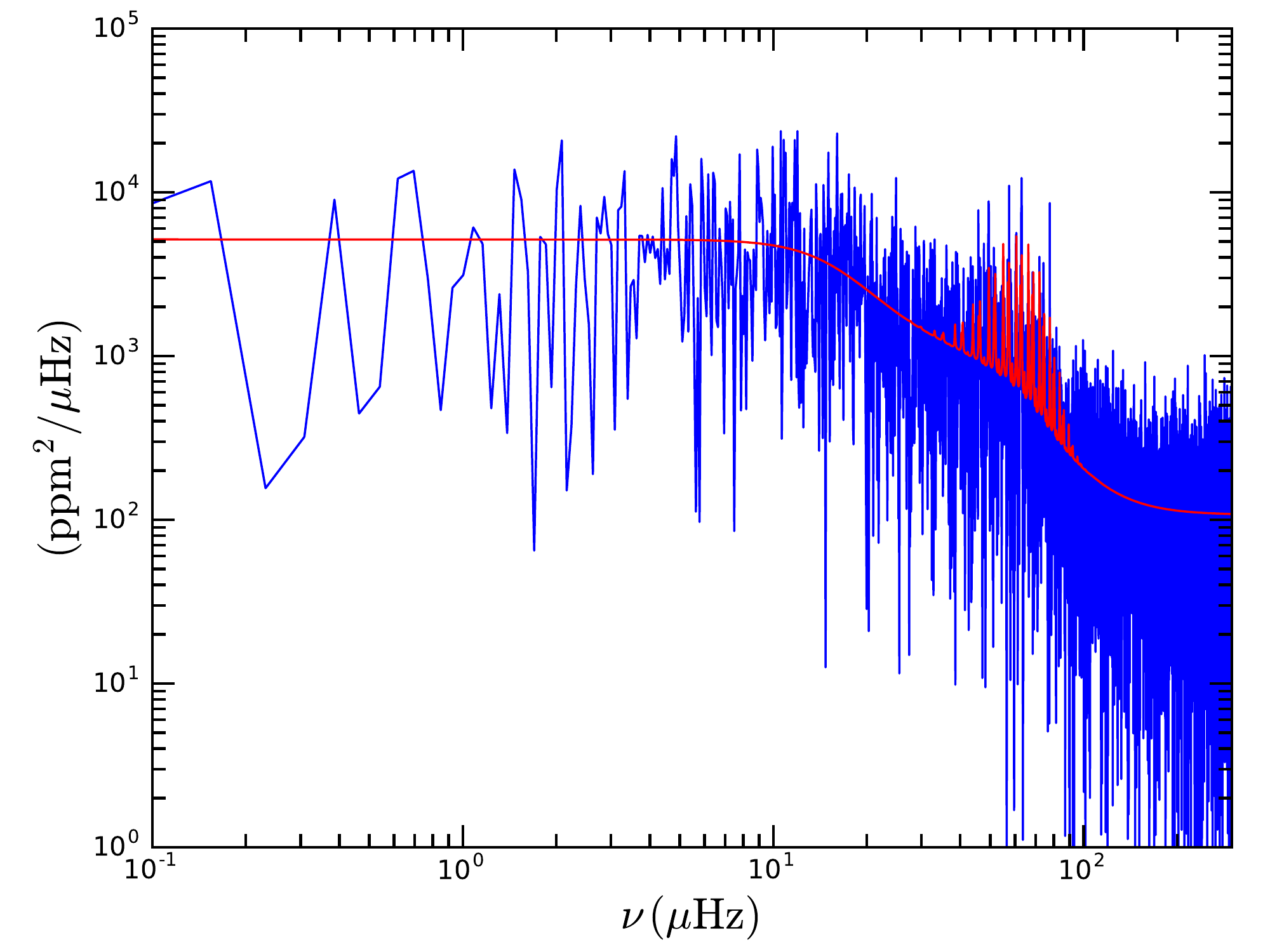}
 \caption{The red curve corresponds to the theoretical spectrum pattern and the blue one illustrates
the simulated power spectrum for a given realisation generated with the SLS for a CoRoT-type star.}
 \label{fig_simus}
\end{figure}

The SLS generates light-curves following an input model which features the oscillations, the granulation and the white noise
to the best of our present knowledge as based on the CoRoT and $\kepler$ experiments.
The activity signal is not considered because, contrarily to pulsations or granulation, we lack  any prescription 
about how it behaves with stellar mass or evolution stage so far. 

Besides these inputs, the SLS mimics the stochastic nature of these phenomena thus producing a spectrum (and the associated light-curve) representative of a 
given observational realization. 
This is achieved by applying an artificial random dispersion, following a $\chi^2$ statistics with two degrees of freedom statistics,
around the theoretical mean model (see Fig. \ref{fig_simus}).

The $\ell=0,2,3$ oscillations components are simulated following the UP \citep{2011A&A...525L...9M}, just as in our MLEUP-model 
(see Sect. \ref{oscillations_model}). 
The dipole $\ell=1$ mixed modes component is considered following \cite{2012A&A...540A.143M} and as detailed in  \ref{full_simulations}.
Finally, the granulation is modeled with two components, following the model F 
of  \cite{2014A&A...570A..41K}, as described in Sect. \ref{background_model}.
 

For both satellites, simulations take into account  the sampling time ($dt$), 
the typical white noise level (see  \ref{full_simulations}), and the attenuation factor as described in Sect. \ref{global_model}.


\subsection{Biases and internal error corrections} \label{biases_correction}

The study of synthetic light-curves (\ref{full_simulations}) allowed us to estimate biases and to characterize them quantitatively (Tab. \ref{tableau_coefficient_simus}).
They are corrected for in the analysis of real data  in Sect. \ref{application}.

This study also reveals  that internal errors sometimes significantly underestimate real errors. 
This can be understood since formal errors derived form the Hessian matrix are at best lower limits for the error estimates, according to Kramer-Rao theorem \citep[see][and references therein]{Kendall1967}.
We therefore established conservative correction functions summarized in Tab. \ref{tableau_correction_erreur_interne} that are  applied to the analysis of real data  in Sect. \ref{application}.

\section{Application to large sets of stars} \label{application}

We applied the MLEUP analysis on a large set of stars observed with CoRoT and $\kepler$.
Here we present and comment the results of this analysis.

\subsection{Observations} \label{observations}

\subsubsection{Target selection} \label{target_selection}

The CoRoT space mission was dedicated to seismology and  the detection of exoplanets \citep{2006ESASP1306...33B}.
In 18 runs, from 2006 to 2013, CoRoT observed about 150,000 stars located in two opposite directions in our galaxy.
We analysed all the ready-to-use CoRoT legacy 
data\footnote{CoRoT legacy data archive: \url{http://idoc-corot.ias.u-psud.fr}} \citep{2016cole.book.....C} 
for which the observations lasted longer than 50 days in order to get a sufficient signal-to-noise ratio 
and  frequency resolution to detect the oscillations.
This represents a total number of 113,677 CoRoT stars. 
The CoRoT legacy data were processed as described in \cite{2016cole.book...41O}.
When stars were observed several times,
we concatenated their light-curves (assuming that there is no temporal coherence) by joining them together and adjusting their average level of intensity.

The second space mission, $\kepler$, a NASA spacecraft \citep{2010AAS...21510101B}, 
was launched in 2009 and observed the same galactic region during more than four years, divided in 17 quarters.
We analysed all long cadence ($dt = 29.42$ min) $\kepler$ data\footnote{$\kepler$ data archive: \url{http://archive.stsci.edu/kepler/}}
for which the observations lasted longer than 10 days.
These data were corrected  as documented in 
the $\kepler$ Data Processing Handbook\footnote{The $\kepler$ Data Processing Handbook website: 
\url{https://archive.stsci.edu/kepler/manuals/KSCI-19081-001_Data_Processing_Handbook.pdf}}.
Then, from the effective temperature $\Teff$ extracted from different catalogues (see Sect. \ref{scaling_relations} for more detail),
we selected the stars in the temperature range $3800$ K $\leq~\Teff~\leq~5700$ K.
Outside this range, we assume they are not red giants. 
As was done for CoRoT data, we concatenated light-curves of stars observed several times.
Gaps smaller than $10^4$ s were filled using a linear interpolation. 
This dataset represents 207,610 stars, most of which were observed during all the duration of the mission.

\subsubsection{Rejection of outliers} \label{Rejection_outliers}

After analysing data from both satellites separately, we kept the results which  satisfy the following criteria:
\begin{itemize}
 \item The fit of the three seismic indices must be properly converged.
 \item $\Alim = 8.0 \leq \Amax \leq \Alim = 700$ (see Sect. \ref{simulations_parameters} for more details). 
 The upper limit was defined in order to avoid false detections caused by possible artefacts in the spectrum.
 \item The signal-to-noise ratio (SNR) is defined as: $SNR = \Henv/B_\mathrm{max}$, with $B_\mathrm{max}$ the height of the background at the frequency $\numax$.
 We keep results with a \textit{SNR} between: $1.0 \leq SNR \leq 70$. The upper limit is used for the same reasons as for $\Amax$.
 \item We restrict the  frequency interval of $\numax$ to  $2~<~\numax~<~100~\mu$Hz  for CoRoT and to  $2~<~\numax~<~250~\mu$Hz for $\kepler$. 
 Indeed, as shown by the simulations (see Sect.~\ref{simulations} and  \ref{full_simulations}), for higher values of $\numax$, 
 the dispersion of parameters becomes too high, especially for $\numax$ and $\Dnu$, and is poorly represented by the internal errors.
 The lower limit is due to the presence of some peaks below $2~\mu$Hz that we suspect to be artefacts for $\kepler$ 
 and due to the limited resolution for CoRoT.
 \item Values of the granulation slope $\agran$ are limited to $\agran < 5.0$,
 because higher slopes generally correspond  to a bad fit, typically due to an artefact. 
  \item For each parameter, the ratio between the internal errors and the associated values must be less than 50\%.
 \item Outliers are removed using the following combination of seismic indices:
 \begin{equation}
   \left( \frac{\Dnu - \Dnu^\mathrm{SR}}{\alpha_\pm \Dnu^\mathrm{SR}}\right )^{2} + 
   \left( \frac{\Henv - \Henv^\mathrm{SR}}{\beta_\pm \Henv^\mathrm{SR}}\right)^{2} \leqslant 1 \, ,
 \end{equation}
 with $\alpha_+ = \alpha_- = 0.30$, $\beta_+ = 3$ and $\beta_- = 1$. 
 $\beta_+ \neq \beta_-$ because the distribution of results for the parameters $\Henv$ is asymmetric.
 $X^\mathrm{SR}$ is the value of the parameter $X$ corresponding to the scaling relation. 
 The scaling laws are initially taken from the literature (see Tab. \ref{relation_echelle_ref}) 
 and iteratively based on the considered data sample. 

 \end{itemize}

For the sub-sample, for which we have both the seismic indices and granulation parameters, we used these additional criteria:
\begin{itemize}
 \item The fit of both granulation parameters must be properly converged.  
 \item Values of the granulation slope $\agran$ are also limited to $\agran \geq 1.0$ since $\agran~<1.0$ does not give physical results.
 \item Outliers are removed using the following combination of seismic indices (see above) and granulation parameters:
 \begin{equation}
   \left( \frac{\Pgran - \Pgran^\mathrm{SR}}{\beta_\pm \Pgran^\mathrm{SR}}\right )^{2} + 
   \left( \frac{\taugran - \taugran^\mathrm{SR}}{\beta_\pm \taugran^\mathrm{SR}}\right )^{2} \leqslant 1 \, ,
 \end{equation}
 with $\beta_+ = 3$ and $\beta_- = 1$ (the distribution of $\Pgran$ and $\taugran$ are also asymmetric).\\
\end{itemize}

Finally, we yield 20,122 stars (2943 CoRoT stars and 17,179 $\kepler$ stars)  for which we extracted the seismic indices. They form the  $S_\mathrm{s}$ dataset. Besides, for 17,109 stars (806 CoRoT stars, and 16,303 $\kepler$ stars) we obtained both the seismic indices and the granulation parameters ($S_\mathrm{s+g}$ dataset). 

As noted recently \citep[see e.g.][]{2016ApJ...827...50M}, some stars in KIC have been unclassified or misclassified. 
Here among the 15,626 $\kepler$ stars identified as red giants in the $\kepler$ Input Catalogue \citep[KIC,][]{2011AJ....142..112B},
we detect oscillations for 13,277 stars.
In addition, we found 3902 new oscillating red-giant stars not identified in KIC as red giants.

\subsection{Results for the various parameters} \label{scaling_relations}

We present in this subsection the  results obtained with MLEUP for the selected CoRoT and $\kepler$  datasets, and for the various parameters.

For some $\kepler$ targets, we have additional informations allowing us to enrich and improve our understanding of the results:
\begin{itemize}
 \item Evolutionary stages: \cite{2016A&A...588A..87V}
 have determined the evolutionary stages of  more than five thousands $\kepler$ stars classified as red giants
 using an automatic measurement of the gravity period spacing of dipole modes $\Delta\Pi_1$.
 Thanks to their results, we are able to discern between stars belonging to the RGB from those in the clump for about 25\% of our selected $\kepler$  dataset.
 However, this method is limited to $\numax \geq 35~\mu$Hz for RGB stars and to $\numax \geq 25~\mu$Hz for clump stars. 
 Indeed, it is difficult to automatically measure the $\Delta\Pi_1$ at low frequency (i.e. for evolved stars)
 since the mixed modes are less visible and therefore more difficult to detect \citep[also see][]{2014A&A...572A..11G}. \\
\item
Mass and radius estimates: 
 Combining $\Teff$, $\Dnu$ and $\numax$ allows us to calculate the seismic mass, 
 radius and luminosity from the following scaling relations \citep[e.g.][]{2010A&A...509A..77K}:
 \begin{equation}
 \frac{M}{\mathrm{M_\odot}} \propto \left( \frac{\numax}{\nu_{\mathrm{ref}}} \right)^3 \left( \frac{\Dnu}{\Dnu_{\mathrm{ref}}} \right)^{-4} \left( \frac{\Teff}{T_\odot} \right)^{3/2} \, ,
 \label{eq_masse}
 \end{equation}
 \begin{equation}
  \frac{R}{\mathrm{R_\odot}} \propto \left( \frac{\numax}{\nu_{\mathrm{ref}}} \right) \left( \frac{\Dnu}{\Dnu_{\mathrm{ref}}} \right)^{-2} \left( \frac{\Teff}{T_\odot} \right)^{1/2} \, ,
  \label{eq_rayon}
  \end{equation}
 and 
  \begin{equation}
 \frac{L}{\mathrm{L_\odot}} \propto \left( \frac{\numax}{\nu_{\mathrm{ref}}} \right)^2 \left( \frac{\Dnu}{\Dnu_{\mathrm{ref}}} \right)^{-4} \left( \frac{\Teff}{T_\odot} \right)^{5} \, .
  \label{eq_luminosite}
  \end{equation}
  These relations are normalised with the reference values given in \cite{2013A&A...550A.126M}: 
 $\Dnu_{\mathrm{ref}} = 138.8~\mu$Hz, $\nu_{\mathrm{ref}} = 3104~\mu$Hz and $T_\odot=5777$ K.\\
The effective temperatures $\Teff$ are extracted from several catalogues. 
 First, we took $\Teff$ from both spectroscopic surveys APOGEE DR12 \citep{2016arXiv160802013S} and LAMOST DR2 \citep{2016yCat.5149....0L}. 
 We got 7205 and 1809 effective temperatures respectively.
 Then, from the photometric Str\"{o}mgren survey for Asteroseismology and Galactic Archaeology (SAGA) \citep{2014ApJ...787..110C}, we got 377 effective temperatures .
 Finally, from \cite{2017ApJS..229...30M}, which is an updated of the \cite{2014ApJS..211....2H} catalogue reporting $\Teff$ for stars 
 observed by $\kepler$ for  quarters Q1 to Q17 (DR25), we obtained 7788 more effectives temperatures. 
 Altogether, these catalogues provide us with effective temperatures for the entire set of  selected $\kepler$  targets. 

\end{itemize}

The analysis of the seismic indices and the fundamental parameters is done using the dataset $S_\mathrm{s}$, and for the granulation parameters, we use the $S_\mathrm{s+g}$ dataset (see Sect. \ref{Rejection_outliers}).

\subsubsection{Height of the Gaussian envelope $\Henv$} \label{Henv}

\begin{figure*}
  \centering
\includegraphics[scale=0.39, trim=0.37cm 0cm 1.cm 0.7cm, clip=True]{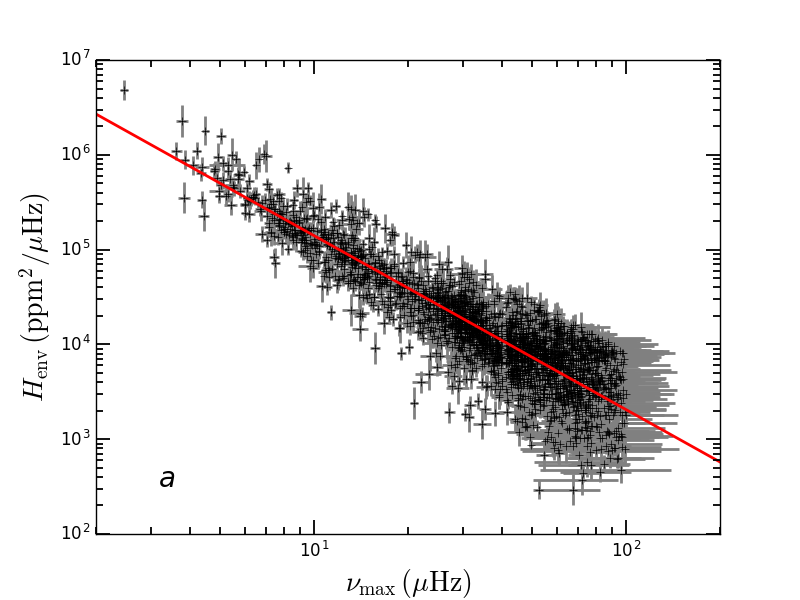}
\includegraphics[scale=0.39, trim=0.37cm 0cm 1.cm 0.7cm, clip=True]{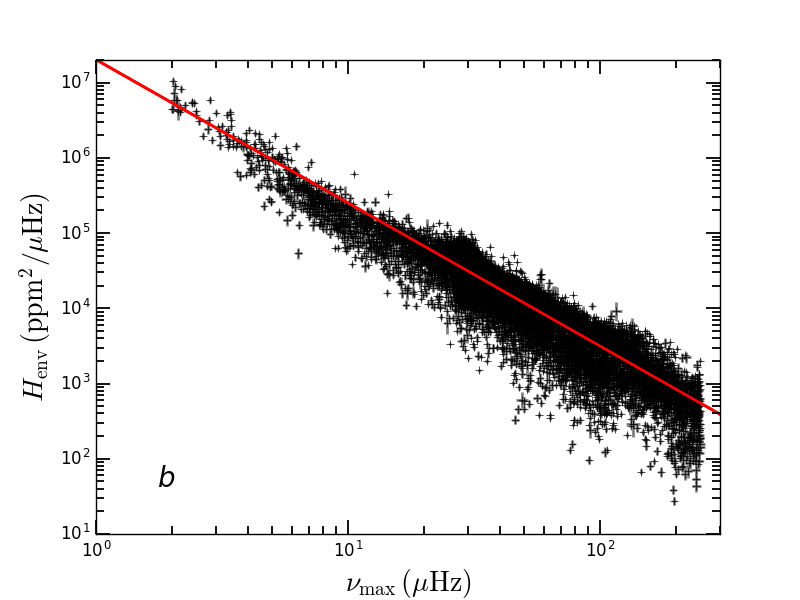}
\includegraphics[scale=0.39, trim=0.37cm 0cm 1.cm 0.7cm, clip=True]{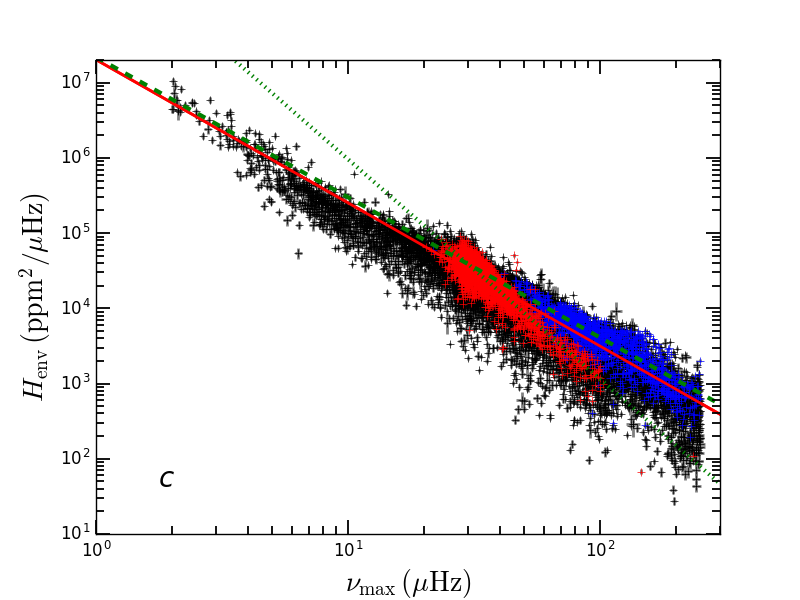}
\includegraphics[scale=0.39, trim=0.37cm 0cm 1.cm 0.7cm, clip=True]{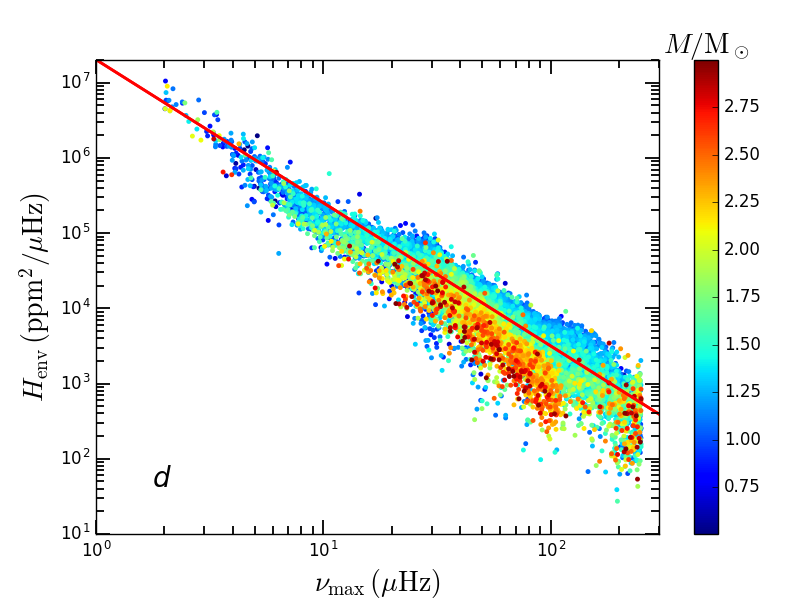}
\caption {Height of the Gaussian envelope $\Henv$ obtained with the CoRoT (\textit{a}) and $\kepler$ (\textit{b}) datasets $S_\mathrm{s}$.
Black crosses represent the values obtained, with their error bars in grey.
The red line is the deduced scaling relations. \newline
\textit{c:} Same as figure  \textit{b} with the 1333 RGB stars in blue and the 3152 clump stars in red.
The green dashed line and the dotted one represent respectively the scaling relation obtained with the RGB and clump stars. \newline
\textit{d:} Same as figure  \textit{b} with the stellar mass indicated via the colour code. 
For better visibility of the mass variation, only stars with a mass  in the range $0.5 \leqslant M/\mathrm{M_\odot} \leqslant 3.0$ are plotted.
\label{fig_resultat_Henv}}
\end{figure*}

Overall, the values of $\Henv$ obtained with CoRoT and $\kepler$ data are comparable (cf. Fig. \ref{fig_resultat_Henv}a and b). 
The dispersion is larger in the case of CoRoT as  expected and as it had been previously suggested by the simulations (cf. Sect.~\ref{simulations} and \ref{full_simulations}).
This is mainly due to shorter observation time $T$ in the case of CoRoT.

Concerning $\kepler$ results, the $\Henv$ distribution shows a broadening below the deduced scaling relation for $\numax \gtrsim 30~\mu$Hz.
Figure \ref{fig_resultat_Henv}c, with the information on the evolutionary stage for some stars, 
reveals that this broadening is mainly due to clump stars (red crosses) which exhibit an envelope height which decreases significantly 
faster with $\numax$ than in RGB stars (blue crosses).
The RGB component is close to the reference because the majority of stars without information on the evolutionary stage belongs to the RGB, 
especially below $\numax = 30~\mu$Hz.
This dependency of $\Henv$ on the evolutionary stage has already been observed by \cite{2011A&A...532A..86M,2012A&A...537A..30M} on a much smaller sample. 
The information on the mass (cf. Fig. \ref{fig_resultat_Henv}d) confirms the distribution of the clump component since,
according to the theory of the stellar evolution, 
higher-mass stars belong essentially to the clump and its extension, the secondary clump \citep[e.g.][]{1994sse..book.....K}.

\subsubsection{Mean large separation $\Dnu$} \label{Dnu}

\begin{figure*}
\centering
\includegraphics[scale=0.39, trim=0.37cm 0cm 1.cm 0.7cm, clip=True]{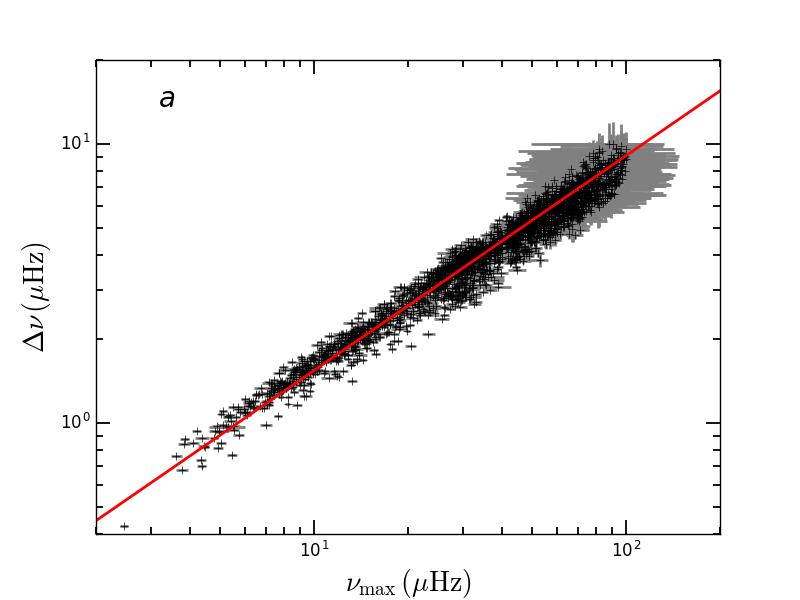}
\includegraphics[scale=0.39, trim=0.37cm 0cm 1.cm 0.7cm, clip=True]{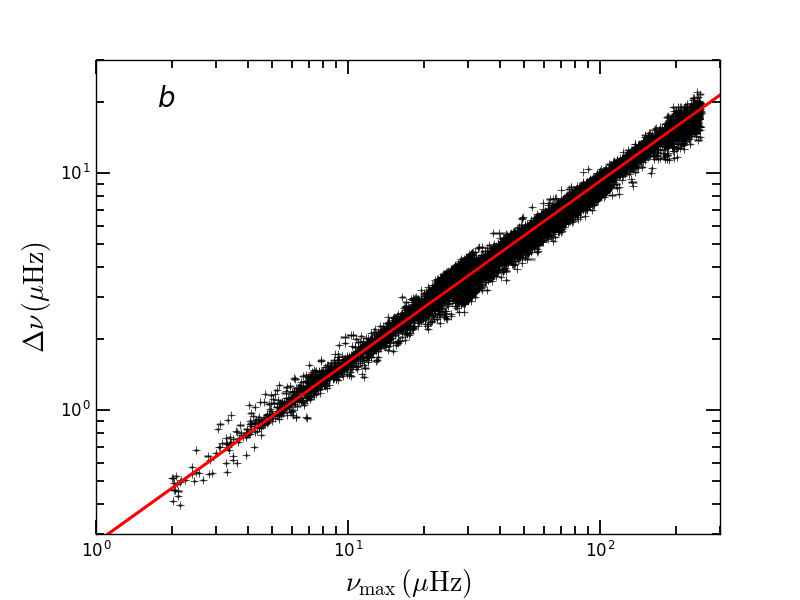}
\includegraphics[scale=0.39, trim=0.37cm 0cm 1.cm 0.7cm, clip=True]{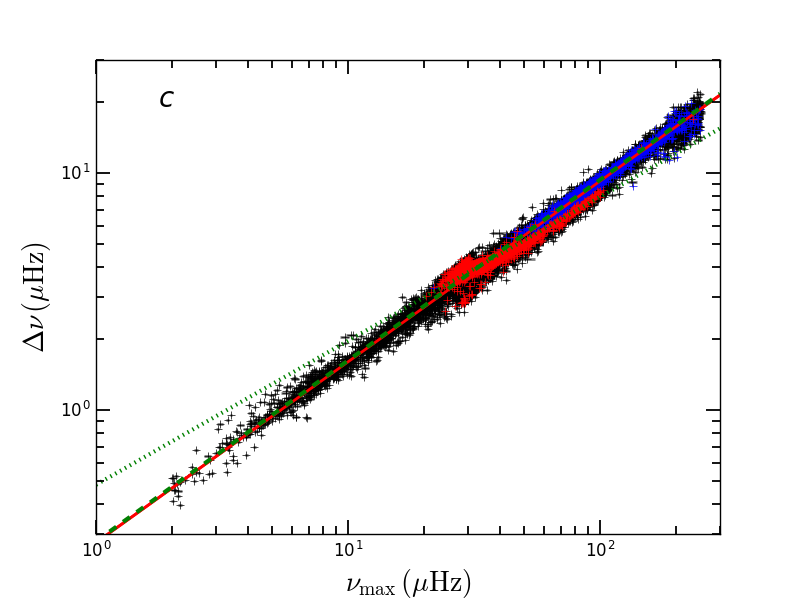}
\includegraphics[scale=0.39, trim=0.37cm 0cm 1.cm 0.7cm, clip=True]{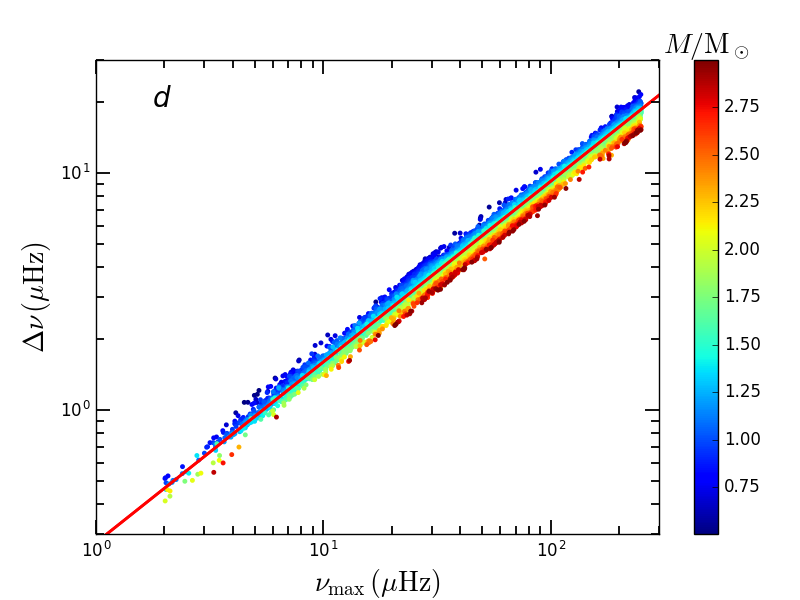}
\caption {Same as Fig. \ref{fig_resultat_Henv} but for the mean large separation $\Dnu$. \label{fig_resultat_Dnu}}
\end{figure*}

For CoRoT and $\kepler$, the dispersion for the parameter $\Dnu$ is quite small (see Fig. \ref{fig_resultat_Dnu}a et b), 
in agreement with the simulations (cf. Sect.~\ref{simulations} and \ref{full_simulations}).  
Interestingly, as for $\Henv$, the information on the evolutionary stage reveals a clear difference between 
the clump (in red) and the RGB (in blue) (cf. Fig. \ref{fig_resultat_Dnu}c).
This may be explained for example by the difference in mass range covered by the RGB and clump stars as we discuss in Sect.~\ref{param_mass}.
Consequently, the $\Dnu$ scaling laws depend slightly on the evolutionary stage. 
To our knowledge, this result has never been observed before. 
Figure \ref{fig_resultat_Dnu}d illustrates the distribution in mass of our sample. We must stress that the apparent gradient in mass is a direct consequence of the scaling law used to estimate masses (Eq.~\ref{eq_masse}). 


\subsubsection{Granulation effective timescale $\taueff$} \label{taueff}

\begin{figure*}
\centering
\includegraphics[scale=0.39, trim=0.37cm 0cm 1.cm 0.7cm, clip=True]{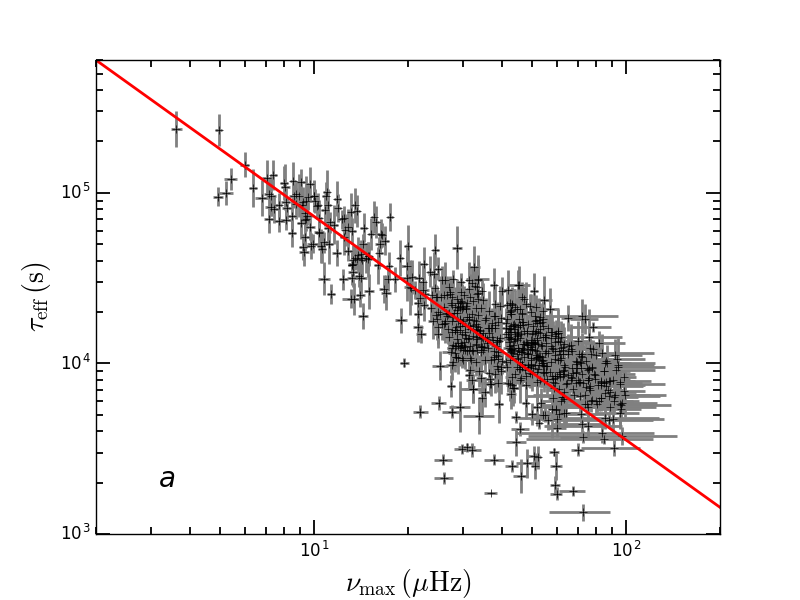}
\includegraphics[scale=0.39, trim=0.37cm 0cm 1.cm 0.7cm, clip=True]{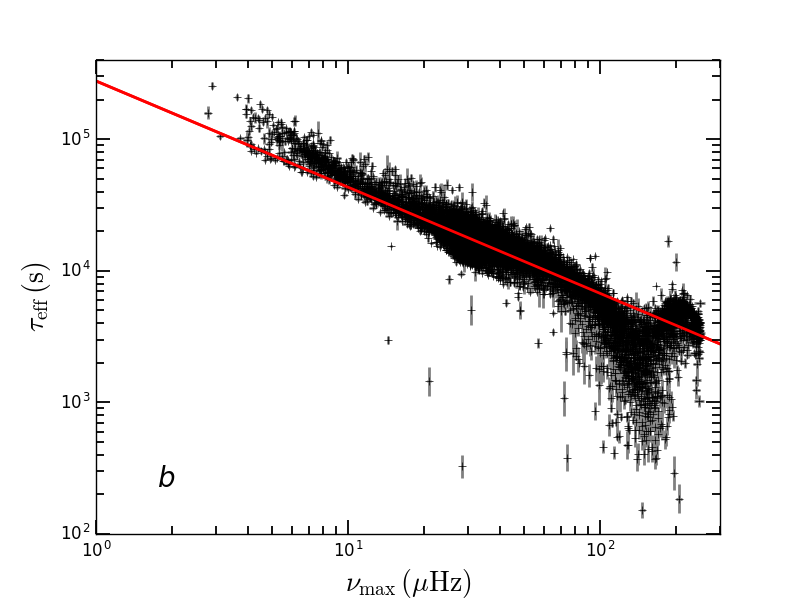}
\includegraphics[scale=0.39, trim=0.37cm 0cm 1.cm 0.7cm, clip=True]{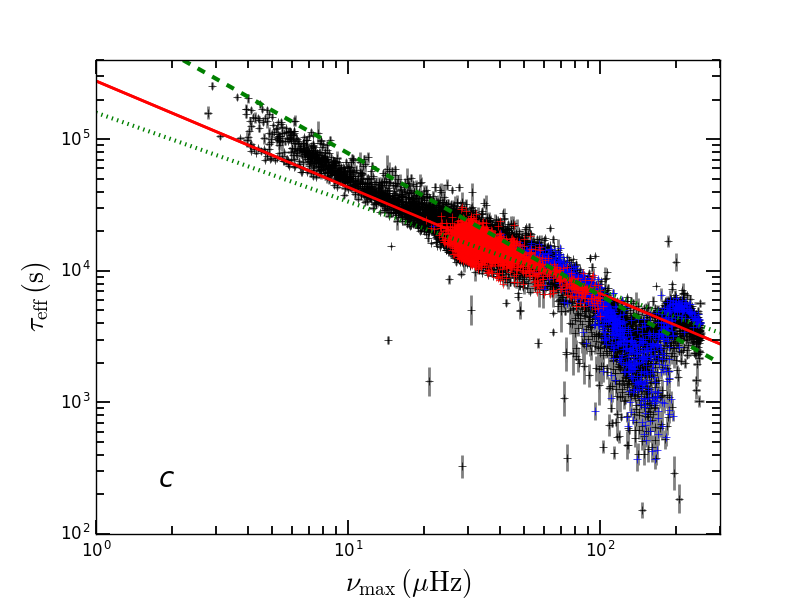}
\includegraphics[scale=0.39, trim=0.37cm 0cm 1.cm 0.7cm, clip=True]{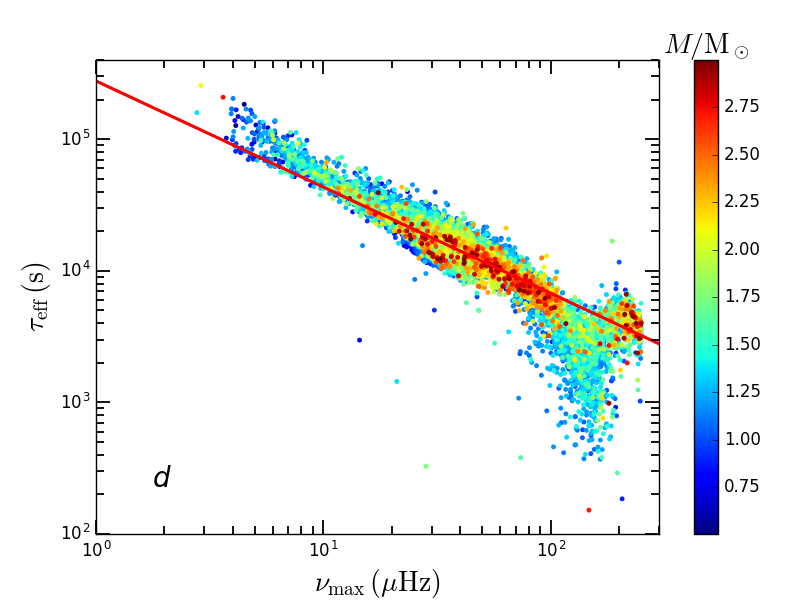}
\caption {Granulation effective timescale $\taueff$ obtained with the CoRoT (\textit{a}) and $\kepler$ (\textit{b}) datasets $S_\mathrm{s+g}$.
Black crosses represent the obtained values, with their error bars in grey.
The red line is the deduced scaling relations. 
In the case of $\kepler$, the  power law representations  were deduced from results with $\numax<100~\mu$Hz. \newline
\textit{c:} Same as figure  \textit{b} with the 1256 RGB stars (in blue) and the 3142 clump stars (in red).
The green dashed line and the dotted one represent respectively the scaling relation obtained with the RGB and clump stars with $\numax<100~\mu$Hz. \newline
\textit{d:} Same as figure  \textit{b} with the stellar mass indicated via the colour code. 
For better visibility of the mass variation, only stars with a mass  in the range $0.5 \leqslant M/\mathrm{M_\odot} \leqslant 3.0$ are plotted.}
\label{fig_resultat_taueff}
\end{figure*}

Results obtained for the parameter $\taueff$ 
have a larger dispersion in the case of CoRoT (cf. Fig. \ref{fig_resultat_taueff}a) than in the case of $\kepler$.
As suggested by the simulations (cf. Sect.~\ref{simulations} and \ref{full_simulations}), this is mainly due to shorter observation time $T$ in the case of CoRoT.
Considering the $\kepler$ results, one can see in Fig. \ref{fig_resultat_taueff}b two characteristic structures.
The first one is in the interval $20<\numax<100~\mu$Hz, where one can see a greater dispersion for $\taueff$.
Figure \ref{fig_resultat_taueff}c reveals that this broadening is due to the clump component  
{ even if   the two components  almost fully overlap each other, 
thereby emphasizing that $\taueff$ depends poorly on the evolutionary stage.}
This is not surprising since it depends essentially on the surface parameters of the star. 
The second structure is in the interval $100<\numax<200~\mu$Hz, where values of $\taueff$ are significantly below the scaling relation.
This feature is explained by the values taken by the slope $\agran$ of the Lorentzian function which fits the granulation 
(see Sect. \ref{resultats_agran} for more details).

\subsubsection{Granulation characteristic amplitude $\sigma_\mathrm{g}^2$} \label{sigma2}

\begin{figure*}
\centering
\includegraphics[scale=0.39, trim=0.37cm 0cm 1.cm 0.7cm, clip=True]{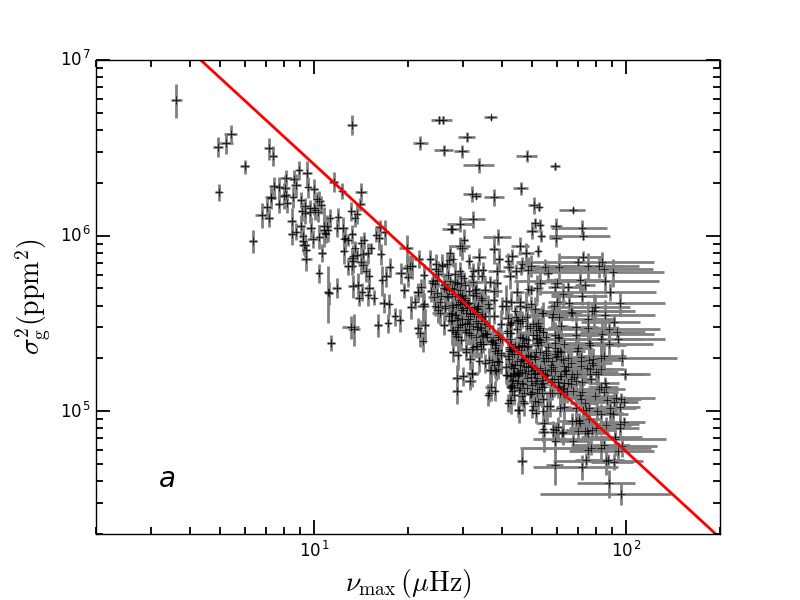}
\includegraphics[scale=0.39, trim=0.37cm 0cm 1.cm 0.7cm, clip=True]{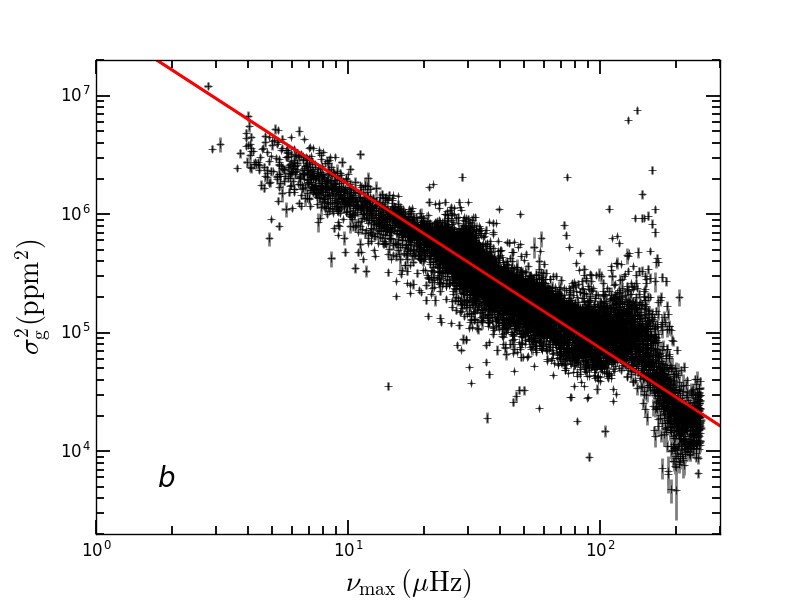}
\includegraphics[scale=0.39, trim=0.37cm 0cm 1.cm 0.7cm, clip=True]{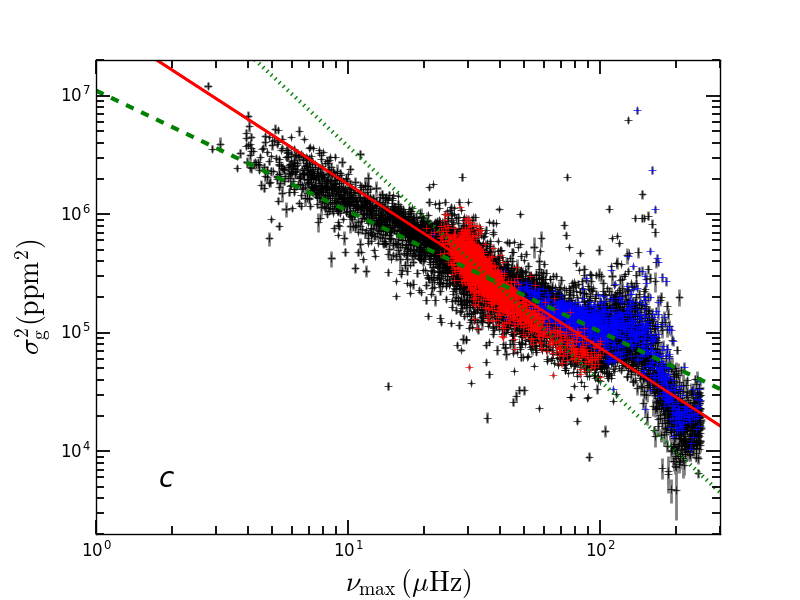}
\includegraphics[scale=0.39, trim=0.37cm 0cm 1.cm 0.7cm, clip=True]{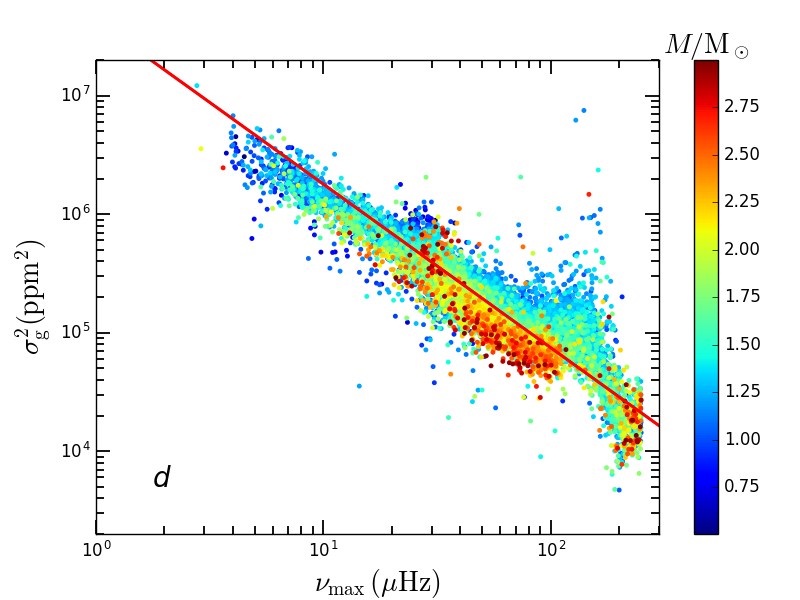}
\caption{Same as Fig. \ref{fig_resultat_taueff}, but for the granulation characteristic amplitude $\sigma_\mathrm{g}^2$.  }
\label{fig_resultat_sigma2}
\end{figure*}

\begin{figure}
\centering
\includegraphics[scale=0.38, trim=0cm 0cm 1.5cm 1cm, clip=True]{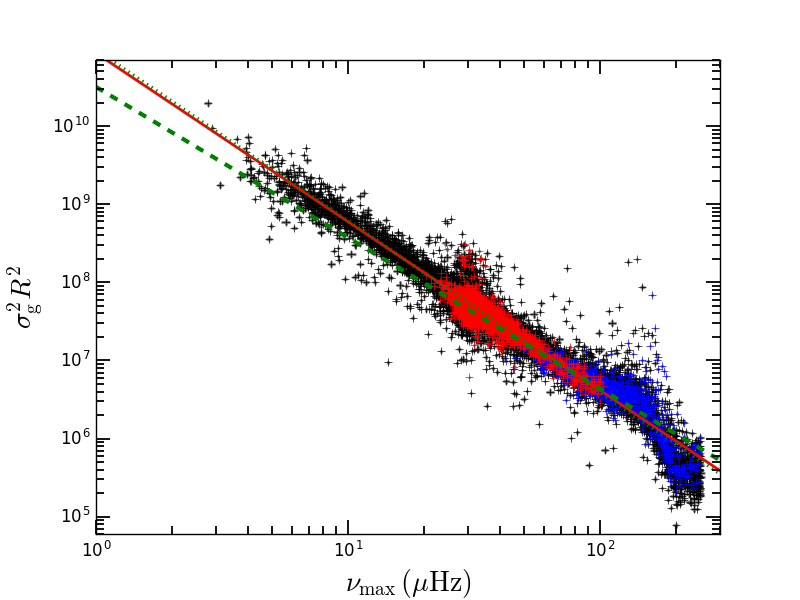}
\caption {Granulation characteristic amplitude $\sigma_\mathrm{g}^2$ multiplied by the seismic radius $R^2$ as a function of $\numax$ 
obtained with the $\kepler$ dataset $S_\mathrm{s+g}$.
Red crosses represent clump stars and the blue ones, RGB stars. 
The red line is the scaling relation determined with a least-square fit considering the internal errors on both axes.} 
   \label{fig_sigmaR_numax}
\end{figure}


Overall, results for $\sigma_\mathrm{g}^2$ are comparable for the CoRoT and $\kepler$ datasets (cf. figure \ref{fig_resultat_sigma2}a and b), 
with a stronger dispersion for CoRoT as expected.
In the case of $\kepler$ results, structures can once again be distinguished. 
The structure located in the interval $100<\numax<200~\mu$Hz is associated with the behaviour of the slope $\agran$ (see Sect. \ref{resultats_agran}) as was the case for $\taueff$.
The bulge from $\numax \simeq 30$ to $\simeq100~\mu$Hz is associated with the clump component as revealed in figure \ref{fig_resultat_sigma2}c 
by  the evolutionary stage as well as in figure \ref{fig_resultat_sigma2}d with the mass distribution.
As for the parameters $\Henv$ and $\Dnu$, we find that $\sigma_\mathrm{g}^2$ depends significantly on the evolutionary stage 
{(cf. Fig. \ref{fig_resultat_sigma2}c and d).} 
This dependency of $\sigma_\mathrm{g}^2$ with the evolutionary stage has already been observed by \cite{2012A&A...544A..90H} on a much smaller sample.

\subsubsection{Granulation slope $\agran$} \label{resultats_agran}

\begin{figure*}
\centering
\includegraphics[scale=0.39, trim=0.37cm 0cm 1.cm 0.7cm, clip=True]{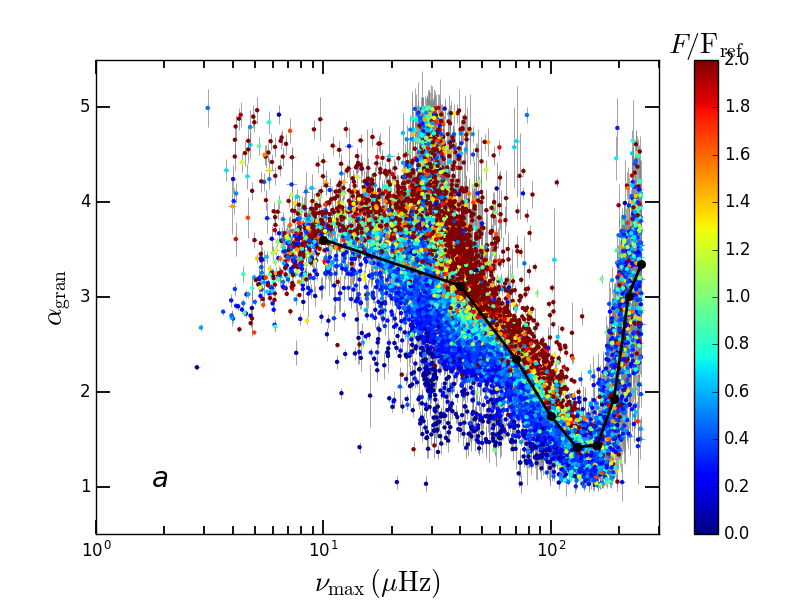}
\includegraphics[scale=0.5, trim=0.37cm 0cm 1.cm 0.7cm, clip=True]{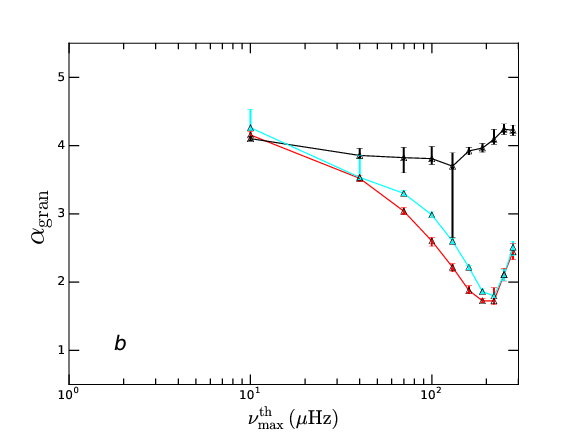}
\caption{\textit{a:} Granulation slope  $\agran$ obtained with the $\kepler$ dataset $S_\mathrm{s+g}$.
The colour code represents the stellar flux normalised by the flux of a 12.5   magnitude of  star and the error bars are in grey.
Black dots are the median of $\agran$ results in a box of $30~\mu$Hz around the corresponding $\numax$ of the simulations. \newline
\textit{b:} Results of the $\kepler$-type simulations for RGB stars.
Triangles represent the median of $\agran$, error bars are the $\pm 1 \sigma$  dispersion  and colours indicate different simulation parameters.
Red corresponds to the simulations described in Section \ref{simulations_parameters},
 cyan represents simulations with a magnitude 8.0 (instead of 12.0) and  black corresponds to the simulations with only one granulation component. }
   \label{fig_resultat_agran}
\end{figure*}

The distribution of the results obtained with the $\kepler$ dataset (Fig. \ref{fig_resultat_agran}a) 
is qualitatively consistent with the simulations (red line in Fig. \ref{fig_resultat_agran}b).
$\agran$ is about 4 at $\numax=10~\mu$Hz, then it decreases until $\numax \simeq 150-200~\mu$Hz after which it increases again.  
This pattern, observed in both observations and simulations, is an artefact of our method which uses one component to fit the granulation.
As shown by the simulations, the decrease of $\agran$ values can be explained by the difference of lifetime of both granulation components with $\numax$.
This is corroborated by the almost flat curve of the simulations with only one component (black line in Fig. \ref{fig_resultat_agran}b).
Indeed, the more $\numax$ increases, the more the two components move away from each other. 
Therefore, since we adjust only one granulation component, $\agran$ needs to take increasingly low values 
in order to correctly take into account the whole contribution of both components. 
Consequently, $\agran$ can take very low values in the interval $100<\numax<200~\mu$Hz (Fig. \ref{fig_resultat_agran}a),
corresponding physically to an artificially too rapid decrease of the granulation coherence time, 
and thus, induces the low values of $\taueff$ and the high values of $\sigma_\mathrm{g}^2$ observed in sections \ref{taueff} and \ref{sigma2}.

There is also a clear correlation between $\agran$ and the stellar flux 
showing that brighter stars systematically have  lower $\agran$ values in the interval $10\leqslant\numax\leqslant150-200~\mu$Hz, 
as it can be seen in Figure \ref{fig_resultat_agran}a for the observations and in Figure \ref{fig_resultat_agran}b for the simulations (cyan line).
This is due to the fact that the fainter the star is, the higher the noise will be and consequently, the flatter the Lorentzian function giving low $\agran$ will be. 

Finally, from $\numax \simeq 150-200~\mu$Hz, $\agran$ increases (Fig. \ref{fig_resultat_agran}a and b) because of the Nyquist frequency which causes a degeneracy of the fit, thus causing an overestimate 
of the white noise and an underestimate of the power of the granulation at high frequencies.
For the observations, $\agran$ reaches higher values than for simulations because of a broader stellar magnitude range. 

\subsection{Comparison with others published studies} \label{comparaison_methods}

In this section, we first compare the results of our analysis with those available for the same stars in the literature.
Then, we use power law representations to compare the general trends associated with the whole set of stars we analysed.
This comparison is made with the power laws obtained with various methods on different sets of stars published in the litterature. Additionaly, we compare results obtained on CoRoT and $\kepler$ data sets.

\subsubsection{Star by star comparison} \label{starbystar_comparaison}

We compared our $\kepler$ results with those available in the APOKASC catalogue \citep{2014ApJS..215...19P}.
These seismic indices were obtained with the COR method \citep{2010MNRAS.402.2049H}. 
For the 1850 stars in common, we made a one to one comparison between our $\Dnu$ and $\numax$ results and APOKASC values.
In case of $\Dnu$,  our results are within the internal errors. 
However, it can be noticed that uncertainties given by \cite{2014ApJS..215...19P} have been estimated by adding 
in quadrature the formal uncertainty returned by the \cite{2010MNRAS.402.2049H}'s method to the standard deviation of the values returned by all methods.
They are thus already representative of the dispersion between various methods.

In the case of $\numax$, we found a small but significant bias. 
It is negligible at low frequencies ($\numax \sim 10~\mu$Hz) but can reach about 10\% at high frequencies ($\numax > 200~\mu$Hz). 
We attribute this difference to the fact that in our approach we do not smooth the power spectrum in order to preserve the oscillation modes heights 
and the stellar background shape.\\

We also compared our CoRoT results with those derived by \cite{2011A&A...525L...9M}. There are 352 common stars. 
The comparison shows that our results are on the whole  consistent with \cite{2011A&A...525L...9M} values. 
However, the measurements  show a larger dispersion than with the APOKASC data, with higher outliers which cannot be explained by the internal errors.
This larger dispersion can probably be explained by the fact that the analysis performed by \cite{2011A&A...525L...9M} 
were based on the first CoRoT data processing pipeline which has been substantially improved since then \citep{2016cole.book...41O}.
For $\Dnu$, this comparison shows that there is no significant bias between our results and those derived by \cite{2011A&A...525L...9M}. 
For $\numax$, we found that the dispersion can be explained by the internal errors. 
However, a small bias increasing with the frequency is observed and is of the same order as the one  observed with the APOKASC data.

\subsubsection{Power law representations} \label{scaling_relation_discution}

Another way to compare our results with results found in the literature is to consider power law representations of our results as a function of $\numax$. 

We adjusted each stellar index dataset by a power law of the form $y = \alpha (\numax) ^\beta$  using a least-square fit 
while taking into account error bars on the axes ($y$ and $\numax$). We obtained the fits for both CoRoT and $\kepler$ datasets. 
Values of the adjusted coefficients $\alpha$ and $\beta$ are reported in Table\ \ref{relation_echelle_resultats} 
together with  the reduced $\chi^2$ values obtained for  each fit.

The reduced $\chi^2$ values are systematically significantly larger than three. 
This means that the difference between the observations and the power law are statistically 
significant (with a probability higher than $99\%$) and hence cannot be explained by the noise. 
Indeed, the measured indices exhibit complex structures that explain the important departure from those power laws.
This highlights the fact that such representations  do not fully  represent the diversity of the red giant  sample. Accordingly, one must pay attention that the comparison made in terms of power laws  should in principle be made with the  same sample of stars, which is in practice difficult or often not possible.  However, in some cases, such comparisons do reveal large differences which must be attributed to differences in the analysis methods as  will be shown hereafter.  

The power laws derived from the CoRoT data are all found to significantly depart from those derived from the $\kepler$ data. 
These departures may be explained by differences in the instrumental response function of both instruments (e.g. bandwidths) but more likely from the fact that the observed population of star are not exactly the same \citep[also see][]{2010ApJ...723.1607H}.  
This latter hypothesis is supported by the fact that the clump and the RGB stars exhibit very different structures. 

We now turn to the power law representations published in the literature (cf. Tab. \ref{relation_echelle_ref} for some references). 
Several studies have focussed on the $\Dnu-\numax$ relation 
on $\kepler$ data \citep[e.g.][]{2010ApJ...723.1607H,2012A&A...537A..30M} 
and some for CoRoT data \citep[e.g.][]{2010A&A...517A..22M, 2009A&A...506..465H, 2009MNRAS.400L..80S}.
The $\Dnu-\numax$   power law derived from  $\kepler$ data set is consistent with the one by  \cite{2012A&A...537A..30M}, 
which was determined by an average of power law obtained by various methods. 
Concerning the $\sigma_\mathrm{g}^2-\numax$ and the $\taueff-\numax$ power law representations,
our $\kepler$  power law and those derived by \cite{2014A&A...570A..41K} have similar slopes $\beta$. However, considering error bars, the difference remains statistically significant. 
Furthermore, our power laws for $\Henv$ strongly (and significantly) differ from the one deduced by \cite{2012A&A...537A..30M}. 
Nonetheless, \cite{2012A&A...537A..30M} do not measure exactly the same quantity as we do. 
Indeed, unlike our method,  these authors smooth the spectrum before fitting it by a Gaussian function.

Except for $\taueff$, all the stellar indices result in power law representation significantly different between the clump and the RGB stars. 
The values of the  $\alpha$ and $\beta$ coefficients are given in table \ref{relation_echelle_resultats} for both evolutionary stages. 
 This shows again how sensitive are the coefficients of the power law w.r.t.  stellar samples with different proportions of RGB and clump stars. Thus part of the differences with the literature found here  are necessarily due to difference in the stellar samples.

For the $\Dnu-\numax$ power law,  the difference between the clump and the RGB sets can either be due to difference in the mass distribution between  the two populations or to  big differences in the core structures between clump and RGB stars (cf. Sect. \ref{param_mass}). 
For  $\sigma_\mathrm{g}^2-\numax$ relation, the difference between the  clump and  RGB stars is essentially due to a radius dependence of $\sigma_\mathrm{g}^2$. 
Indeed, \cite{2006A&A...445..661L} showed that $\sigma_\mathrm{g}^2 \propto 1/R^2$. 
Our results confirm this power law as shown in figure \ref{fig_sigmaR_numax}, where values of $\sigma_\mathrm{g}^2 R^2$ 
for RGB and clump stars follow very similar power laws.
($\beta_\mathrm{clump}=-2.187 \pm 0.002$ and $\beta_\mathrm{RGB}=-1.929 \pm 0.004$).
Based on all the stars with $\numax<100~\mu$Hz of the dataset $S_\mathrm{s+g}$, 
we deduced the following power law while taking into account the internal errors on both axes
\begin{equation} \label{eq_sigmaR_rayon}
 \sigma_\mathrm{g}^2 R^2 = (8.59 \pm 0.02)10^{10}~\numax^{-2.1574 \pm 0.0006}  \; .
\end{equation} 
Finally, concerning the $\Henv-\numax$ relation, the dependence on  evolutionary stage is also found to be statically significant but has not yet been explained.

\subsection{Stellar parameters inferred from seismic indices} \label{param_fondamentaux}

 In order to illustrate the results obtained with this large sample of stars, we estimated the stellar seismic masses, radii and luminosities via  equations (\ref{eq_masse}), (\ref{eq_rayon}) and (\ref{eq_luminosite}) respectively, using the seismic indices from the $\kepler$ dataset $S_\mathrm{s}$ and the effective temperatures from different catalogs 
(see Sect. \ref{scaling_relations} for more details).

In the following sub-sections, we comment on  these results in the light of  stellar evolution theory.

\begin{figure*}
\centering
\includegraphics[scale=0.39, trim=0.37cm 0cm 1.cm 0.7cm, clip=True]{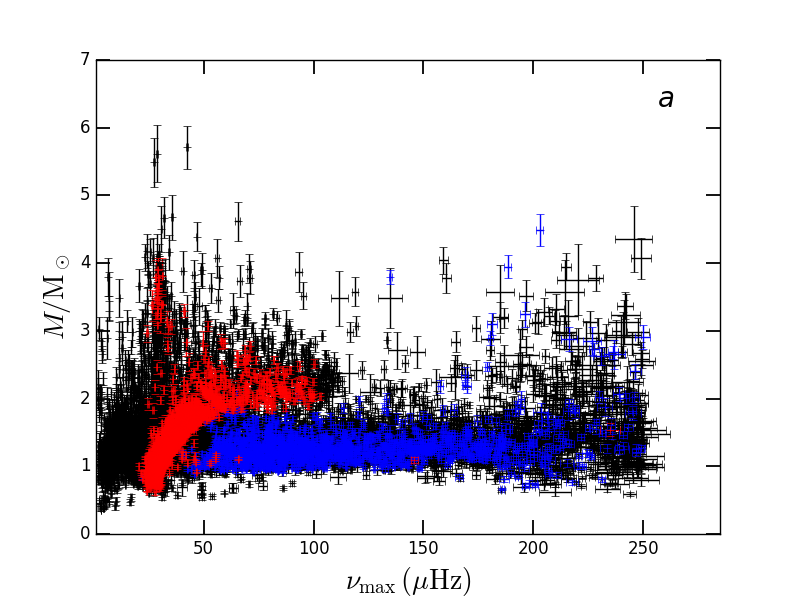}
\includegraphics[scale=0.39, trim=0.37cm 0cm 1.cm 0.7cm, clip=True]{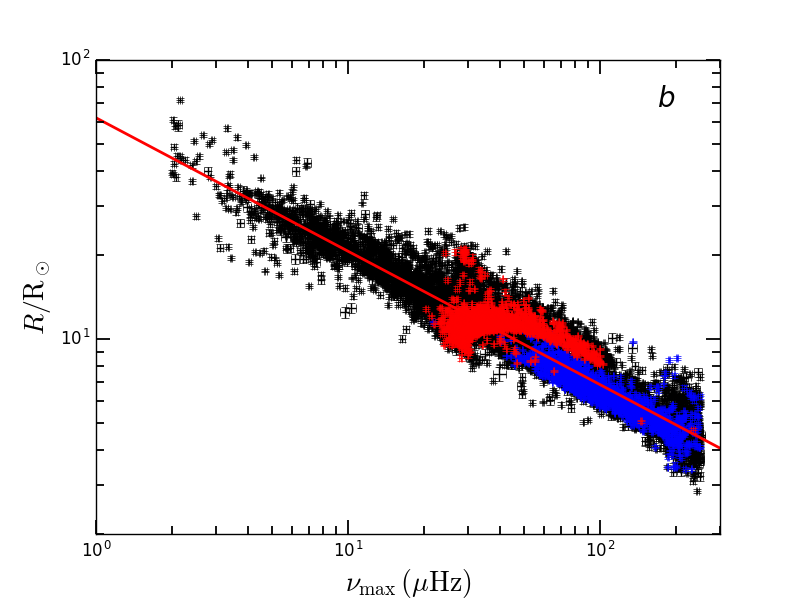}
\caption {Mass (\textit{a}) and radius (\textit{b}) distribution as a function of $\numax$ obtained with the $\kepler$ dataset $S_\mathrm{s}$.
Red crosses represent clump stars and the blue ones, RGB stars. 
The red line (right panel) is the scaling relation determined with a least-square fit considering the internal errors on both axes.}
   \label{fig_resultat_M_R}
   
   
\includegraphics[scale=0.39, trim=0.37cm 0cm 1.cm 0.7cm, clip=True]{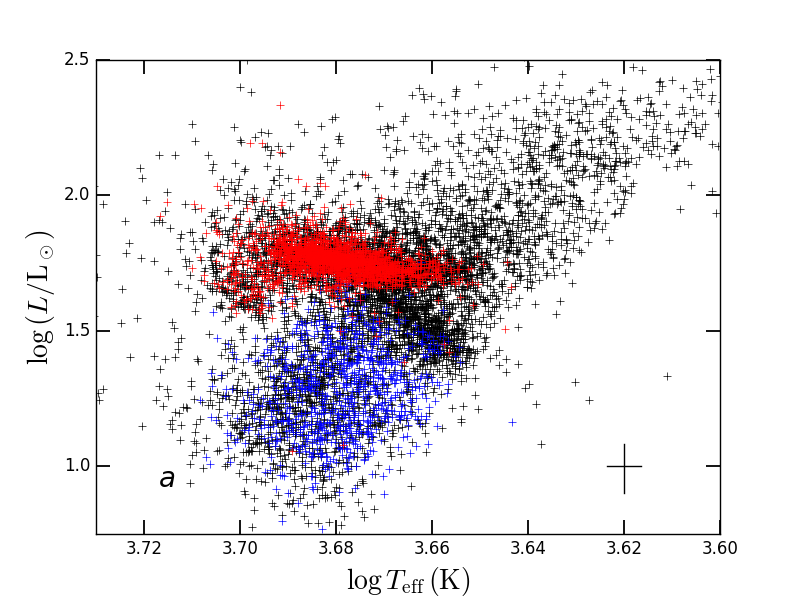}
\includegraphics[scale=0.39, trim=0.37cm 0cm 1.cm 0.7cm, clip=True]{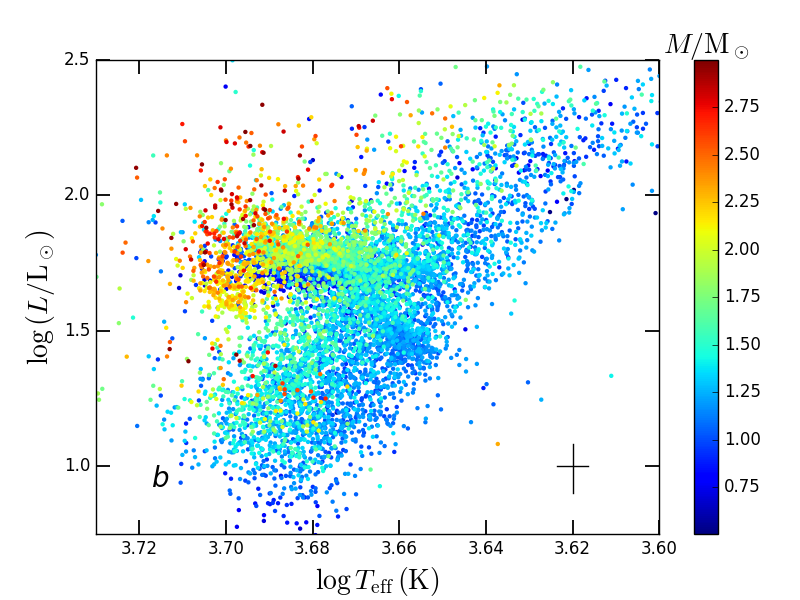}
\caption{Hertzsprung--Russell diagram with the information on the evolutionary stage (\textit{a}) and the mass (\textit{b}) 
for the $\kepler$ dataset $S_\mathrm{s}$.\newline
\textit{a:} Red crosses represent clump stars and the blue ones, RGB stars. \newline
\textit{b:} The colour code represents the stellar mass.
For better visibility of the mass variation, only stars with a mass included in the range $0.5 \leqslant M/\mathrm{M_\odot} \leqslant 3.0$ are represented. 
The black cross indicate the median 1-$\sigma$ error bars in both axes. In order to better reveal the fine structures, we only plotted stars for which the effective temperature was determined by the spectroscopy surveys APOGEE and LAMOST.}
 \label{fig_HR_diagram}
\end{figure*}

\subsubsection{Mass distribution} \label{param_mass}

Figure \ref{fig_resultat_M_R}a presents the mass distribution as a function of $\numax$. 
We can clearly see that  RGB stars (in blue) mainly have  masses between $\sim1$ and $2~\mathrm{M_\odot}$. 
Indeed, less massive stars are non-existent on the RGB because their lifetimes on the main sequence exceed the age of the Universe, while more-massive stars evolve faster and follow different evolutionary tracks.
Clump stars (in red) cover a wider mass range than RGB stars, from $\sim0.5$ to $3.0~\mathrm{M_\odot}$.
Clump stars with low masses ($<1~\mathrm{M_\odot}$) are stars which have travelled all along the RGB until the tip where they suffer strong mass loss.
Masses higher than $\sim 2~\mathrm{M_\odot}$ belong to the so-called secondary clump \citep{1999MNRAS.308..818G}, close to the primary clump in the HR diagram.
Typically, these stars have switched to the helium reactions in a non-degenerate core.\\


In this context, the different slopes found for  $\Dnu$-$\numax$ relation between clump  and  RGB stars in section \ref{Dnu} can have several causes. On one hand, as already commented, this can be due to difference in mass distribution between the two samples. 
However, to demonstrate this argument, one would require independent mass measurements. 
The different slopes could also be explained by the very different structures of clump and RGB stars. 
Indeed, scaling relations assume homologous structures \citep[see e.g.][]{2013ASPC..479...61B}.  Due to big differences in the core structures between clump and RGB stars, we may expect a difference in scaling relations, as suggested by \citet{2012MNRAS.419.2077M}.

\subsubsection{Radius distribution} \label{param_rayon}

In the radius-$\numax$ diagram (Fig. \ref{fig_resultat_M_R}b),  clump stars exhibit a small radius range.
This is consistent with the fact that they have a rather constant radius during this evolutionary stage while RGB stars experience a strong increase 
of their radius during their ascent of the RGB up to the tip. \\

There is a strong correlation between the radius and $\numax$. 
The scaling relation obtained from the $\kepler$ dataset $S_\mathrm{s}$ (cf. Tab. \ref{relation_echelle_resultats})
is consistent with the value measured by \cite{2010A&A...517A..22M} and the theoretical value \citep[][c.f. Tab.~\ref{relation_echelle_ref}]{2010A&A...517A..22M}.

\subsubsection{Hertzsprung--Russell diagram} \label{param_HR}

Figure \ref{fig_HR_diagram} shows the location of our sample in the Hertzsprung--Russell (HR) diagram.
On the right and left panels, one can recognize the red giant branch from $\log(\Teff)~\sim~3.6$ to $\sim~3.75$ K. 
From classical observing techniques, it is not possible to distinguish the stars belonging to the red clump 
from those ascending the red giant branch 
because of their overlap in luminosity and temperature. 
However, this is possible thanks to the seismic constraints related to the physical conditions in the stellar cores \citep[e.g.][cf. Sect. \ref{scaling_relations}]{2016A&A...588A..87V}.
In Fig. \ref{fig_HR_diagram}a, all stars identified as RGB stars (in blue) are at the bottom of the RGB, corresponding to relatively young red-giant stars. 
Regarding stars identified as clump stars (in red), their position is consistent with the red clump in the HR diagram.
However, some of them are above ($\log(L/\mathrm{L_\odot}) > 2.1$). Those stars are probably leaving the red clump on their way to the asymptotic giant branch (AGB). 
Another feature is highlighted by  the figure \ref{fig_HR_diagram}a: the presence of the bump as predicted by models of  stellar evolution \citep[e.g.][]{2012A&A...543A.108L} 
can be seen as an over-density of stars on  the RGB below the clump around $\log L \sim 1.5$. 
Note also that a similar signature of the presence of the bump  and secondary clump has  recently been revealed  by \cite{2018A&A...609A.116R} using the Gaia Data Release 1. 

In the figure \ref{fig_HR_diagram}b, 
one notes different extension of the RGB stars in the HR diagram depending on mass. Indeed, low stellar masses extend toward lower temperature and to a less extent lower luminosity than the higher ones. 
This is again in qualitative agreement with what is obtained with models which show 
that the higher mass stars start the RGB at higher luminosity and follow   hotter tracks \citep[e.g.][]{2012A&A...543A.108L}.

The quantitative analysis of these observations is out of the scope of the present study.  
Nevertheless, they  confim that seismic indices coupled with classical observational constraints open new perspective for quantitative comparison with theoretical stellar evolution models.

\section{Conclusion} \label{conclusion}

The method MLEUP developed and described here allows analysing automatically and homogeneously large datasets of light-curves. It extracts  simultaneously 
the fundamental seismic indices of the oscillations ($\Dnu$, $\numax$ and $\Henv$) 
and the parameters characterizing the granulation ($\taueff$ and $\sigma_\mathrm{g}$) of red-giant solar type pulsators.

The performances of MLEUP were first evaluated
using sets of simulated light-curves representative for both evolutionary stages, RGB and clump, and both CoRoT and $\kepler$ observation conditions.
These tests were used to characterize biases on the values  and associated uncertainties obtained with MLEUP (cf. Sect. \ref{simulations} and \ref{full_simulations}). 
These biases were then used to correct the measurements obtained with real data (cf. Tab. \ref{tableau_coefficient_simus}).

We applied MLEUP to all CoRoT data with a duration of observation larger than 50 days and to all long cadence $\kepler$ data.
We successfully extracted the seismic indices for  2943 CoRoT stars and 17,179 $\kepler$ stars, 
increasing significantly the number of $\kepler$ stars known as oscillating red giants.
We were able to extract simultaneously the seismic indices and the granulation parameters for 806 CoRoT stars and 16,303 $\kepler$ stars.
To our knowledge, the number of seismic indices and granulation parameters derived by MLEUP is significantly higher than any previously published analyses.  
Those indices and parameters are available in the \textit{Stellar Seismic Indices} (SSI) database.

For some $\kepler$ targets, we have additional informations. 
The asymptotic period spacing $\Delta\Pi_1$ is available for $\sim 25\%$ of our $\kepler$  datasets \citep{2016A&A...588A..87V}, 
allowing us to distinguish the RGB stars from those of the red-clump.
The effective temperature, taken from a combination of different catalogs, is available for the whole  $\kepler$  dataset  (see Sect. \ref{scaling_relations} for more details).
Thanks to those additional constraints we were able to deduce power law representations for both evolutionary stages individually (cf. Tab. \ref{relation_echelle_resultats}) and 
we estimated the mass, radius and luminosity of numerous stars via scaling relations combining the effective temperature, $\numax$ and $\Dnu$.
Our results firmly establish trends previously observed with less objects, 
such as the dependency of $\Henv$ and $\sigma_\mathrm{g}^2$ with the evolutionary stage \citep{2011A&A...532A..86M, 2012A&A...537A..30M, 2012A&A...544A..90H}. 
We also revealed an other dependency which has never been observed to our knowledge: 
the faster variation of $\Dnu$ with $\numax$ for stars belonging to the secondary clump than for those of the RGB.
Based on theoretical scaling relations, we showed that the dependency with the evolutionary stage in the case of $\sigma_\mathrm{g}^2$ is essentially due to a radius dependence. 
Concerning the $\Henv$ trend, it is not well understood and call for dedicated theoretical studies.   \\

By now, the MLEUP method has been optimized for red-giant stars because CoRoT and $\kepler$ have detected solar type oscillations for several tens of thousands of such object
in comparison to the few hundred main-sequence solar type pulsators.
However, it should be possible to adapt MLEUP to analyse sub-giant and main-sequence stars.
Indeed, the oscillations spectra of those solar type pulsators contain less mixed modes than those of red giants.
Consequently, it should be easier to fit the Universal Pattern for these kind of stars.
Likewise, MLEUP could be adapted to evolutionary stages later than red giants, such as asymptotic giant branch (AGB), with $\numax$ below $1~\mu$Hz.
In this respect, data from OGLE \citep[e.g.][]{2013A&A...559A.137M} offer a great perspective, but the methods will have to be adapted to handle these very long time series with low duty cycles compared to CoRoT or $\kepler$ data.

In the future, TESS \citep[\textit{Transiting Exoplanet Survey Satellite},][]{2015JATIS...1a4003R}
and PLATO \citep[\textit{PLAnetary Transits and Oscillation of stars}][]{2014ExA....38..249R} 
will provide data for a large number of bright main-sequence and sub-giant objects
for which the extraction of seismic indices and granulation parameters will be possible.
Thereby, a method such as MLEUP will be valuable to analyse automatically all these data.

\section*{Acknowledgements}  \label{Acknowledgements}

This paper is based on data from the CoRoT Archive. 
The CoRoT space mission has been developed and operated by the CNES, 
with contributions from Austria, Belgium, Brazil, ESA (RSSD and Science Program), Germany, and Spain.
This paper also includes data collected by the $\kepler$ mission. 
Funding for the $\kepler$ mission was provided by the NASA Science Mission directorate.
The authors acknowledge the entire $\kepler$ and CoRoT team, whose efforts made these results possible. 
We acknowledge financial support from the SPACEInn FP7 project (SPACEInn.eu) and from 
the \textquotedblleft Programme  National  de  Physique Stellaire\textquotedblright\ (PNPS,  INSU,  France) of CNRS/INSU.
We thank Carine Babusiaux and Laura Ruiz-Dern for having provided us spectroscopic measurements 
of effective temperatures together than their expertise in this domain. 
We thank Daniel Reese for improving the text in many places. Finally, we thank the referee for her/his valuable comments" 




\appendix

\section{Detailed presentation of the tests on synthetic spectra} \label{full_simulations}

\begin{figure*}
\centering
\includegraphics[scale=0.38]{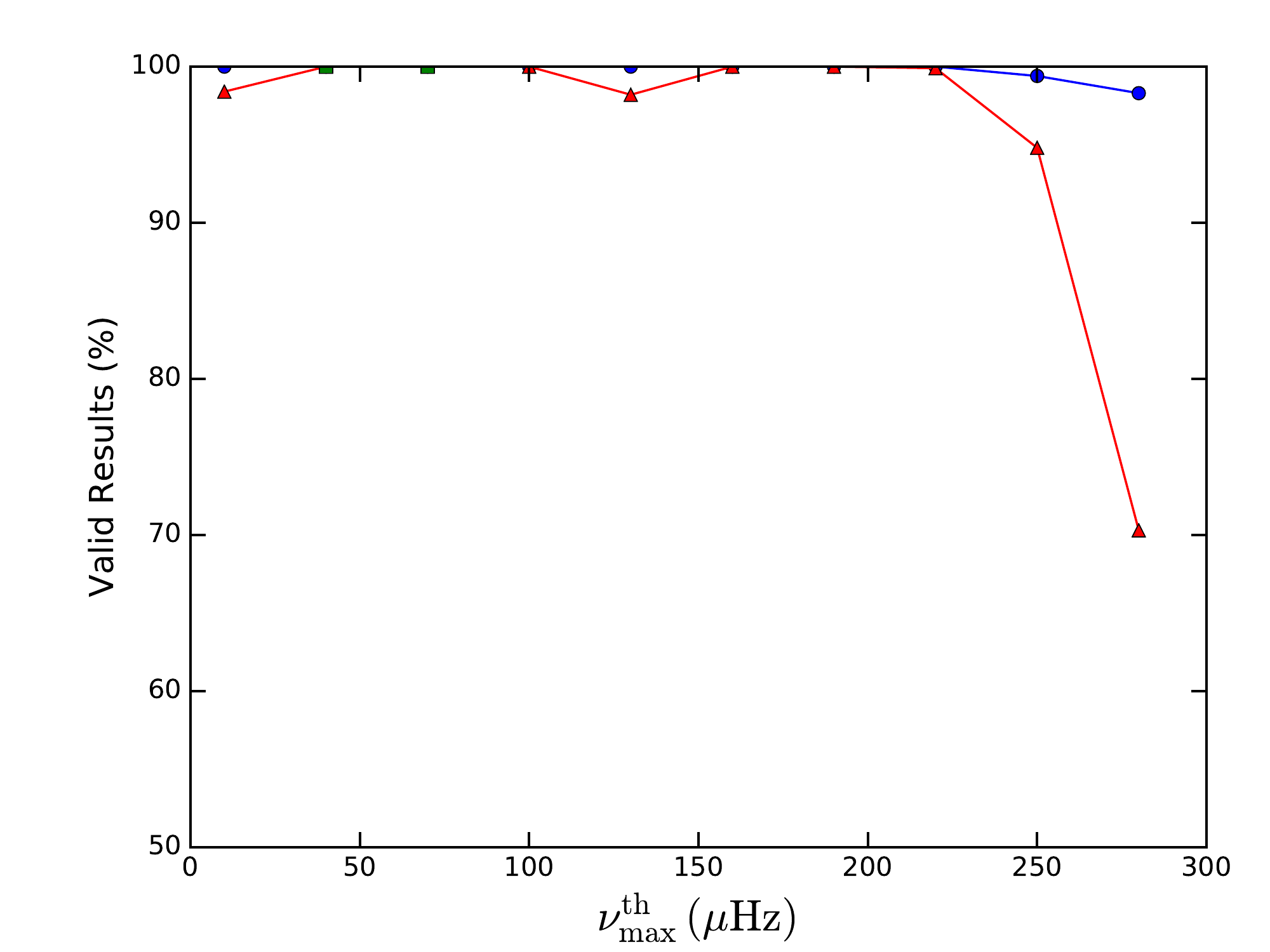}
\includegraphics[scale=0.38]{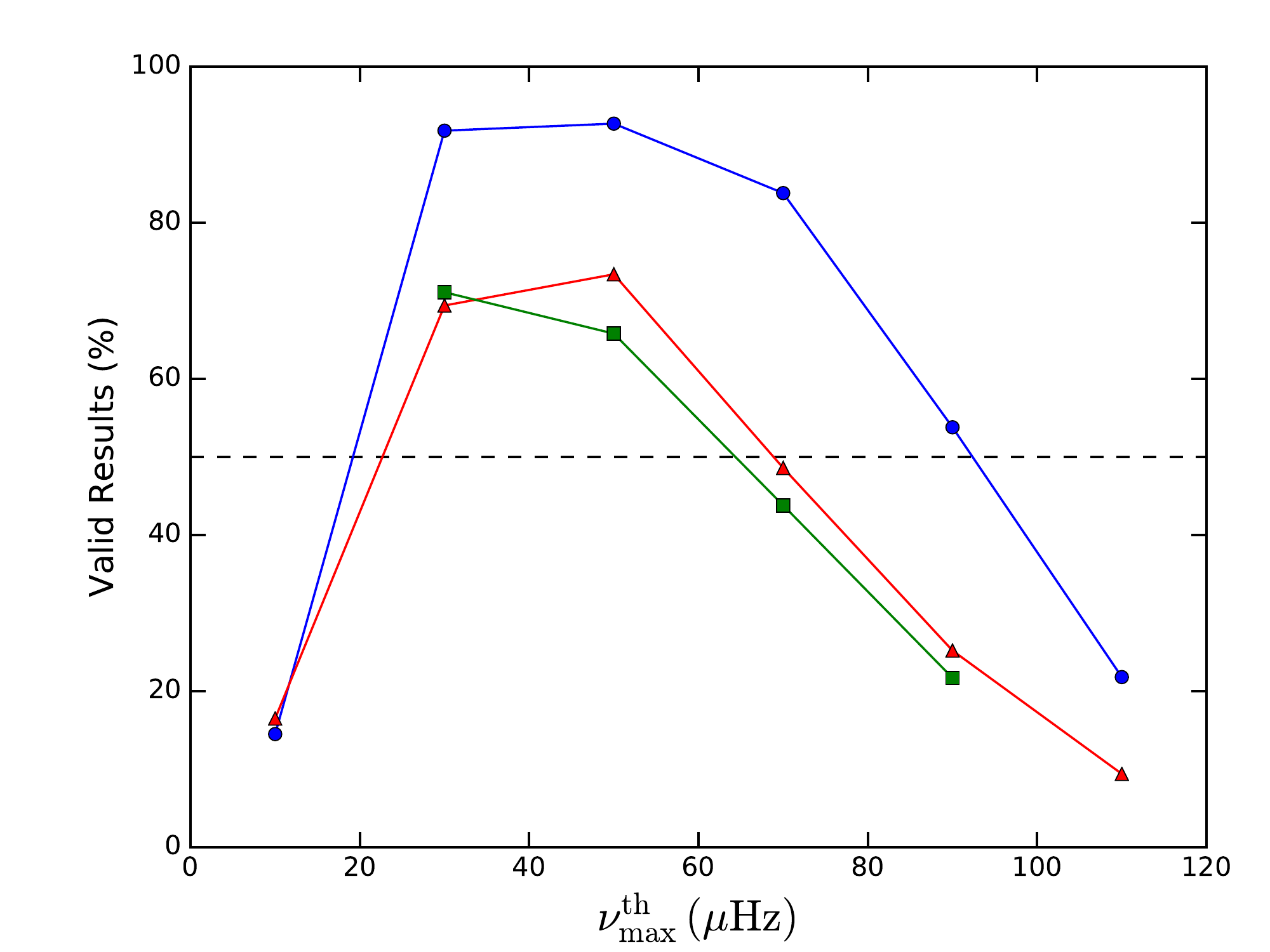}
\caption{Figures representing the percentage of valid results (\textit{VR}) as a function of $\numax$ 
obtained with $\kepler$-type simulations (left) and CoRoT-type simulations (right).\newline
Red triangles correspond to RGB simulations, blue circles to simulations without mixed modes and green squares to clump simulations.}
   \label{fig_MLEUP_VR}
 \end{figure*}

In this section, we present a detailed description of the sets of synthetic light-curves  used to test
the MLEUP method, and we illustrate the statistical analysis of these tests in a series of figures 
(Figures \ref{fig_MLEUP_VR} to \ref{fig_MLEUP_sigma}).

\subsection{The set of synthetic light-curves}

For each instrument CoRoT and $\kepler$, we produced a large number of synthetic light-curves, 
spanning evolution
on the red-giant branch (RGB) and on the clump. Indeed clump and RGB synthetic spectra with
similar $\numax$ values differ only by their mixed mode patterns.
Mixed modes are characterized by two variables: $\Delta \Pi$, the asymptotic gravity-mode period spacing, and $q$, the dimensionless coupling coefficient.
Both are taken from \cite{2016A&A...588A..87V}. 
In the case of RGB stars, $\Delta \Pi$ depends on $\Dnu$ as follows: $\Delta \Pi~=~43~\Dnu^{0.25}$ \citep{2016A&A...588A..87V}.
For clump stars, regardless of $\numax$, $\Delta \Pi$ is fixed to a value (270~s) representative
of the observed distribution 
\citep{2012A&A...540A.143M, 2016A&A...588A..87V}.
Moreover, the $\Dnu$ range for clump stars is limited to $ 3~\mu$Hz$~<~\Dnu~<~9~\mu$Hz, which corresponds roughly to $ 25~\mu$Hz$~<~\numax~<~100~\mu$Hz.

In order to quantify the specific influence of the mixed mode pattern on the MLEUP analysis, 
we also produced synthetic light-curves without mixed modes (only pure pressure modes).

For each instrument considered here, and in each of the above cases, 
sets of light-curves have been produced 
with  typical  observation times  ($T=1000$ days for $\kepler$ and $150$ days for CoRoT), 
sampling rates ($dt=1740$ s and $512$ s, respectively)
 and white noise levels ($2040~\mathrm{ppm.s}^{1/2}$ and $11605~\mathrm{ppm.s}^{1/2}$, respectively)
which correspond to typical star magnitudes ($V=12$ and $V=13$ for $\kepler$ and CoRoT respectively).

Then, in order to take into account the stochastic nature of the observed phenomena and 
to measure the variation between different realizations, called the realization noise, we simulated 1000 synthetic light-curves for each set of parameters.

\subsection{Statistical analysis of the results obtained for synthetic light-curves with MLEUP}

Since no activity was included in simulated light-curves, we do not take it into account in our MLEUP model ($N=1$ in Eq. (\ref{equation_BG+UP})).
Figures \ref{fig_MLEUP_VR} to \ref{fig_MLEUP_sigma} illustrate the results obtained with MLEUP for the CoRoT and $\kepler$-type simulations.\\

\subsubsection{Percentage of valid results (\textit{VR})}

For each set of light-curves, we first inspected the percentage of valid results.
Results are considered valid when they meet the  following two criteria: 
first, fits must be properly converged and secondly, the height of the envelope autocorrelation function $\Amax$ computed during the last step
must be higher or equal to the rejection threshold   $\Alim$, fixed at $\Alim = 8.0$ for a probability $P=99\%$ (cf. 6\textsuperscript{th} step, Sect. \ref{step_7}).

These results are illustrated in Fig. \ref{fig_MLEUP_VR} which, for each satellite and for each 
frequency $\numax$, displays the parameter $VR$, i.e. the percentage of valid results for 1000 realizations.

For $\kepler$ simulations (Fig. \ref{fig_MLEUP_VR} left), we obtain excellent performances, 
with $VR \sim 100\%$ until $\numax=250~\mu$Hz. 
Then, \textit{VR} decreases to $\sim 70\%$ at $\numax=280~\mu$Hz, 
which is very close to the Nyquist frequency ($\nu_{\mathrm{Nyq}} \simeq 287~\mu$Hz).
Mixed modes influence \textit{VR} for $\numax$ higher than $\sim 250~\mu$Hz.

For CoRoT simulations (Fig. \ref{fig_MLEUP_VR} right), \textit{VR} has a bell-shaped.
The difference with $\kepler$ simulations can be explained by the difference in the duration of the observations ($T$) and in
 the signal-to-noise level, 
which are both more favorable for $\kepler$.
This interpretation is supported by the fact that the $H_0$ test ($\Amax \geq \Alim $) is found to be the main 
rejection criteria here.

We get more than 50\% of valid results in the frequency range $20 \lesssim \numax \lesssim 70~\mu$Hz for RGB and clump simulations.
The difference between both evolutionary stages is small, about 5\% lower for clump simulations.
However, the difference is much higher (a factor of 1.3 to 2) compared to simulations without mixed modes.
This result is not surprising since the presence of mixed modes disturbs the regularity of modes and therefore the performance of MLEUP. \\

\subsubsection{Biases and error estimates}   \label{Biases_error_estimates}

\begin{figure*}
\centering
\begin{annotatedFigure}
{\includegraphics[scale=0.35, trim=0cm 0cm 0cm 0cm, clip=True]{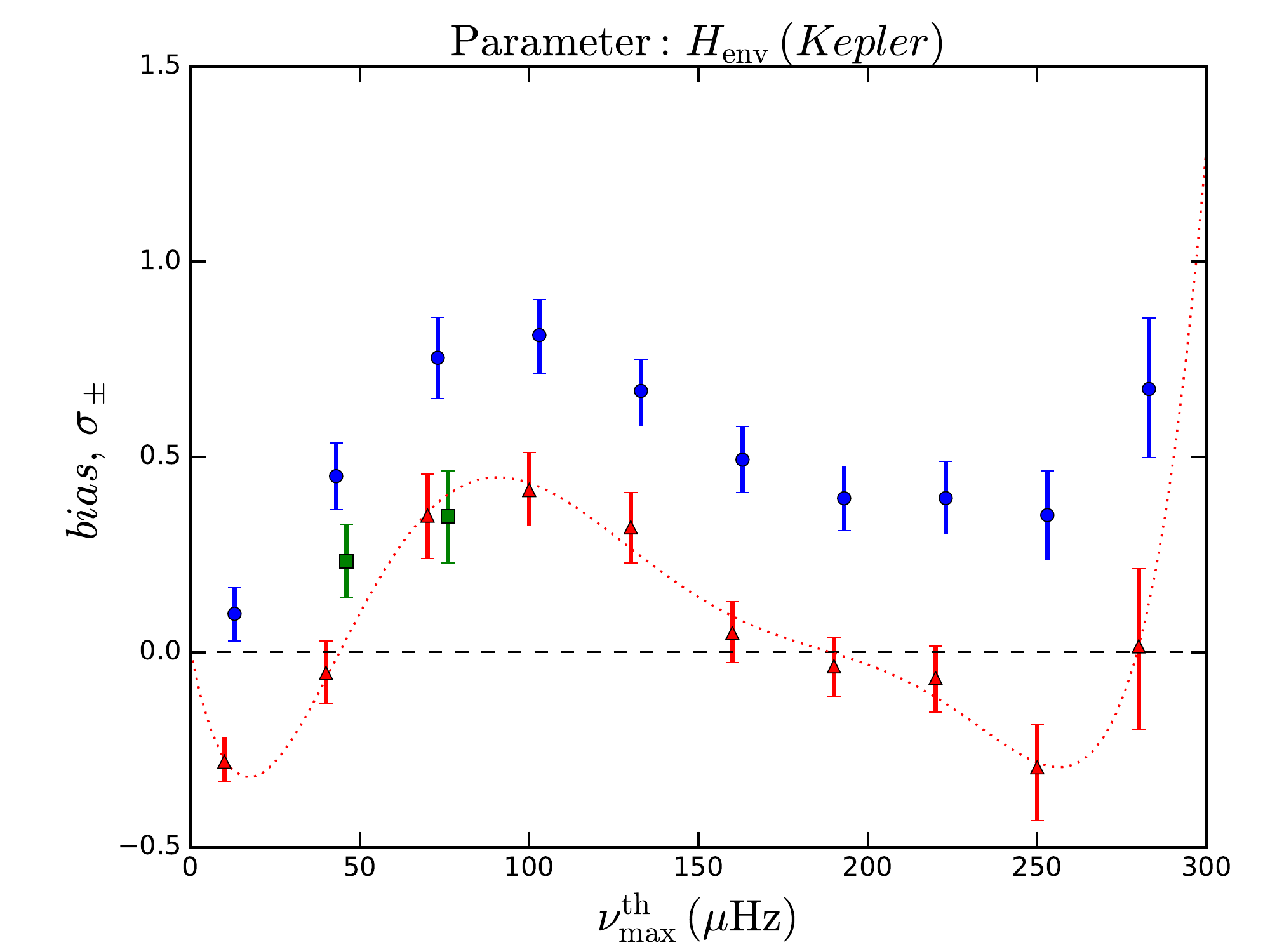}}
	\annotatedFigureBox{-0.02,0.5105}{0.013,0.5107}{a}{-0.04,0.51}
\end{annotatedFigure}
\includegraphics[scale=0.35, trim=0cm 0cm 0cm 0cm, clip=True]{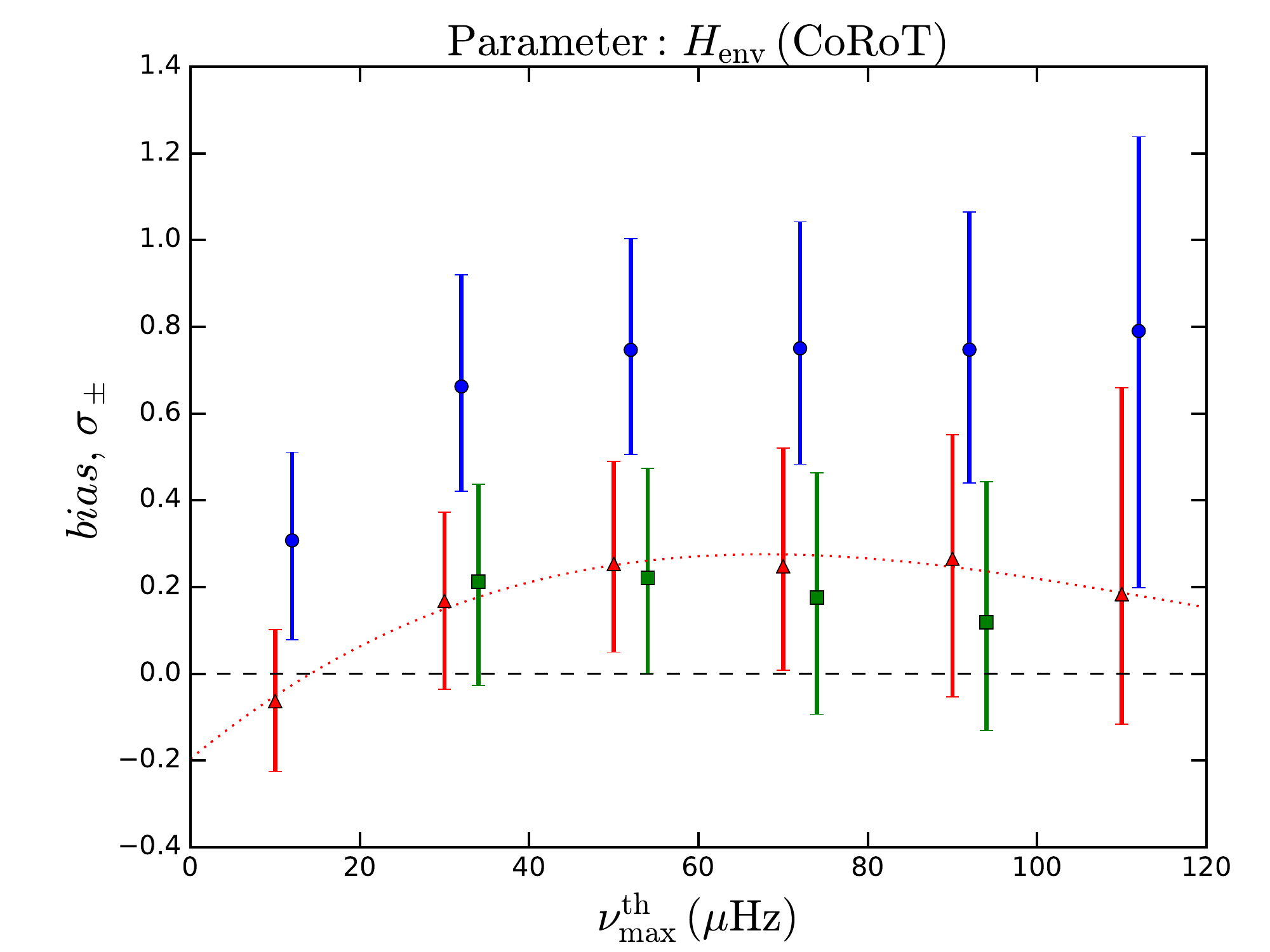}
\begin{annotatedFigure}
	{\includegraphics[scale=0.35, trim=0cm 0cm 0cm 0cm, clip=True]{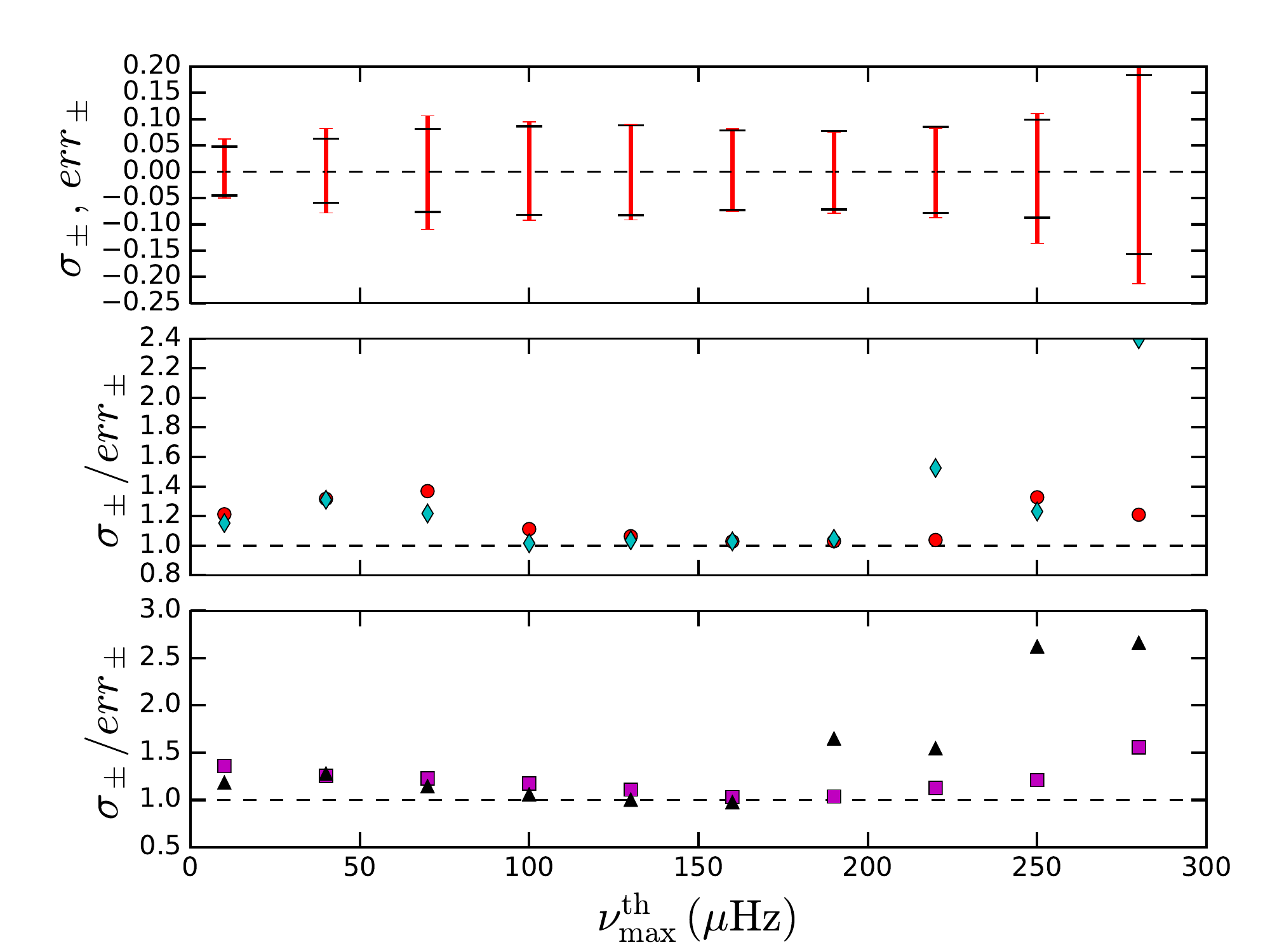}}
	\annotatedFigureBox{-0.03,0.8000}{-0.007,0.7947}{b}{-0.04,0.8}
	\annotatedFigureBox{-0.05,0.5333}{-0.015,0.5333}{c}{-0.04,0.52}
	\annotatedFigureBox{-0.05,0.400}{-0.025,0.200}{d}{-0.04,0.24}{white}{white}{white}{white}
\end{annotatedFigure}
\includegraphics[scale=0.35, trim=0cm 0cm 0.5cm 0cm, clip=True]{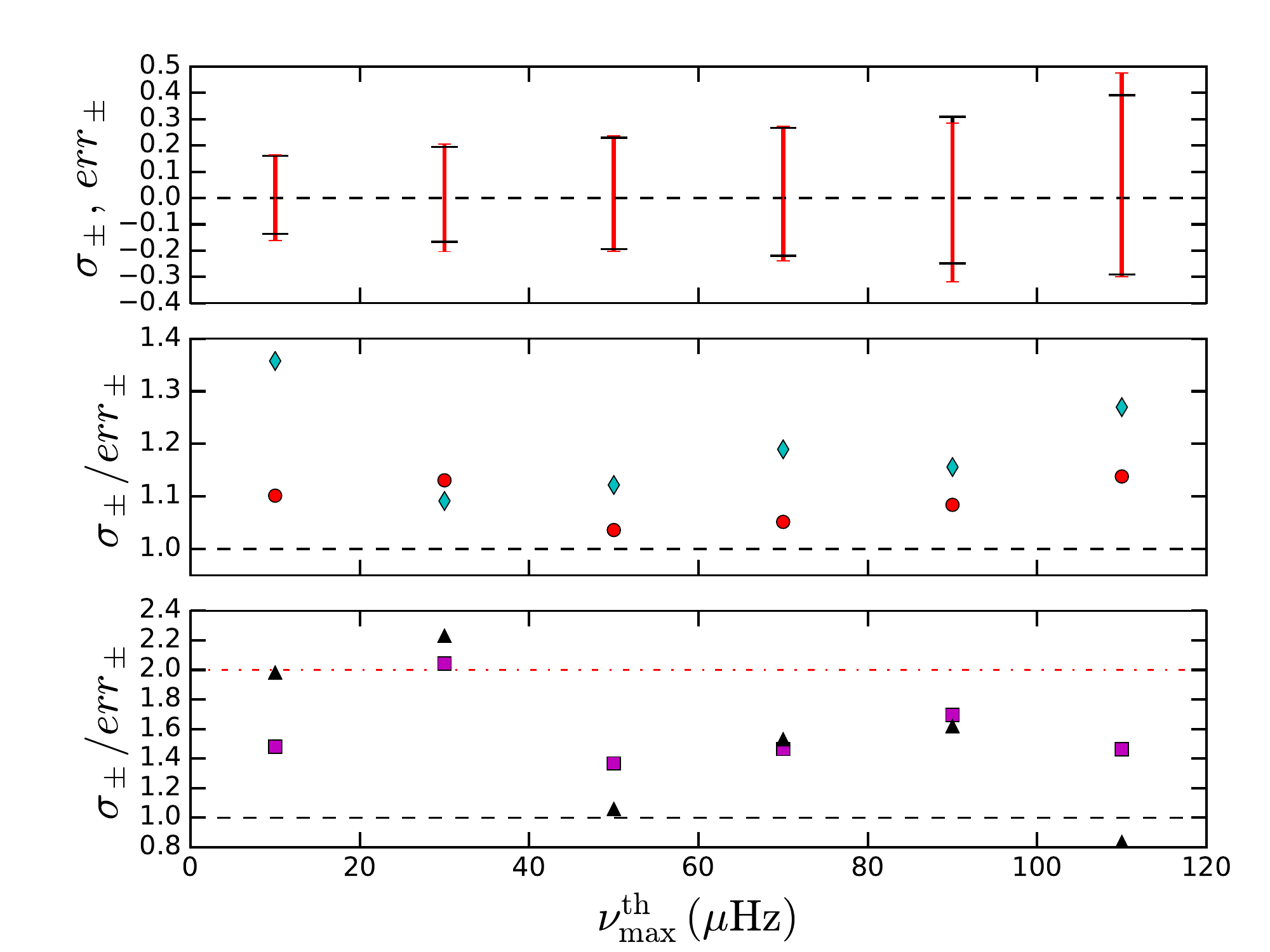}
 \caption{Figures showing the results obtained for the index $\Henv$ 
  as a function of the frequency $\numax$ for the $\kepler$-type (left) and CoRoT-type (right) simulations.\newline
 \textit{a:} Blue circles represent the indicator \textit{bias} for simulations without mixed modes, 
 red triangles for RGB simulations and green squares for clump simulations, 
 obtained with typical values (case (i) of Tab. \ref{tableau_observation_condition}).
 The three types of simulations are slightly shifted in frequency for better visibility.
 Error bars correspond to the  dispersion $\sigma_\pm$
 and the red dotted line is polynomial fit to  the bias in the RGB simulations (coefficients given in Tab. \ref{tableau_coefficient_simus}). \newline
 \textit{b:} In red, the dispersion $\sigma_\pm$ and in black, the median of internal errors $err_\pm$ for RGB simulations with typical values (case (i)).\newline
 \textit{c:} Ratio $\sigma_{\pm}/err_{\pm}$ between the dispersion and the internal errors for RGB simulations. 
 When the ratio is close to 1, the internal errors are considered representative of the dispersion.
 Red dots correspond to the case (i). Cyan diamond are for the case (ii) 
 The red dash-dot line shows the correction function applied to the internal errors (cf. Tab. \ref{tableau_correction_erreur_interne}).   \newline
 \textit{d:} Same as (\textit{c}), with magenta squares for the case (iii) and black triangles for the case (iv).} 
 \label{fig_MLEUP_Henv}
 \end{figure*}

\begin{figure*}
\centering
\begin{annotatedFigure}
{\includegraphics[scale=0.35, trim=0cm 0cm 0cm 0cm, clip=True]{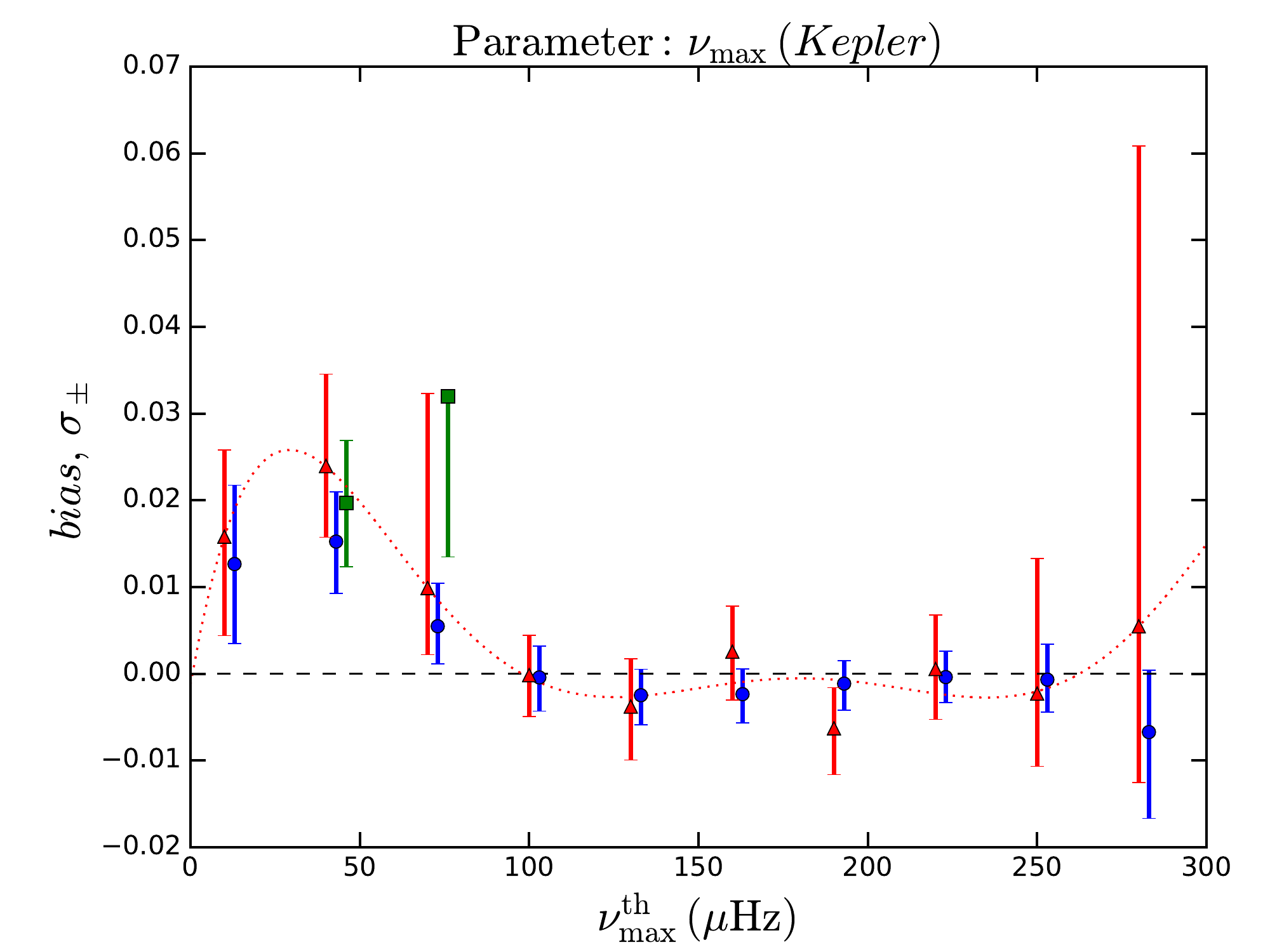}}
	\annotatedFigureBox{-0.02,0.5105}{0.013,0.5107}{a}{-0.04,0.51}
\end{annotatedFigure}
\includegraphics[scale=0.35, trim=0cm 0cm 0cm 0cm, clip=True]{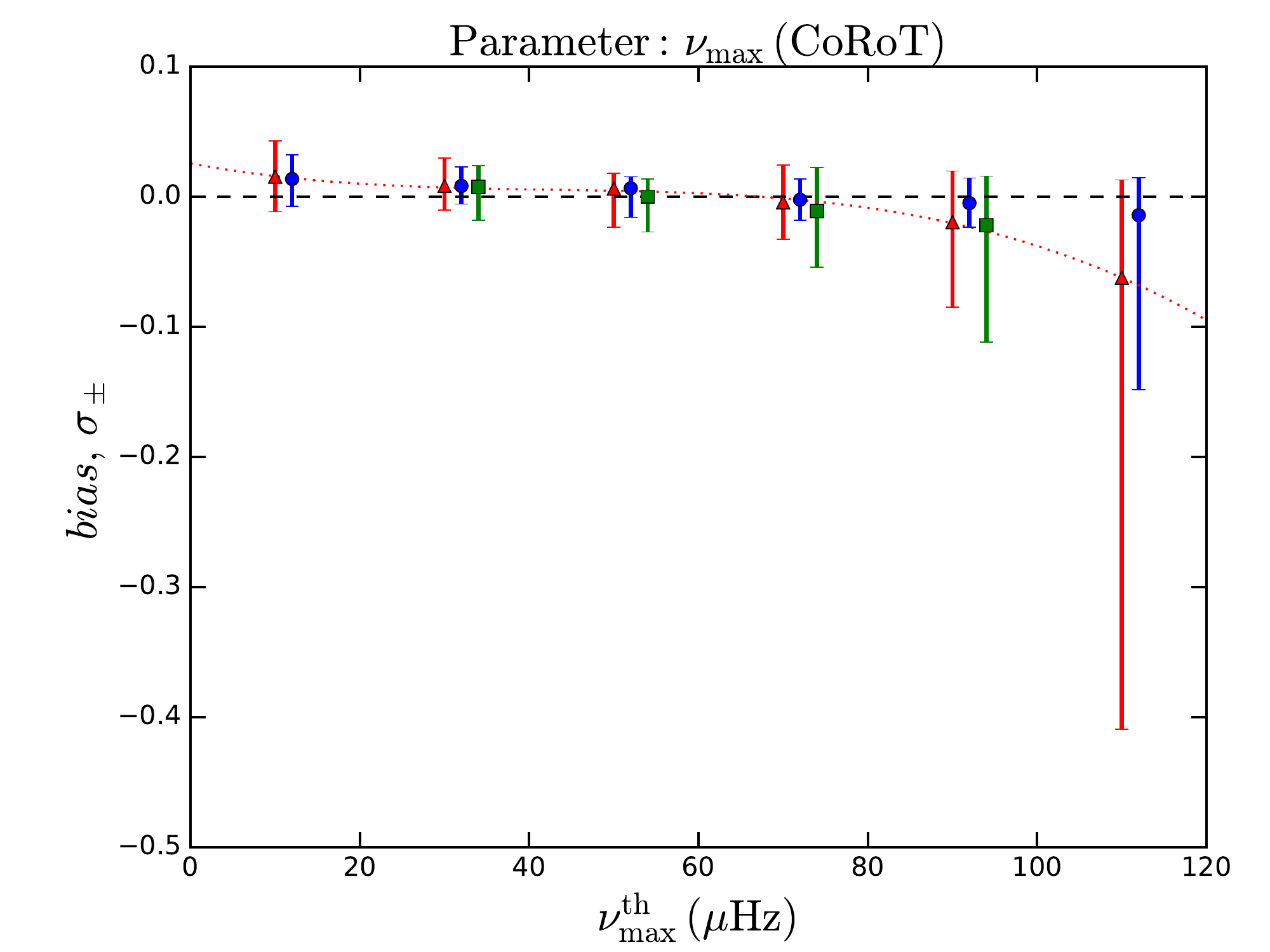}
\begin{annotatedFigure}
	{\includegraphics[scale=0.35, trim=0cm 0cm 0cm 0cm, clip=True]{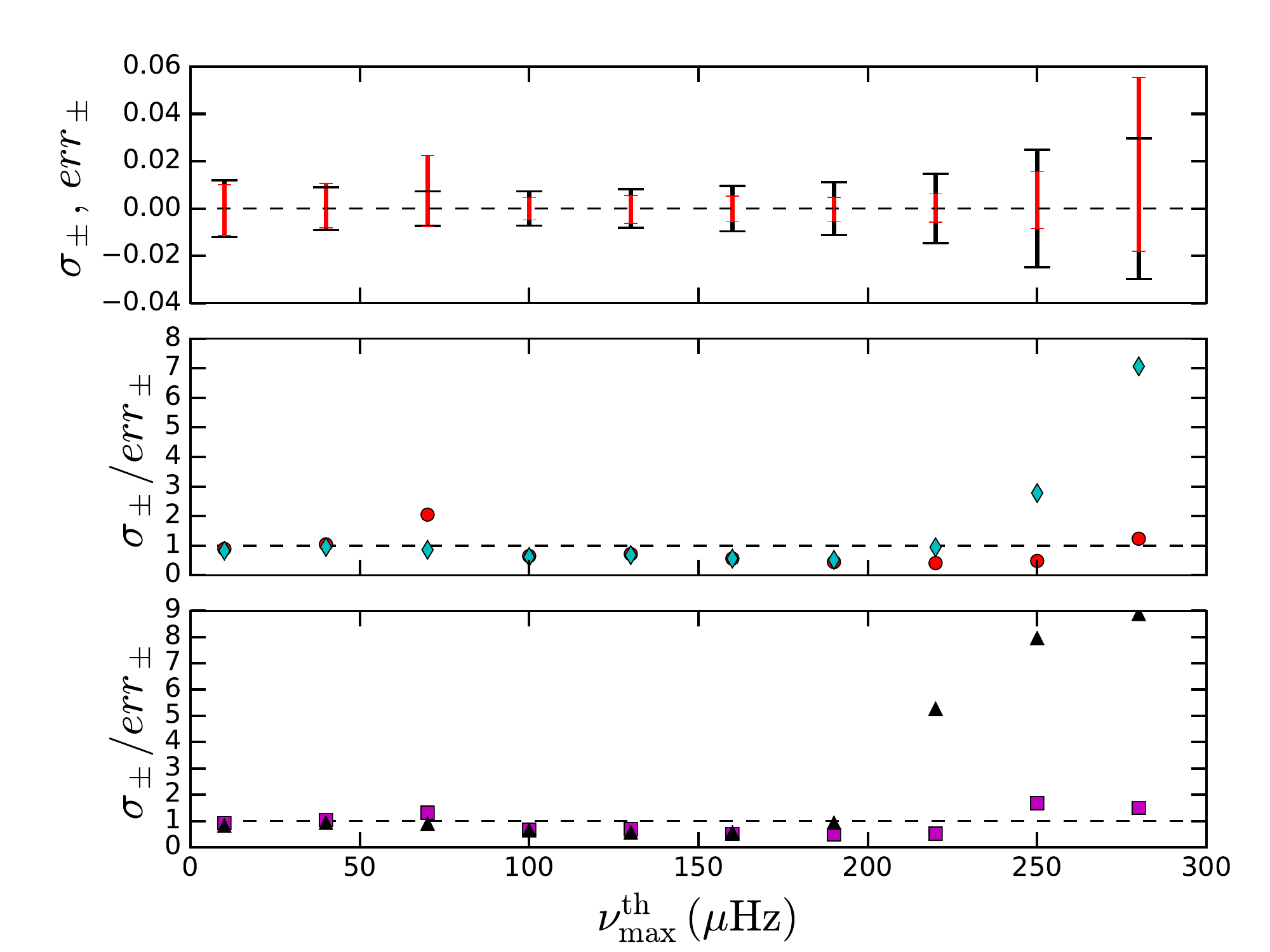}}
	\annotatedFigureBox{-0.03,0.8000}{-0.007,0.7947}{b}{-0.04,0.8}
	\annotatedFigureBox{-0.05,0.5333}{-0.015,0.5333}{c}{-0.04,0.52}
	\annotatedFigureBox{-0.05,0.400}{-0.025,0.200}{d}{-0.04,0.24}{white}{white}{white}{white}
\end{annotatedFigure}
\includegraphics[scale=0.35, trim=0cm 0cm 0.5cm 0cm, clip=True]{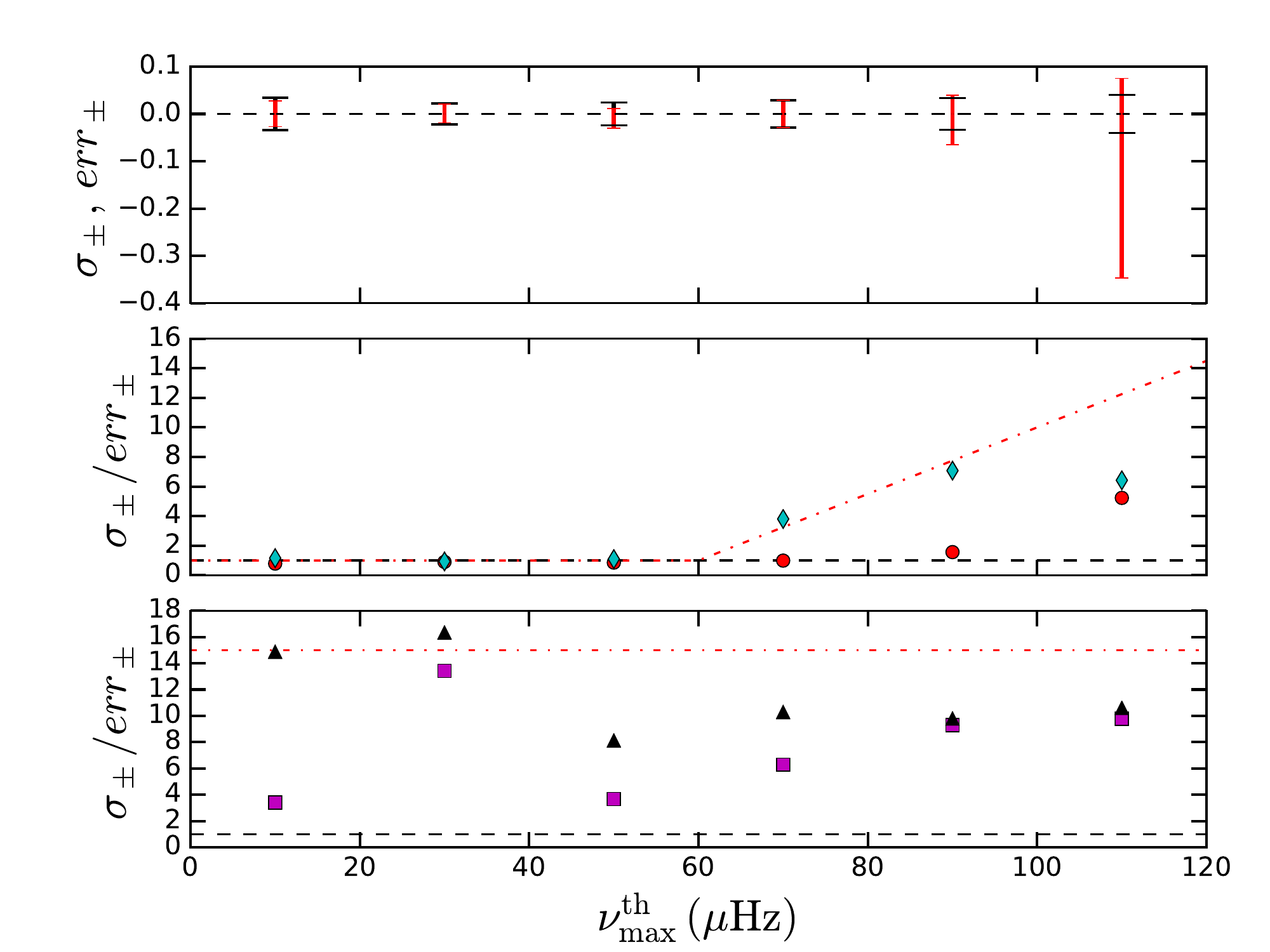}
\caption{
Same as Fig. \ref{fig_MLEUP_Henv} but for the index $\numax$.
} 
 \label{fig_MLEUP_numax}
\end{figure*}

\begin{figure*}
\centering
\begin{annotatedFigure}
{\includegraphics[scale=0.35, trim=0cm 0cm 0.5cm 0cm, clip=True]{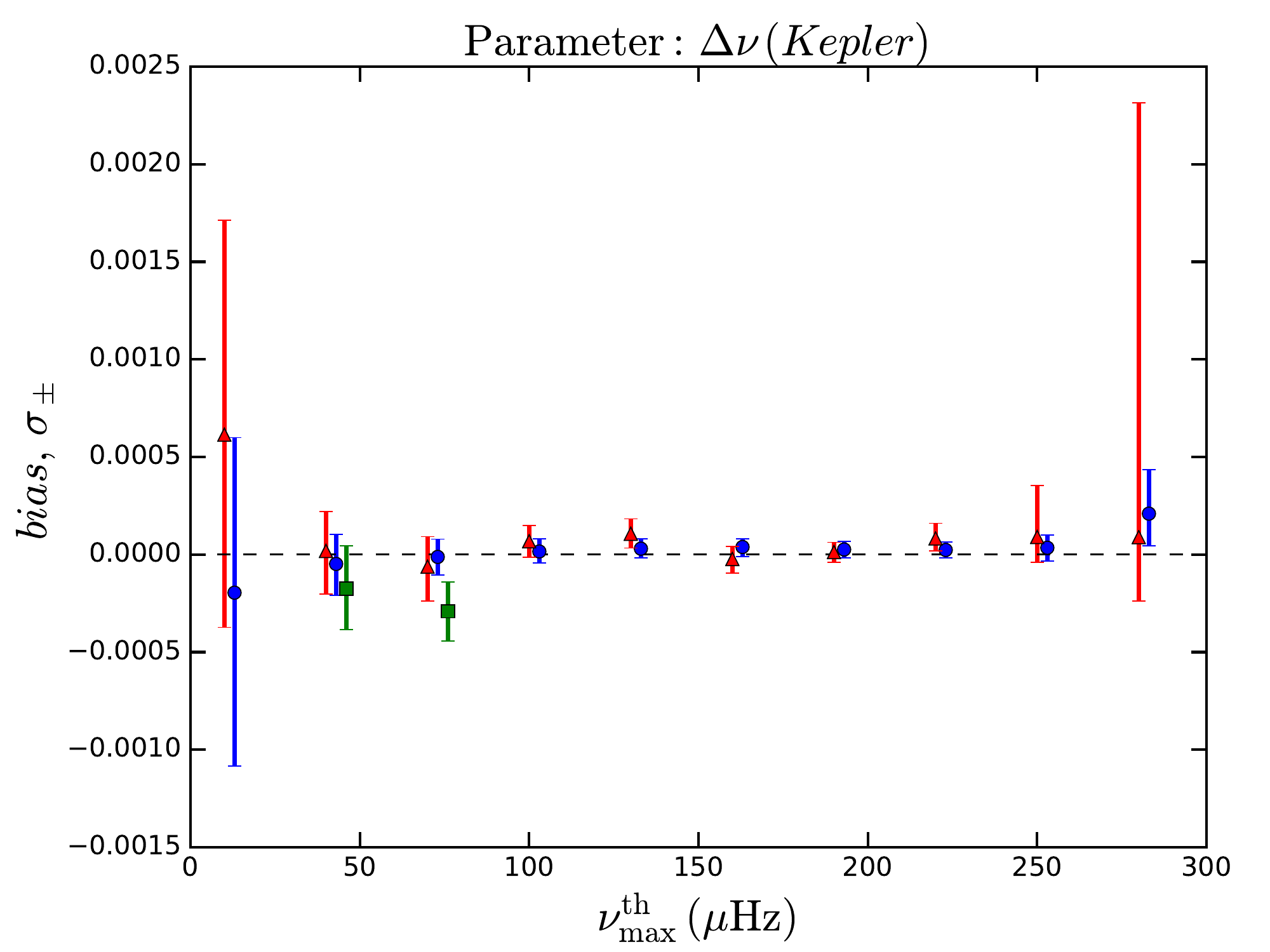}}
	\annotatedFigureBox{-0.02,0.5105}{0.013,0.5107}{a}{-0.04,0.51}
\end{annotatedFigure}
\includegraphics[scale=0.35, trim=0cm 0cm 0cm 0cm, clip=True]{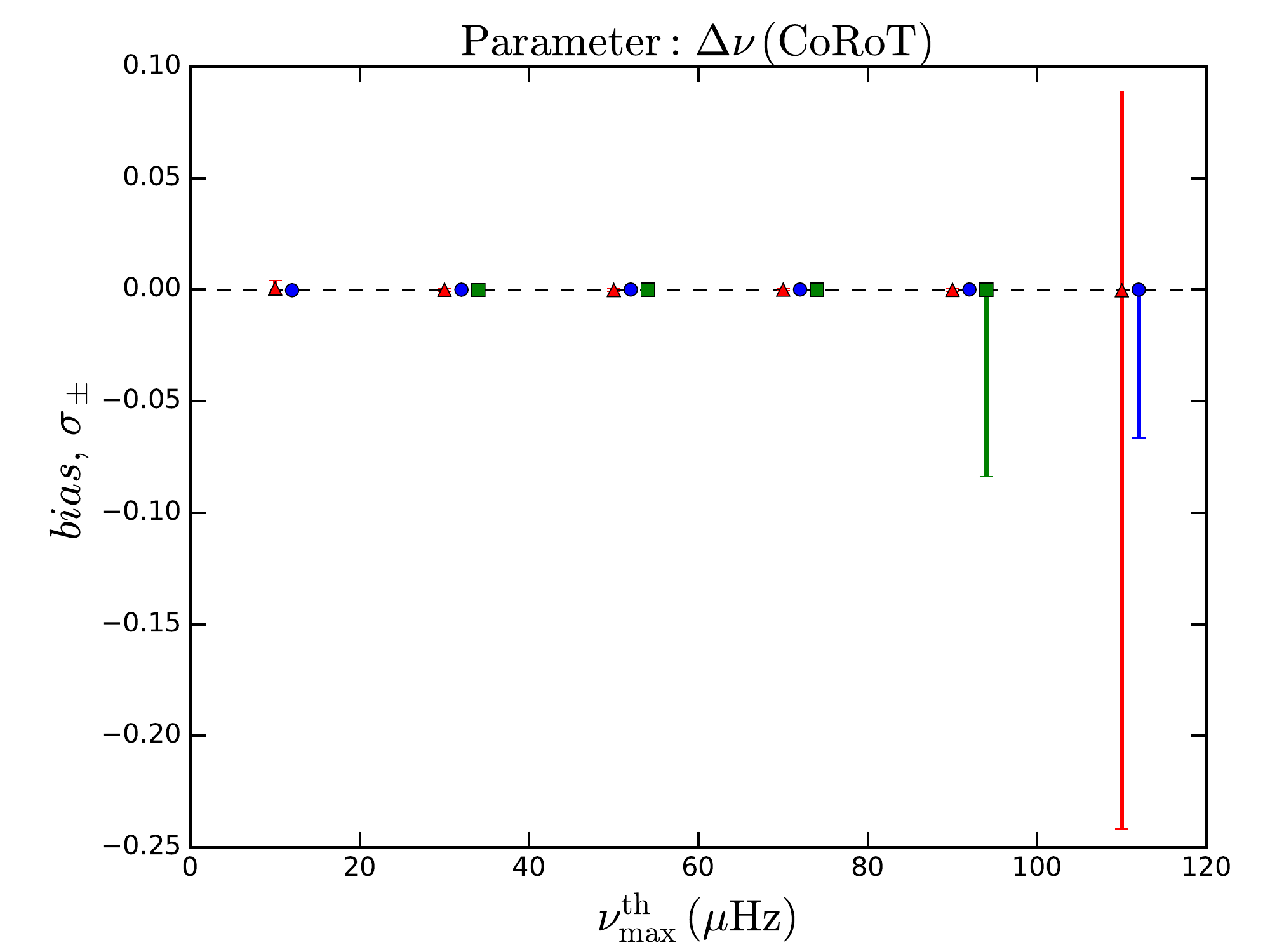}
\begin{annotatedFigure}
	{\includegraphics[scale=0.35, trim=0cm 0cm 0cm 0cm, clip=True]{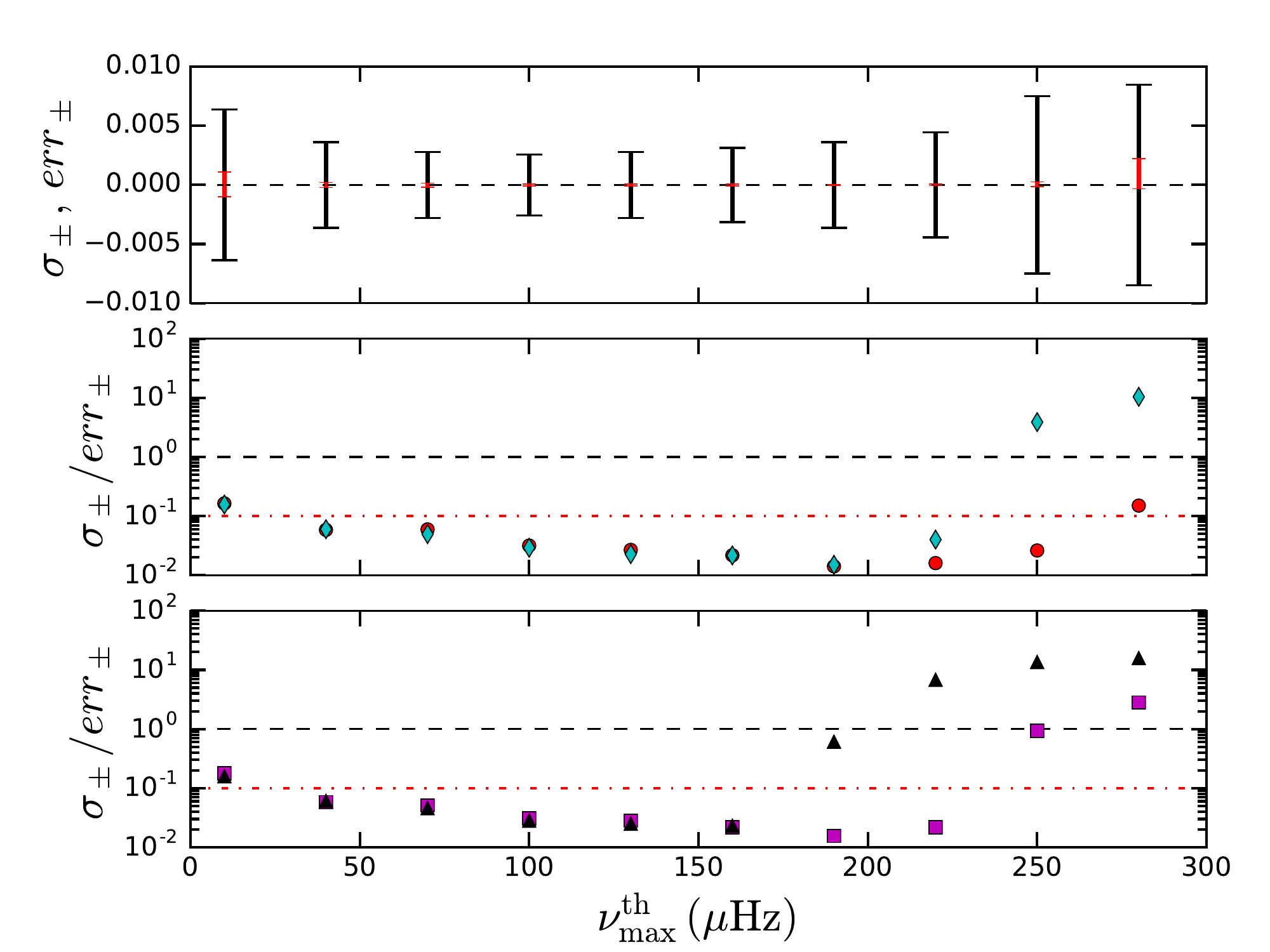}}
	\annotatedFigureBox{-0.03,0.8000}{-0.007,0.7947}{b}{-0.04,0.8}
	\annotatedFigureBox{-0.05,0.5333}{-0.015,0.5333}{c}{-0.04,0.52}
	\annotatedFigureBox{-0.05,0.400}{-0.025,0.200}{d}{-0.04,0.24}{white}{white}{white}{white}
\end{annotatedFigure}
\includegraphics[scale=0.35, trim=0cm 0cm 0.5cm 0cm, clip=True]{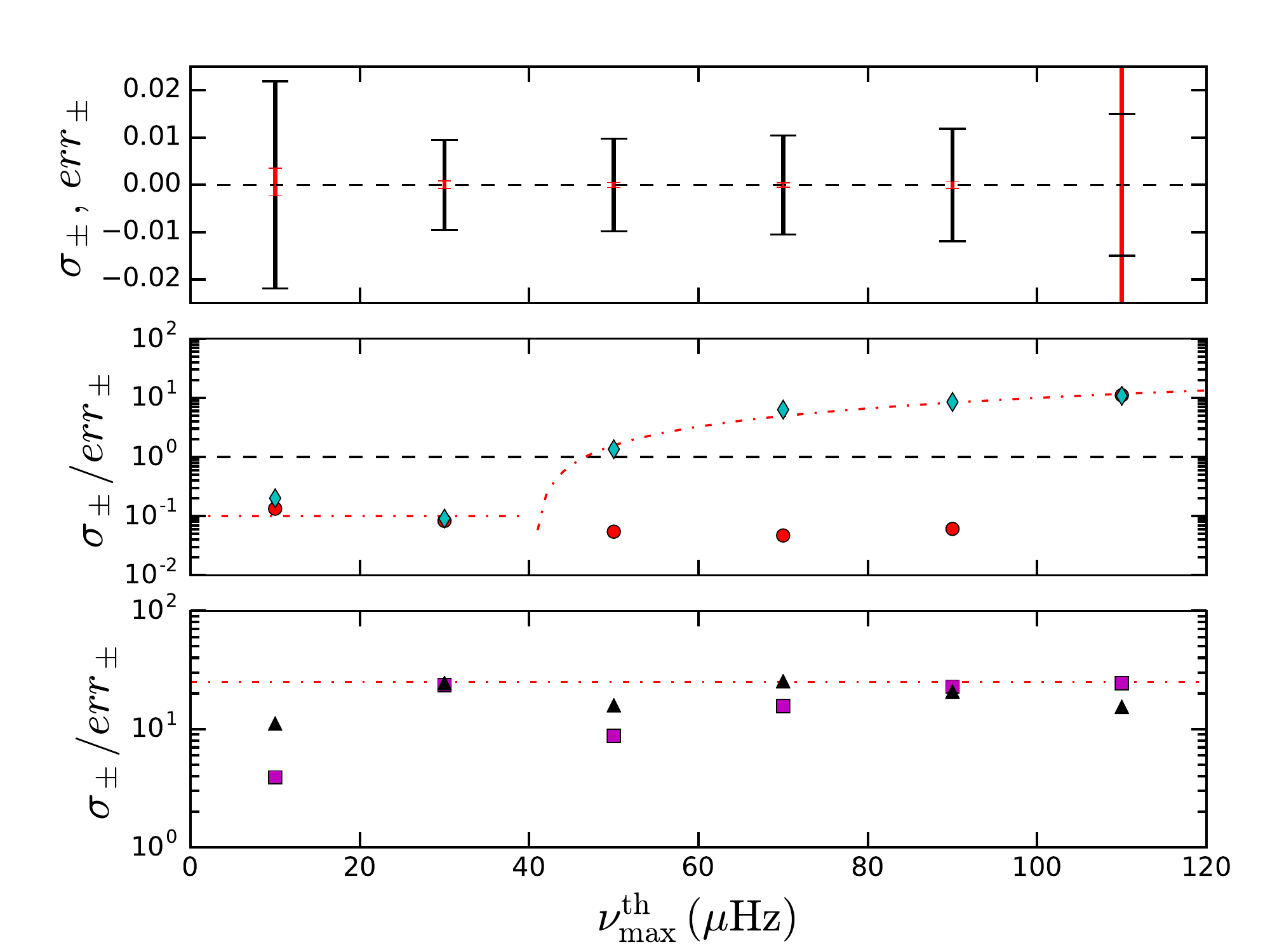}
 \caption{
Same as Fig. \ref{fig_MLEUP_Henv} but for the index $\Dnu$.
} 
 \label{fig_MLEUP_Dnu}
\end{figure*}

\begin{figure*}
\centering
\begin{annotatedFigure}
{\includegraphics[scale=0.35, trim=0cm 0cm 0cm 0cm, clip=True]{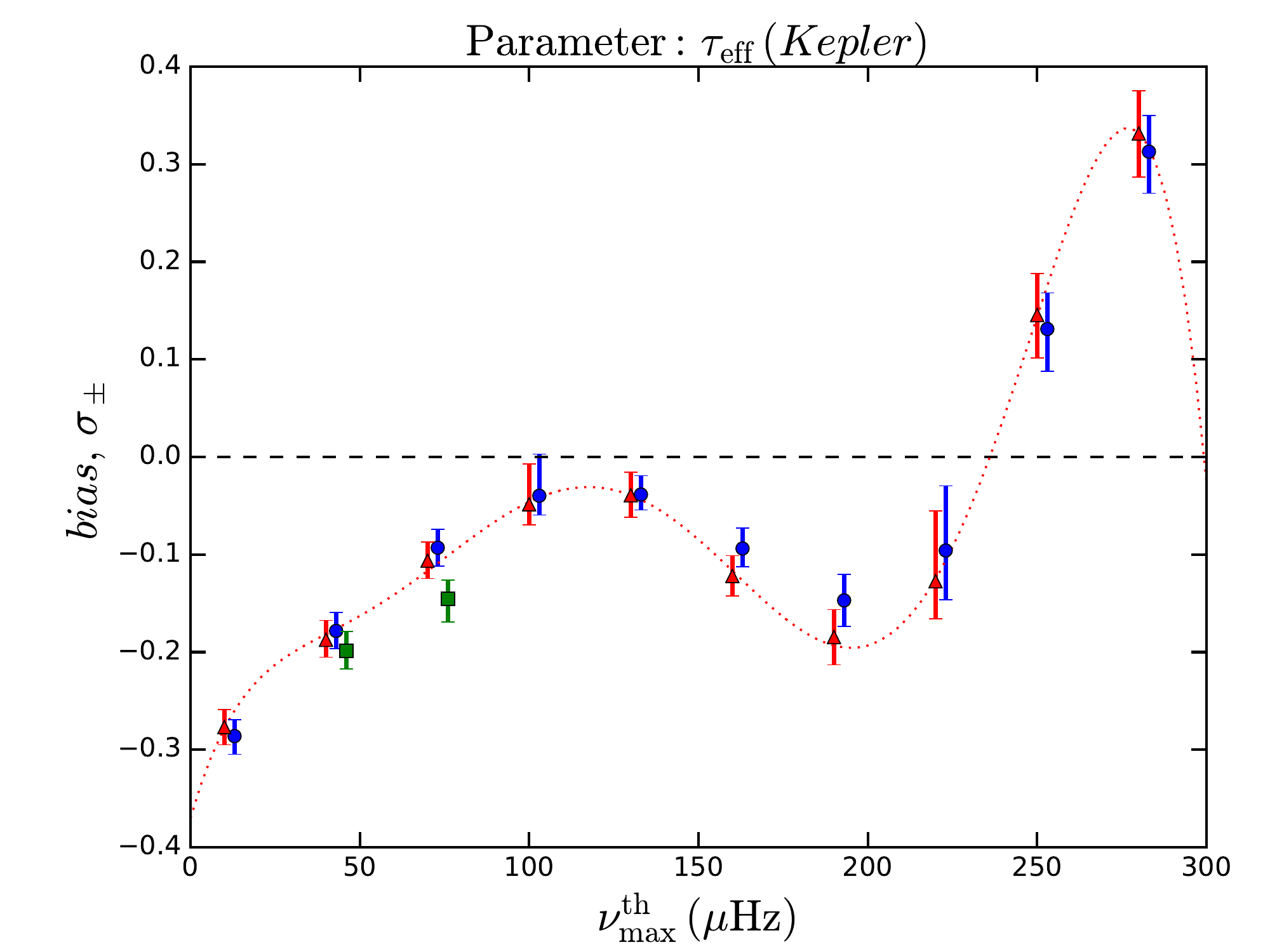}}
	\annotatedFigureBox{-0.02,0.5105}{0.013,0.5107}{a}{-0.04,0.51}
\end{annotatedFigure}
\includegraphics[scale=0.35, trim=0cm 0cm 0cm 0cm, clip=True]{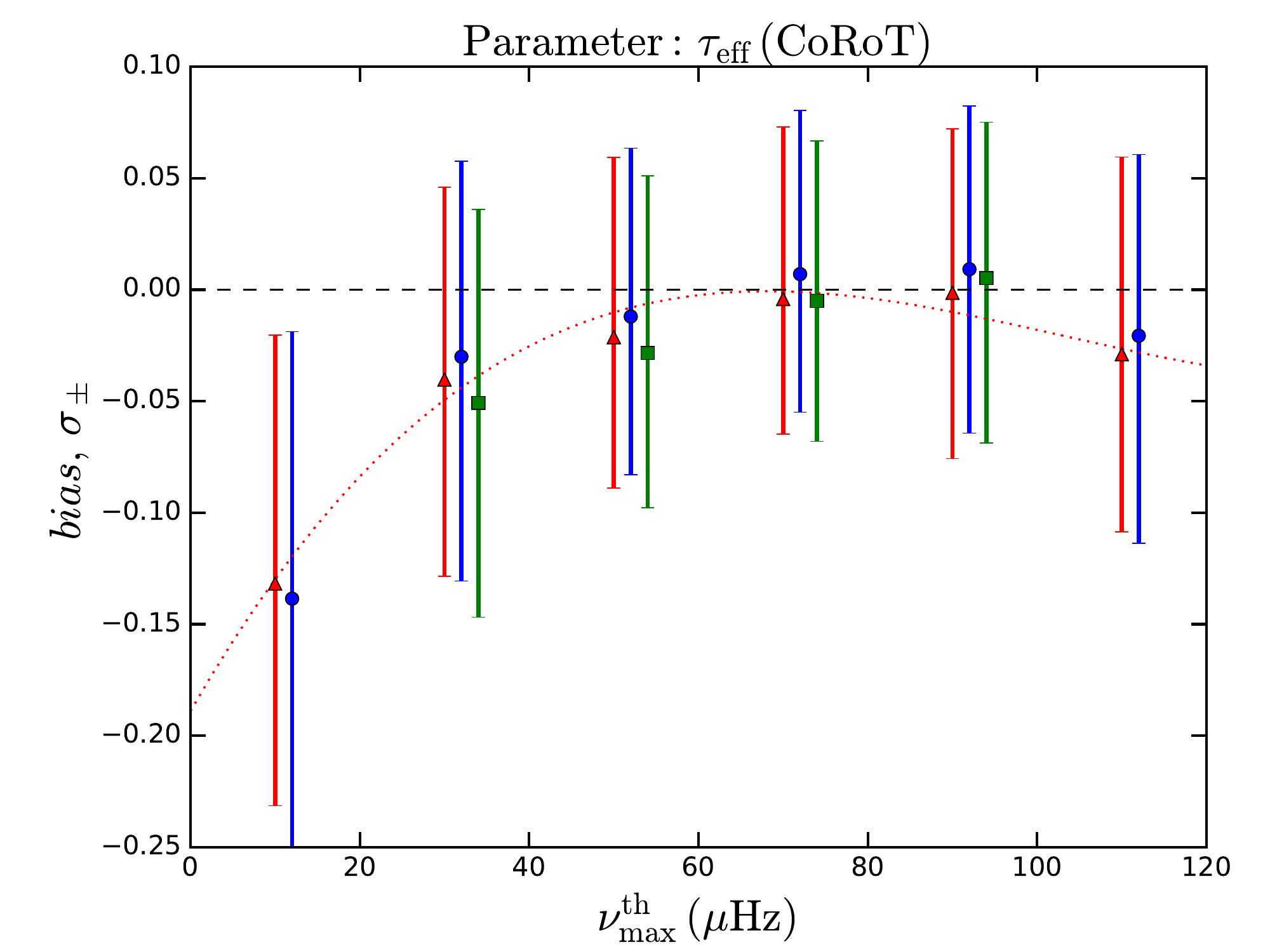}
\begin{annotatedFigure}
	{\includegraphics[scale=0.35, trim=0cm 0cm 0cm 0cm, clip=True]{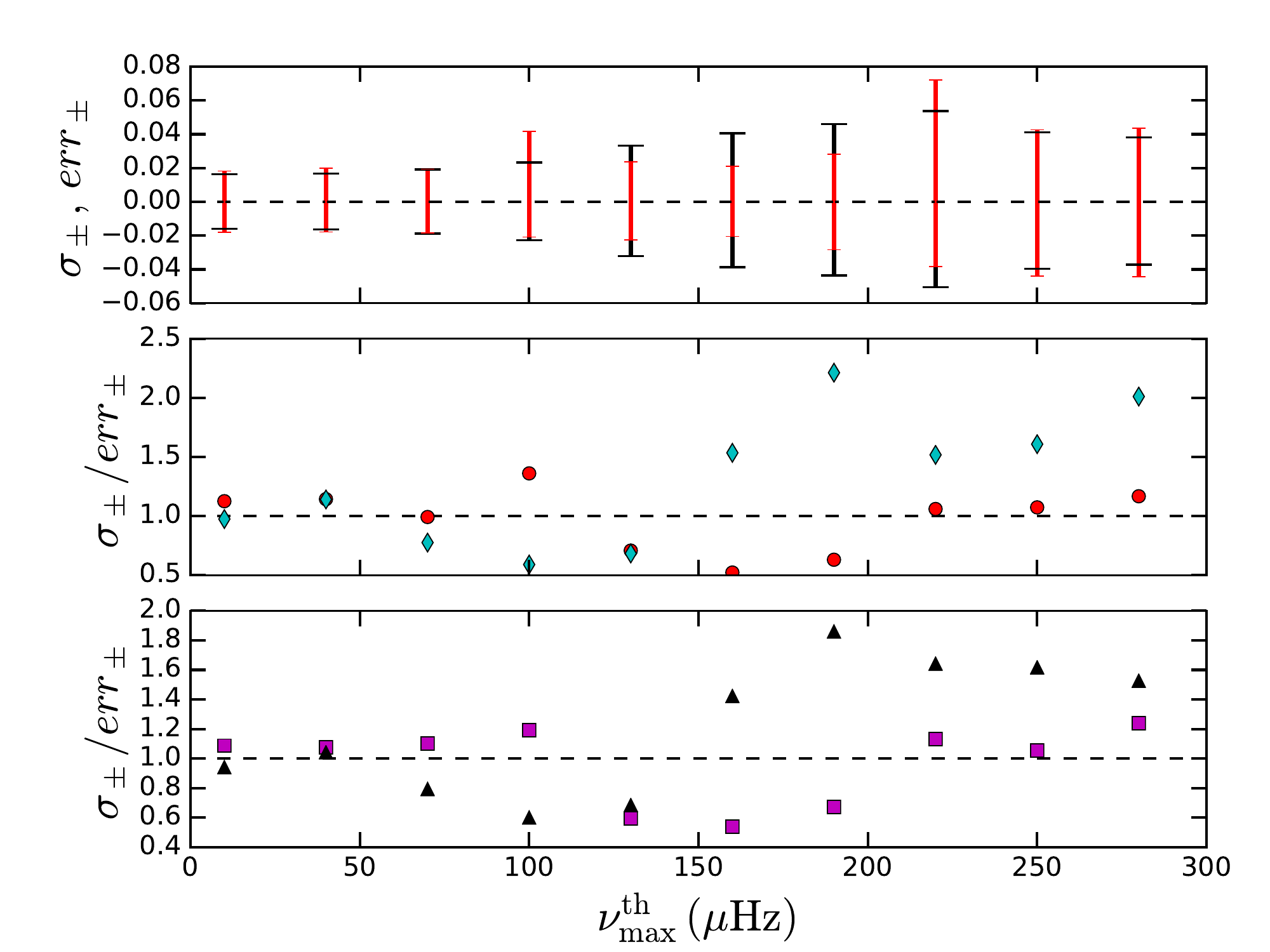}}
	\annotatedFigureBox{-0.03,0.8000}{-0.007,0.7947}{b}{-0.04,0.8}
	\annotatedFigureBox{-0.05,0.5333}{-0.015,0.5333}{c}{-0.04,0.52}
	\annotatedFigureBox{-0.05,0.400}{-0.025,0.200}{d}{-0.04,0.24}{white}{white}{white}{white}
\end{annotatedFigure}
\includegraphics[scale=0.35, trim=0cm 0cm 0.5cm 0cm, clip=True]{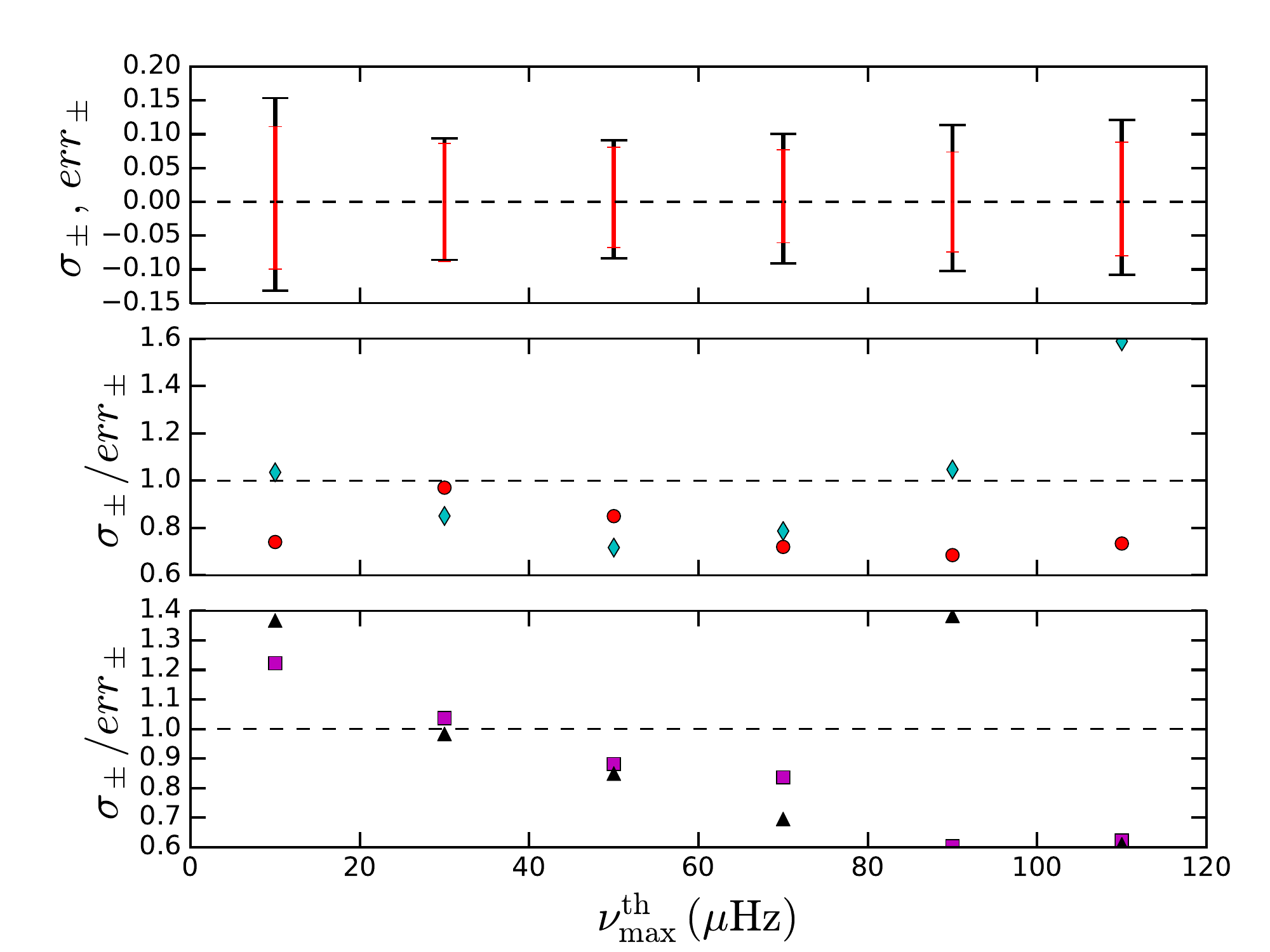}
 \caption{
Same as Fig. \ref{fig_MLEUP_Henv} but for the parameter $\taueff$.
} 
 \label{fig_MLEUP_taueff}
\end{figure*}

\begin{figure*}
\centering
\begin{annotatedFigure}
{\includegraphics[scale=0.35, trim=0cm 0cm 0cm 0cm, clip=True]{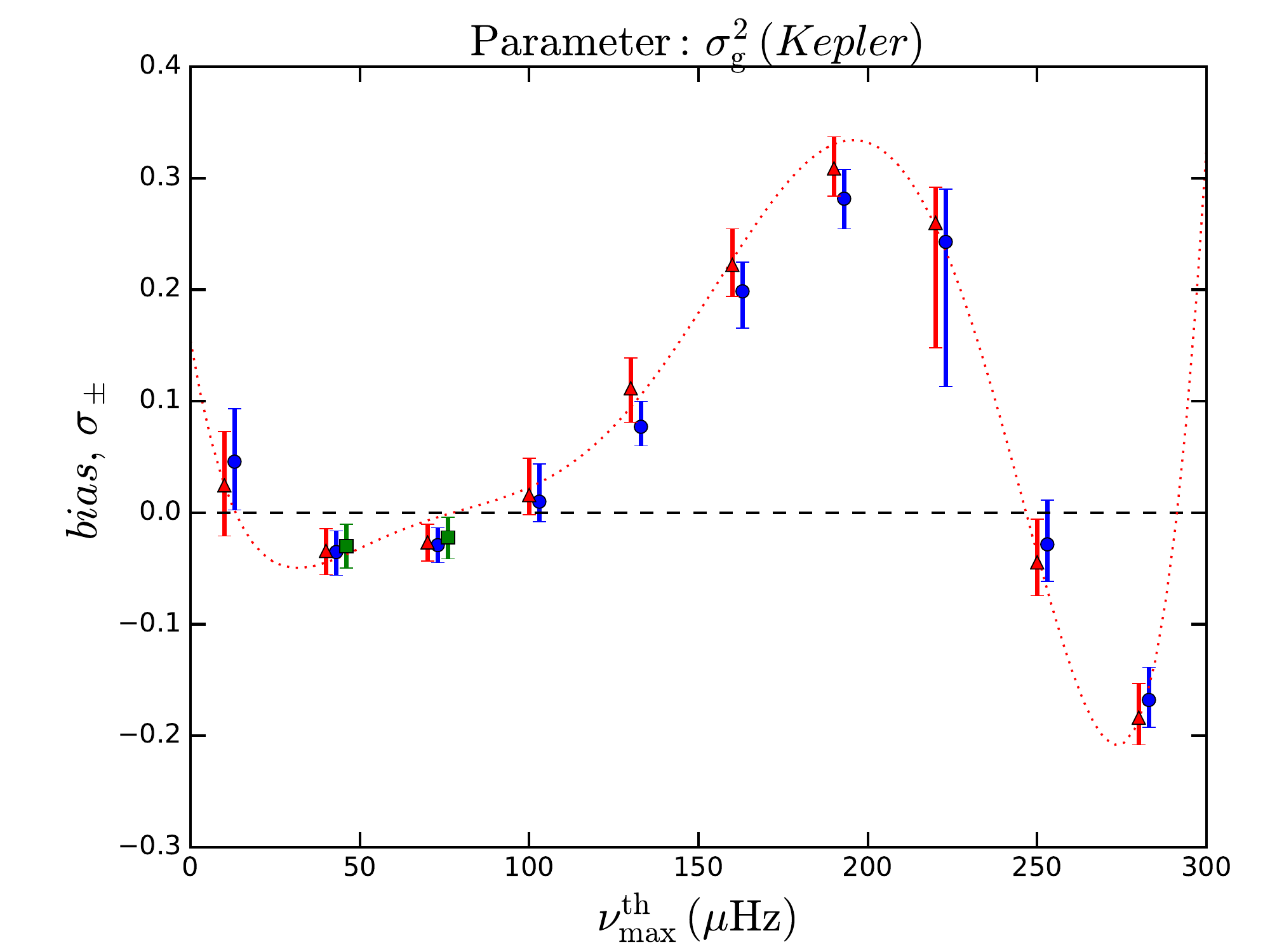}}
	\annotatedFigureBox{-0.02,0.5105}{0.013,0.5107}{a}{-0.04,0.51}
\end{annotatedFigure}
\includegraphics[scale=0.35, trim=0cm 0cm 0cm 0cm, clip=True]{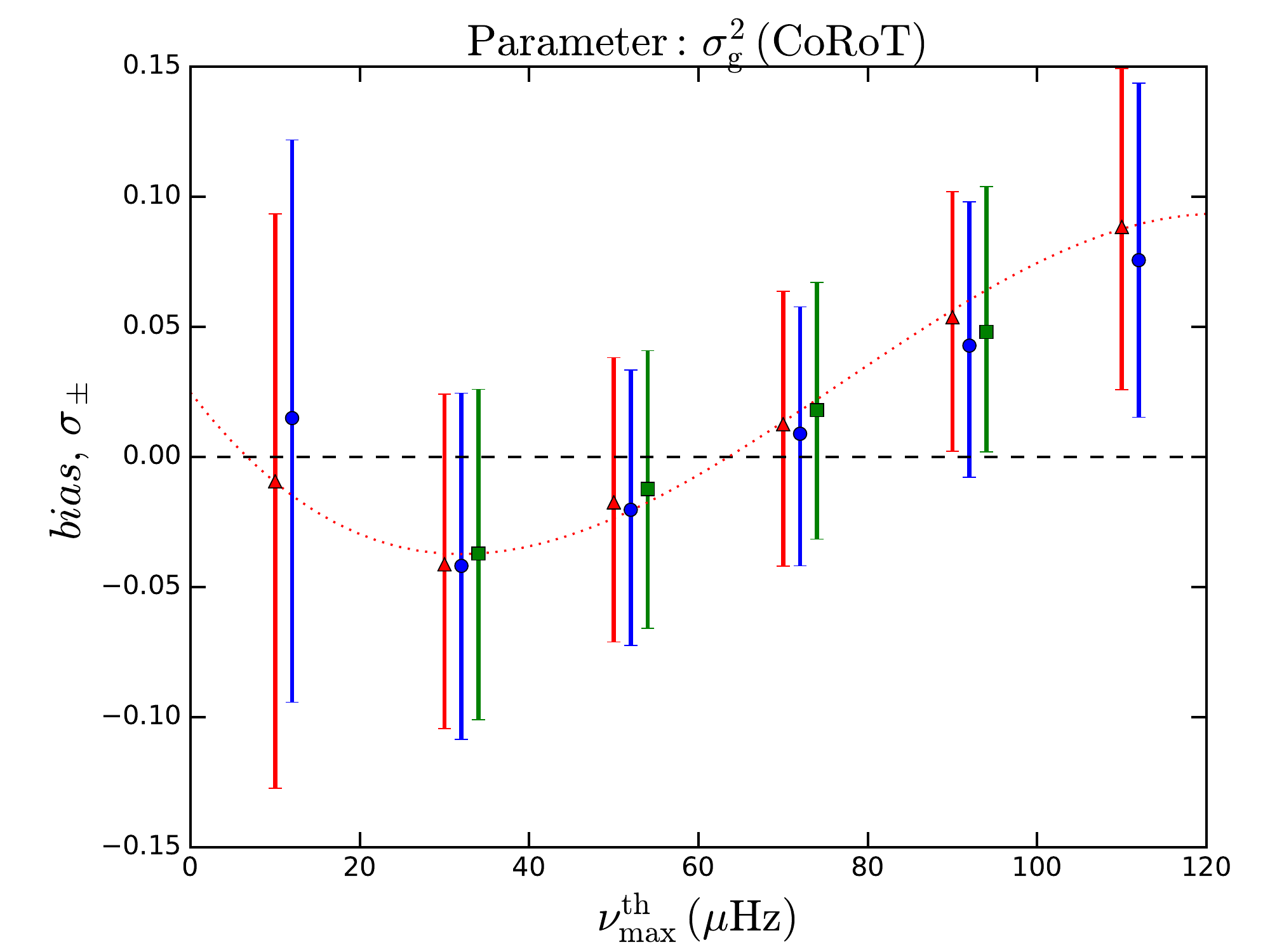}
\begin{annotatedFigure}
	{\includegraphics[scale=0.35, trim=0cm 0cm 0cm 0cm, clip=True]{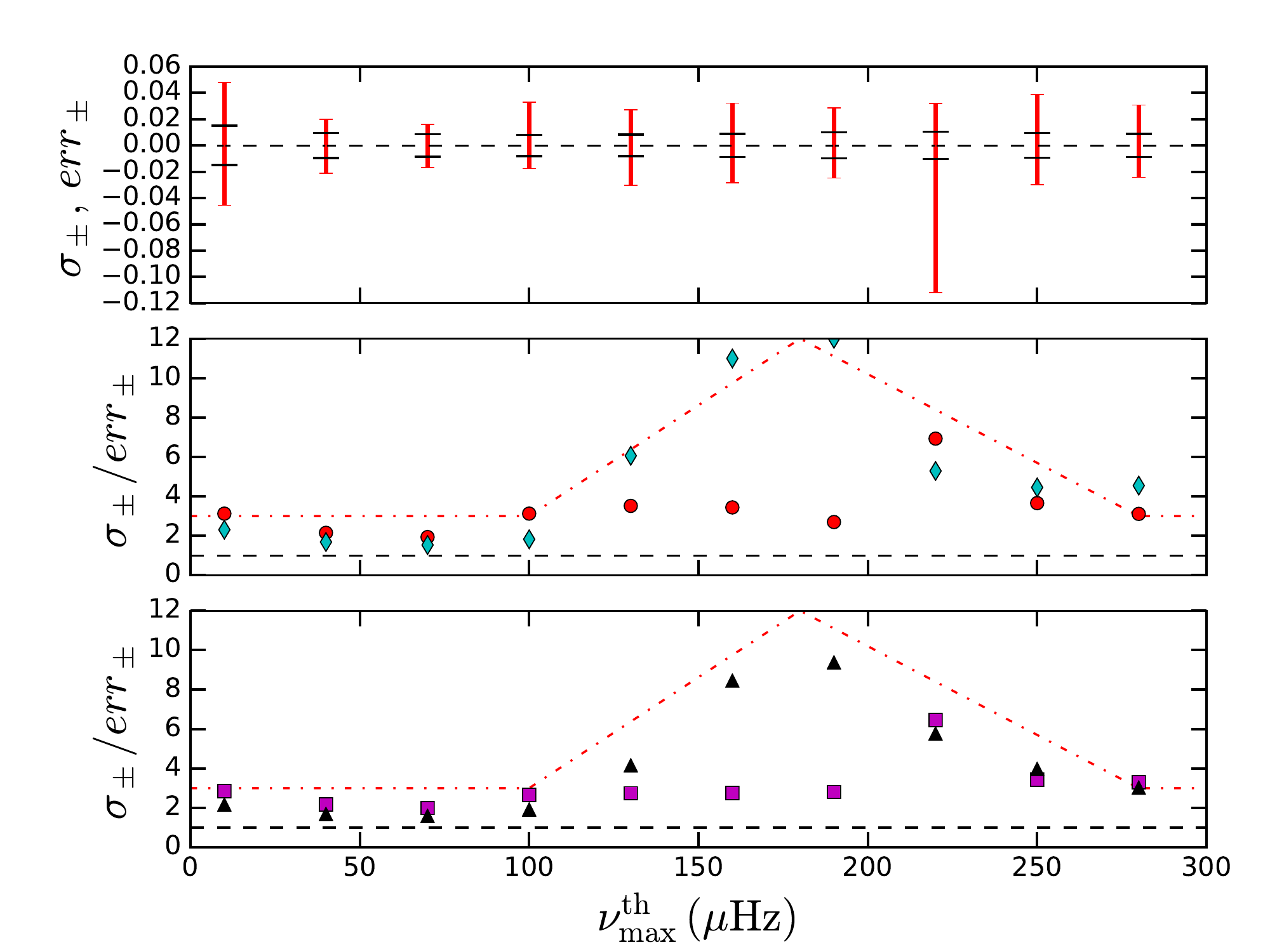}}
	\annotatedFigureBox{-0.03,0.8000}{-0.007,0.7947}{b}{-0.04,0.8}
	\annotatedFigureBox{-0.05,0.5333}{-0.015,0.5333}{c}{-0.04,0.52}
	\annotatedFigureBox{-0.05,0.400}{-0.025,0.200}{d}{-0.04,0.24}{white}{white}{white}{white}
\end{annotatedFigure}
\includegraphics[scale=0.35, trim=0cm 0cm 0.5cm 0cm, clip=True]{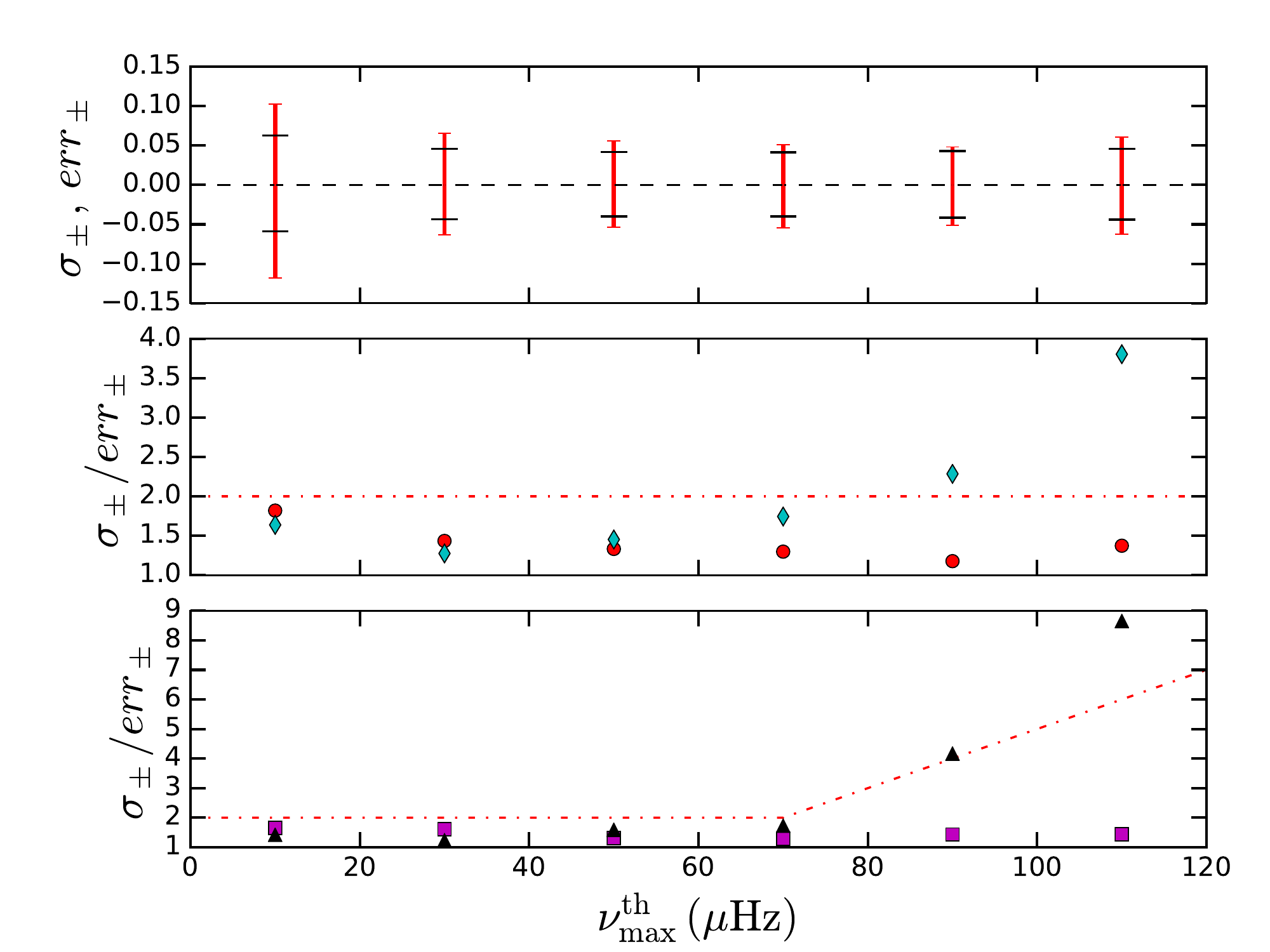}
 \caption{
Same as Fig. \ref{fig_MLEUP_Henv} but for the parameter $\sigma_\mathrm{g}^2$.
} 
 \label{fig_MLEUP_sigma}
\end{figure*}

For each seismic index or granulation parameter, we investigated bias and error estimates 
associated with our analysis pipeline following three indicators: 
\begin{enumerate}
  \item Bias: $bias = (\mathrm{median}(X)-X_{\mathrm{th}})/X_{\mathrm{th}}$,
  with $X$, the measured value of a given parameter at the frequency $\numax$ and $X_{\mathrm{th}}$, the theoretical input value.
   For each parameter, and typical duration of the observations and typical magnitudes, the bias is illustrated in 
panels a ($\kepler$ on left and CoRoT on right) of Figures
\ref{fig_MLEUP_Henv} to \ref{fig_MLEUP_sigma}. 
We tested different observation durations and magnitudes as described in Tab. \ref{tableau_observation_condition}.
Since bias is not found to show significant
variation with the observation duration nor with the magnitude, only results obtained for typical $T$ and $V$ values are presented in these panels. 

Bias is generally found to be significant for the various seismic indices and granulation parameters.
Surprisingly enough, in the case of $\Henv$, 
the bias is smaller for synthetic light-curves including mixed modes  
(Fig. \ref{fig_MLEUP_Henv}). For other parameters, the influence of mixed modes appears to be negligible.

Figures \ref{fig_MLEUP_Henv} to \ref{fig_MLEUP_sigma} also reveal that the bias is not significantly different for RGB  and clump stars.
For each parameter, a polynomial fit to the bias (as a function of $\numax$) is obtained 
(see red dotted line represented in panels a
of Figures \ref{fig_MLEUP_Henv} to \ref{fig_MLEUP_sigma} and coefficients given in 
Tab. \ref{tableau_coefficient_simus}). These polynomial functions are 
used to correct results obtained with MLEUP on real data in Sect.~\ref{application}.

  \item Dispersion: $\sigma_\pm = |\mathrm{q_\pm}-\mathrm{median}(X)|/X_{\mathrm{th}}$,
  with $\mathrm{q_\pm}$, the percentile at $\pm 34\%$ around the median.
  The comparison between $\sigma_+$ and $\sigma_-$ provides information on the symmetry 
of the distribution of results. This dispersion is illustrated for each seismic 
index or granulation parameter (here again for typical $T$ and $V$ values) 
in  panels $b$ of Figures
\ref{fig_MLEUP_Henv} to \ref{fig_MLEUP_sigma}, together with internal errors defined as,
 
  \item Internal errors: $err_\pm = \mathrm{median}(\delta X)/X_{\mathrm{th}}$, 
  with $\delta X$, the internal errors determined from the Hessian matrix.

  The ratio $\sigma_{\pm}/err_{\pm}=(\sigma_{+}+\sigma_{-})/(err_{+}+err_{-})$ tells us whether the internal errors given by the MLEUP method
  are representative of the real dispersion of the results. 
  This is illustrated for each parameter and for each satellite in panels c and d of figures
  \ref{fig_MLEUP_Henv} to \ref{fig_MLEUP_sigma}, 
  for different observation time $T$ and magnitude $V$ 
  taken from Tab. \ref{tableau_observation_condition}.

 Internal errors are generally found to be representative of the dispersion of the 
results for typical durations and magnitudes. 
However, departures from this behaviour can be found for shorter 
durations or higher magnitudes. When this is the case, we propose a conservative calibration
of the error estimate illustrated by the red dash-dot lines in  panels c and d of Figures
\ref{fig_MLEUP_Henv} to \ref{fig_MLEUP_sigma} and summarized in Tab. \ref{tableau_correction_erreur_interne}.
This correction of the error estimates based on internal errors is also used
to correct results obtained with MLEUP on real data in Sect. \ref{application}.
  
\end{enumerate}

\begin{center}
\begin{table*}
\centering
    \caption{Simulated light-curve parameters for RGB type-stars for four observation conditions representative     of our CoRoT and $\kepler$ datasets:
(i):  Typical values are based on average values from the datasets, (ii): the value of $T_\mathrm{obs}$ taken for the low duration case corresponds to the tenth percentile of 
the duration distribution in our data sets,  (iii): the value of $V$ mag taken for the high magnitude case corresponds to the 90th percentile of the magnitude 
distribution, (iv): Least favorable case, with $T$ from the case (ii) and $V$ from the case (iii). \label{tableau_observation_condition}}
\begin{tabular}{r | c c | c c}
 \toprule
                                    & \multicolumn{2}{c|}{CoRoT}                 & \multicolumn{2}{c}{$\kepler$}  \\
            Cases                   & Observation time (day)   & $V$ mag         & Observation time (day)      & $V$ mag       \\
    \midrule
    (i)             & $T_\mathrm{obs}=150$     & $V=13$          & $T_\mathrm{obs}=1000$       & $V=12$\\
    (ii)               & $T_\mathrm{obs}=80$      & $V=13$          & $T_\mathrm{obs}=400 $       & $V=12$\\
    (iii)             & $T_\mathrm{obs}=150$     & $V=15$          & $T_\mathrm{obs}=1000$       & $V=14$\\
    (iv)    & $T_\mathrm{obs}=80$      & $V=15$          & $T_\mathrm{obs}=400 $       & $V=14$\\
\bottomrule
\end{tabular}    
\end{table*}
\end{center}

\begin{sidewaystable}
 \caption{Conservative correction functions of the internal errors given by MLEUP for both satellites, CoRoT and $\kepler$, 
 and for the following parameters: $\Henv$, $\numax$, $\Dnu$ and $\sigma_\mathrm{g}^2$. 
 These corrections have been determined using the synthetic light curves described in appendix \ref{full_simulations}.\label{tableau_correction_erreur_interne} }
\begin{tabular*}{\textheight}{ c | c | c | c | c }
 \toprule
                                     & \multicolumn{2}{c|}{CoRoT}                                                                        & \multicolumn{2}{c}{$\kepler$} \\

                                      & \multicolumn{1}{c}{condition}                                    & correction                     & \multicolumn{1}{c}{condition}     & correction     \\
\midrule
\multirow{1}{*}{$\Henv$}              & $T_\mathrm{obs} \leqslant 100$ d                                 & $err\times2$                   & ---                        & ---     \\
\hline                          
\multirow{2}{*}{$\numax$}             & $T_\mathrm{obs} \leqslant 100$ d                                 & $err\times15$                  & \multirow{2}{*}{---}    &  \multirow{2}{*}{---}      \\
                                      & $T_\mathrm{obs} > 100$ d \& $\numax \geqslant 60~\mu$Hz          & $err\times(0.225~\numax-12.5$)  &                            &     \\

\hline                          
\multirow{3}{*}{$\Dnu$}               & $T_\mathrm{obs} \leqslant 100$ d                                 &  $err\times25$                 &  \multirow{3}{*}{---}   &  \multirow{3}{*}{$err\times0.1$} \\
                                      & $T_\mathrm{obs} > 100$ d \& $\numax < 40~\mu$Hz                  &  $err\times0.1$                &                            &     \\
                                      & $T_\mathrm{obs} > 100$ d \& $\numax \geqslant 40~\mu$Hz          &  $err\times(0.168~\numax-6.83$) &                            &     \\

\hline                          
\multirow{3}{*}{$\sigma_\mathrm{g}^2$}& $T_\mathrm{obs} \leqslant 100$ d  \& $\numax < 70~\mu$Hz         &  $err\times2$                  & $\numax < 100~\mu$Hz                & $err\times3.0$   \\
                                      & $T_\mathrm{obs} \leqslant  100$ d \& $\numax \geqslant 70~\mu$Hz &  $err\times(0.1~\numax-5.0)$      & $100 \leqslant \numax < 180~\mu$Hz  &  $err\times(0.1125~\numax-8.25)$    \\
                                      & $T_\mathrm{obs} > 100$ d                                         &  $err\times2$                  & $\numax \geqslant 180~\mu$Hz        &   $err\times(-0.09~\numax-28.2)$    \\
\bottomrule
\end{tabular*}
\end{sidewaystable}

\begin{sidewaystable}
 \caption{Polynomial coefficients (in ascending order) characterizing biases of the MLEUP method for $\kepler$ and CoRoT data, 
 for $\Henv$, $\numax$, $\taueff$ and $\sigma_\mathrm{g}^2$. 
 These coefficients were obtained through a least-squares adjustment of RGB simulations using a polynomial function of degree six for $\kepler$ and three for CoRoT. \label{tableau_coefficient_simus}}
\centering
\begin{tabular*}{\textheight}{ c | c | r r r r r r r r}
\toprule
                          & degrees    &\multicolumn{1}{c}{0}&\multicolumn{1}{c}{1}&\multicolumn{1}{c}{2}&\multicolumn{1}{c}{3}&\multicolumn{1}{c}{4}&\multicolumn{1}{c}{5}&\multicolumn{1}{c}{6}\\
\midrule
\multirow{4}{*}{CoRoT}    & $\Henv$    & $-2.0.10^{-1}$     & $ 1.6.10^{-2}$     & $-1.7.10^{-4}$     & $ 4.8.10^{-7}$    & \multicolumn{1}{c}{---} & \multicolumn{1}{c}{---} & \multicolumn{1}{c}{---} \\
                          & $\numax$   & $ 2.6.10^{-2}$     & $-1.2.10^{-3}$     & $ 2.6.10^{-5}$     & $-2.0.10^{-7}$    & \multicolumn{1}{c}{---} & \multicolumn{1}{c}{---} & \multicolumn{1}{c}{---} \\
                          & $\taueff$  & $-1.9.10^{-1}$     & $ 6.6.10^{-3}$     & $-7.3.10^{-5}$     & $ 2.4.10^{-7}$    & \multicolumn{1}{c}{---} & \multicolumn{1}{c}{---} & \multicolumn{1}{c}{---} \\
                          & $\sigma_\mathrm{g}^2$ & $ 2.5.10^{-2}$     & $-4.3.10^{-3}$     & $ 8.4.10^{-5}$     & $-3.7.10^{-7}$    & \multicolumn{1}{c}{---} & \multicolumn{1}{c}{---} & \multicolumn{1}{c}{---} \\
\midrule
\multirow{4}{*}{$\kepler$}& $\Henv$    & $ 7.0.10^{-4}$     & $-4.3.10^{-2}$     & $ 1.8.10^{-3}$     & $-2.4.10^{-5}$    & $ 1.5.10^{-7}$     & $-4.4.10^{-10}$   & $ 5.0.10^{-13}$\\
                          & $\numax$   & $-1.2.10^{-3}$     & $ 2.3.10^{-3}$     & $-6.2.10^{-5}$     & $ 6.6.10^{-7}$    & $-3.3.10^{-9}$     & $ 8.0.10^{-12}$   & $-7.4.10^{-15}$\\
                          & $\taueff$  & $-3.7.10^{-1}$     & $ 1.2.10^{-2}$     & $-3.6.10^{-4}$     & $ 5.7.10^{-6}$    & $-4.4.10^{-8}$     & $ 1.5.10^{-10}$   & $-1.9.10^{-13}$ \\
                          & $\sigma_\mathrm{g}^2$ & $ 1.5.10^{-1}$     & $-1.8.10^{-2}$     & $ 5.6.10^{-4}$     & $-7.8.10^{-6}$    & $ 5.6.10^{-8}$     & $-1.9.10^{-10}$   & $ 2.3.10^{-13}$  \\
\bottomrule
\end{tabular*}
\end{sidewaystable}

\section{Results from the analysis of CoRoT and $\kepler$ datasets} \label{tableau_resultats}

This section collects additional material that we refer to in Sect.~\ref{application}.

\begin{center}
\begin{table*}
\centering
    \caption{Scaling relations estimated as a power law of $\numax$ ($(\alpha \pm \delta\alpha) \numax^{\beta \pm \delta\beta}$)
    used as references to compare with our own deduced scaling relations (see Tab. \ref{relation_echelle_resultats}).
    They come from theoretical scaling relations, observational studies and 3D hydrodynamical models of the surface layer of the stars.     
    The number in parentheses indicates the reference where these relations are given:
    (1) \cite{2013LNP...865..179B}; (2) \cite{2012A&A...537A..30M}; (3) \cite{2013sf2a.conf...25M}; (4) \cite{2011A&A...529L...8K}; (5) \cite{2011ApJ...741..119M};
    (6) \cite{2014A&A...570A..41K}; (7) \cite{2009CoAst.160...74H}; (8) \cite{2013A&A...559A..40S}; (9) \cite{2006A&A...445..661L}; (10) \cite{2010A&A...517A..22M}. \label{relation_echelle_ref}}
  \begin{tabular}{c | l | l | l }
 \toprule
   Parameter &\multicolumn{1}{c|}{Theoretical}& \multicolumn{1}{c|}{Observations}              & \multicolumn{1}{c}{Models}\\
\midrule
    $\Henv$   & $\propto \numax^{-1,73}$ (1)   & $(2.03\pm0.05)10^7~\numax^{-2.38\pm0.01}$ (2)  & \multicolumn{1}{c}{---}       \\ 
              &                                &                                                &                               \\ 
    $\Dnu$    & $\propto \numax^{0.75}$ (3)    & $(0.276\pm0.002)~\numax^{0.751\pm0.002}$ (2)   & \multicolumn{1}{c}{---}       \\ 
              &                                &                                                &                               \\ 
    $\Pgran$  & $\propto \numax^{-2}$ (4)      & $\numax^{-1.90 \pm 0.01}$ (5)                  & \multicolumn{1}{c}{---}       \\ 
              &                                & $\numax^{-2.1}$ (6)                            & \multicolumn{1}{c}{---}       \\ 
              &                                &                                                &                               \\ 
    $\taueff$ & $\propto \numax^{-1}$ (7)      & $\numax^{-0.89 \pm 0.005}$ (5)                 & $3.28 \times 10^5~\numax^{-0.94}$ (8) \\
              &                                & $(836 \pm 4)~\numax^{-0.886\pm 0.002}$  (6)    &                               \\ 
              &                                &                                                &                               \\     
    $\sigma_\mathrm{g}$& $\propto \numax^{-0.5}$ (9)    & $(3382 \pm 9)~\numax^{-0.609\pm0.002}$ (6)     & $2.42 \times 10^3~\numax^{-0.51}$ (8) \\
              &                                &                                                &                               \\     
$R/\mathrm{R_\odot}$& $\propto \numax^{-0.5}$ (10) & $(56.7 \pm 1.0)~\numax^{-0.48\pm0.01}$ (10)  & \multicolumn{1}{c}{---}       \\
 \bottomrule
  \end{tabular}
\end{table*}
\end{center}

\begin{sidewaystable}
\vspace{8cm}
    \caption{Scaling relations estimated as a power law of $\numax$ ($(\alpha \pm \delta\alpha) \numax^{\beta \pm \delta\beta}$) for the following parameters: 
    the height of the Gaussian envelope $\Henv$, the mean large separation $\Dnu$,  the radius $R/\mathrm{R_\odot}$,  the granulation effective timescale $\taueff$,
    the granulation characteristic amplitude $\sigma_\mathrm{g}^2$, the power of the granulation at the zero frequency $\Pgran$, and the characteristic granulation timescale $\taugran$. These scaling relations given are determined from a least-squares fit taking into account the internal errors of the parameters following both axes.
    The reduced $\chi^2$  is indicated for each relation as well as the number of stars of each dataset used.
    For the seismic indices ($\Henv$ and $\Dnu$) and the radius ($R/\mathrm{R_\odot}$), 
    we used the datasets $S_\mathrm{s}$, for which we successfully extracted  only the seismic indices.     Regarding the granulation parameters ($\taueff$, $\sigma_\mathrm{g}^2$, $\Pgran$ and $\taugran$),      we used the datasets $S_\mathrm{s+g}$, for which we extracted successfully both the seismic indices and the granulation parameters.
    In the case of $\kepler$, we adjusted only results with $\numax<100~\mu$Hz for the granulation parameters (see Sect. \ref{taueff} for more details). 
    Information about the evolutionary stage (clump and RGB stars) is obtained taken from \cite{2016A&A...588A..87V} and the effective temperature ($\Teff$)
    used to derive the fundamental parameters come from different catalogues (see Sect. \ref{scaling_relations} for more detail). \label{relation_echelle_resultats}} 
 \begin{tabular*}{\textheight}{c | c | c c c c | c c c c}
 \toprule
   \multirow{3}{*}{Parameter} & \multirow{3}{*}{Population} & \multicolumn{8}{c}{Scaling relations: $(\alpha \pm \delta\alpha) \numax^{\beta \pm \delta\beta}$ } \\
                              &        & \multicolumn{4}{c|}{$\kepler$}                                                      & \multicolumn{4}{c}{CoRoT} \\
                              &        & Coefficient $\alpha$    & Exponent $\beta$   & reduced $\chi^2$   & Number  &  Coefficient $\alpha$    &  Exponent $\beta$   & reduced $\chi^2$  & Number  \\
\midrule
\multirow{3}{*}{$H_{\mathrm{env}}$}    &  all  & $(2.021 \pm 0.005) 10^{7}$ & $-1.9029 \pm 0.0006$  &  103.3  & 17179        & $(9.6 \pm 0.2) 10^{6}$   & $-1.835 \pm 0.006$  &    6.5 & 2943 \\ 
                                       & Clump & $(8.0 \pm 0.1) 10^{8}$     & $-2.921 \pm 0.003$    &   28.4  & 3152         &  ---                     & ---                 & ---    &  ---      \\ 
                                       &  RGB  & $(2.19 \pm 0.03) 10^{7}$   & $-1.863 \pm 0.003$    &   93.4  & 1333         &  ---                     & ---                 & ---    &  ---      \\  
    \hline
\multirow{3}{*}{$\Delta\nu$}           &  all  & $0.27591 \pm 0.00007$      & $0.76236 \pm 0.00006$ &  159.9  & 17179        & $0.2618 \pm 0.0007$      & $0.7702 \pm 0.0008$ &   10.3 & 2943 \\ 
                                       & Clump & $0.4841 \pm 0.0004$        & $0.6070 \pm 0.0002$   &  105.4  & 3152         &  ---                     & ---                 & ---    &  ---      \\ 
                                       &  RGB  & $0.2794 \pm 0.0003$        & $0.7624 \pm 0.0003$   &  139.2  & 1333         &  ---                     & ---                 & ---    &  ---      \\ 
     \hline
 \multirow{1}{*}{$R/\mathrm{R_\odot}$}  &  all  & $62.06 \pm 0.04$           & $-0.4780 \pm 0.0002$  &   97.8  & 17179        &  ---                     & ---                 & ---    &  ---      \\ 
\hline
 \multirow{3}{*}{$\tau_{\mathrm{eff}}$} &  all  & $(2.770 \pm 0.006) 10^{5}$ & $-0.8065 \pm 0.0006$  &   31.3  & 13789        & $(1.49 \pm 0.06) 10^{6}$ & $-1.31 \pm 0.01$    &   11.5 & 806 \\
                                        & Clump & $(1.59 \pm 0.01) 10^{5}$   & $-0.675 \pm 0.002$    &   25.8  & 3140         &  ---                     & ---                 & ---    &  ---      \\ 
                                        &  RGB  & $(9.4 \pm 0.2) 10^{5}$     & $-1.080 \pm 0.006$    &   15.5  & 666          &  ---                     & ---                 & ---    &  ---      \\ 
     \hline
 \multirow{3}{*}{$\sigma_\mathrm{g}^2$} &  all  & $(4.300 \pm 0.005) 10^{7}$ & $-1.3802 \pm 0.0003$  &  357.1  & 13789        & $(1.10 \pm 0.02) 10^{8}$ & $-1.635 \pm 0.005$  &   80.2 & 806 \\ 
                                        & Clump & $(3.56 \pm 0.01) 10^{8}$   & $-1.978 \pm 0.001$    &  145.3  & 3140         &  ---                     & ---                 & ---    &  ---      \\ 
                                        &  RGB  & $(1.10 \pm 0.01) 10^{7}$   & $-1.017 \pm 0.002$    &  336.3  & 666          &  ---                     & ---                 & ---    &  ---      \\ 
     \hline 
 \multirow{3}{*}{$P_{\mathrm{gran}}$}   &  all  & $(1.306 \pm 0.004) 10^{7}$ & $-1.9180 \pm 0.0008$  &   66.1  & 13789        & $(2.5 \pm 0.1) 10^{7}$   & $-2.04 \pm 0.02$    &    7.7 & 806 \\ 
                                        & Clump & $(9.10 \pm 0.09) 10^{7}$   & $-2.496 \pm 0.003$    &   46.0  & 3140         &  ---                     & ---                 & ---    &  ---      \\ 
                                        &  RGB  & $(4.9 \pm 0.1) 10^{6}$     & $-1.635 \pm 0.007$    &   54.6  & 666          &  ---                     & ---                 & ---    &  ---      \\ 
     \hline
 \multirow{3}{*}{$\tau_{\mathrm{gran}}$}&  all  & $(5.72 \pm 0.01) 10^{4}$   & $-0.5332 \pm 0.0006$  &   72.3  & 13789        & $(9.4 \pm 0.4) 10^{5}$   & $-1.28 \pm 0.01$    &   14.8 & 806 \\ 
                                        & Clump & $(2.20 \pm 0.02) 10^{4}$   & $-0.306 \pm 0.002$    &   51.7  & 3140         &  ---                     & ---                 & ---    &  ---      \\ 
                                        &  RGB  & $(3.62 \pm 0.09) 10^{4}$   & $-0.379 \pm 0.006$    &   40.8  & 666          &  ---                     & ---                 & ---    &  ---      \\ 
\bottomrule
   \end{tabular*}
\end{sidewaystable}



\bibliography{MLE_UP_AN} 

\begin{thebibliography}{}

\bibitem [\protect \citeauthoryear {%
{Andersen}%
\ \protect \BOthers {.}}{%
{Andersen}%
\ \protect \BOthers {.}}{%
{\protect \APACyear {1998}}%
}]{%
Andersen1998}
\APACinsertmetastar {%
Andersen1998}%
\begin{APACrefauthors}%
{Andersen}, B.%
, {Appourchaux}, T.%
, {Crommelnynck}, D.%
, {Frohlich}, D.%
, {Jimenez}, A.%
, {Rabello Soares}, M.%
\BCBL {}\ \BBA {} {Wehrli}, C.%
\end{APACrefauthors}%
\unskip\
\newblock
\APACrefYear{1998},
\newblock
(\BVOL~181; J.~{Provost}\ \BBA {} F.~{Schmider}, \BEDS{}).
\PrintBackRefs{\CurrentBib}

\bibitem [\protect \citeauthoryear {%
{Anderson}%
, {Duvall}%
\BCBL {}\ \BBA {} {Jefferies}%
}{%
{Anderson}%
\ \protect \BOthers {.}}{%
{\protect \APACyear {1990}}%
}]{%
1990ApJ...364..699A}
\APACinsertmetastar {%
1990ApJ...364..699A}%
\begin{APACrefauthors}%
{Anderson}, E\BPBI R.%
, {Duvall}, T\BPBI L., Jr.%
\BCBL {}\ \BBA {} {Jefferies}, S\BPBI M.%
\end{APACrefauthors}%
\unskip\
\newblock
\APACrefYearMonthDay{1990}{{\APACmonth{12}}}{},
\newblock
\unskip
\newblock
\APACjournalVolNumPages{\apj}{364}{}{699-705}.
\newblock
\begin{APACrefDOI} \doi{10.1086/169452} \end{APACrefDOI}
\PrintBackRefs{\CurrentBib}

\bibitem [\protect \citeauthoryear {%
{Appourchaux}%
, {Gizon}%
\BCBL {}\ \BBA {} {Rabello-Soares}%
}{%
{Appourchaux}%
\ \protect \BOthers {.}}{%
{\protect \APACyear {1998}}%
}]{%
1998A&AS..132..107A}
\APACinsertmetastar {%
1998A&AS..132..107A}%
\begin{APACrefauthors}%
{Appourchaux}, T.%
, {Gizon}, L.%
\BCBL {}\ \BBA {} {Rabello-Soares}, M\BHBI C.%
\end{APACrefauthors}%
\unskip\
\newblock
\APACrefYearMonthDay{1998}{{\APACmonth{10}}}{},
\newblock
\unskip
\newblock
\APACjournalVolNumPages{\aaps}{132}{}{107-119}.
\newblock
\begin{APACrefDOI} \doi{10.1051/aas:1998441} \end{APACrefDOI}
\PrintBackRefs{\CurrentBib}

\bibitem [\protect \citeauthoryear {%
{Auvergne}%
\ \protect \BOthers {.}}{%
{Auvergne}%
\ \protect \BOthers {.}}{%
{\protect \APACyear {2009}}%
}]{%
2009A&A...506..411A}
\APACinsertmetastar {%
2009A&A...506..411A}%
\begin{APACrefauthors}%
{Auvergne}, M.%
, {Bodin}, P.%
, {Boisnard}, L.%
\ et al.\end{APACrefauthors}%
\unskip\
\newblock
\APACrefYearMonthDay{2009}{{\APACmonth{10}}}{},
\newblock
\unskip
\newblock
\APACjournalVolNumPages{\aap}{506}{}{411-424}.
\newblock
\begin{APACrefDOI} \doi{10.1051/0004-6361/200810860} \end{APACrefDOI}
\PrintBackRefs{\CurrentBib}

\bibitem [\protect \citeauthoryear {%
{Baglin}%
\ \protect \BOthers {.}}{%
{Baglin}%
\ \protect \BOthers {.}}{%
{\protect \APACyear {2006}}%
}]{%
2006ESASP1306...33B}
\APACinsertmetastar {%
2006ESASP1306...33B}%
\begin{APACrefauthors}%
{Baglin}, A.%
, {Auvergne}, M.%
, {Barge}, P.%
\ et al.\end{APACrefauthors}%
\unskip\
\newblock
\APACrefYearMonthDay{2006}{{\APACmonth{11}}}{},
\newblock
{\BBOQ}\APACrefatitle {{Scientific Objectives for a Minisat: CoRoT}}
  {{Scientific Objectives for a Minisat: CoRoT}}.{\BBCQ}
\newblock
\BIn{} M.~{Fridlund}, A.~{Baglin}, J.~{Lochard}\BCBL {}\ \BOthers {.}\ (\BEDS),
  \APACrefbtitle {ESA Special Publication} {ESA Special Publication}\ \BVOL\
  1306, \BPG~33.
\PrintBackRefs{\CurrentBib}

\bibitem [\protect \citeauthoryear {%
{Baglin}%
\ \protect \BOthers {.}}{%
{Baglin}%
\ \protect \BOthers {.}}{%
{\protect \APACyear {2009}}%
}]{%
2009IAUS..253...71B}
\APACinsertmetastar {%
2009IAUS..253...71B}%
\begin{APACrefauthors}%
{Baglin}, A.%
, {Auvergne}, M.%
, {Barge}, P.%
, {Deleuil}, M.%
, {Michel}, E.%
\BCBL {}\ \BBA {} {CoRoT Exoplanet Science Team}.%
\end{APACrefauthors}%
\unskip\
\newblock
\APACrefYearMonthDay{2009}{{\APACmonth{02}}}{},
\newblock
{\BBOQ}\APACrefatitle {{CoRoT: Description of the Mission and Early Results}}
  {{CoRoT: Description of the Mission and Early Results}}.{\BBCQ}
\newblock
\BIn{} F.~{Pont}, D.~{Sasselov}\BCBL {}\ \BBA {} M\BPBI J.~{Holman}\ (\BEDS),
  \APACrefbtitle {IAU Symposium} {IAU Symposium}\ \BVOL~253, \BPG~71-81.
\newblock
\begin{APACrefDOI} \doi{10.1017/S1743921308026252} \end{APACrefDOI}
\PrintBackRefs{\CurrentBib}

\bibitem [\protect \citeauthoryear {%
{Belkacem}%
}{%
{Belkacem}%
}{%
{\protect \APACyear {2012}}%
}]{%
2012sf2a.conf..173B}
\APACinsertmetastar {%
2012sf2a.conf..173B}%
\begin{APACrefauthors}%
{Belkacem}, K.%
\end{APACrefauthors}%
\unskip\
\newblock
\APACrefYearMonthDay{2012}{{\APACmonth{12}}}{},
\newblock
{\BBOQ}\APACrefatitle {{Determination of the stars fundamental parameters using
  seismic scaling relations}} {{Determination of the stars fundamental
  parameters using seismic scaling relations}}.{\BBCQ}
\newblock
\BIn{} S.~{Boissier}, P.~{de Laverny}, N.~{Nardetto}\BCBL {}\ \BOthers {.}\
  (\BEDS), \APACrefbtitle {SF2A-2012: Proceedings of the Annual meeting of the
  French Society of Astronomy and Astrophysics} {SF2A-2012: Proceedings of the
  Annual meeting of the French Society of Astronomy and Astrophysics}\
  \BPG~173-188.
\PrintBackRefs{\CurrentBib}

\bibitem [\protect \citeauthoryear {%
{Belkacem}%
\ \protect \BOthers {.}}{%
{Belkacem}%
\ \protect \BOthers {.}}{%
{\protect \APACyear {2011}}%
}]{%
2011A&A...530A.142B}
\APACinsertmetastar {%
2011A&A...530A.142B}%
\begin{APACrefauthors}%
{Belkacem}, K.%
, {Goupil}, M\BPBI J.%
, {Dupret}, M\BPBI A.%
, {Samadi}, R.%
, {Baudin}, F.%
, {Noels}, A.%
\BCBL {}\ \BBA {} {Mosser}, B.%
\end{APACrefauthors}%
\unskip\
\newblock
\APACrefYearMonthDay{2011}{{\APACmonth{06}}}{},
\newblock
\unskip
\newblock
\APACjournalVolNumPages{\aap}{530}{}{A142}.
\newblock
\begin{APACrefDOI} \doi{10.1051/0004-6361/201116490} \end{APACrefDOI}
\PrintBackRefs{\CurrentBib}

\bibitem [\protect \citeauthoryear {%
{Belkacem}%
\ \BBA {} {Samadi}%
}{%
{Belkacem}%
\ \BBA {} {Samadi}%
}{%
{\protect \APACyear {2013}}%
}]{%
2013LNP...865..179B}
\APACinsertmetastar {%
2013LNP...865..179B}%
\begin{APACrefauthors}%
{Belkacem}, K.%
\BCBT {}\ \BBA {} {Samadi}, R.%
\end{APACrefauthors}%
\unskip\
\newblock
\APACrefYearMonthDay{2013}{}{},
\newblock
{\BBOQ}\APACrefatitle {{Connections Between Stellar Oscillations and Turbulent
  Convection}} {{Connections Between Stellar Oscillations and Turbulent
  Convection}}.{\BBCQ}
\newblock
\BIn{} M.~{Goupil}, K.~{Belkacem}, C.~{Neiner}\BCBL {}\ \BOthers {.}\ (\BEDS),
  \APACrefbtitle {Lecture Notes in Physics, Berlin Springer Verlag} {Lecture
  Notes in Physics, Berlin Springer Verlag}\ \BVOL~865, \BPG~179.
\newblock
\begin{APACrefDOI} \doi{10.1007/978-3-642-33380-4_9} \end{APACrefDOI}
\PrintBackRefs{\CurrentBib}

\bibitem [\protect \citeauthoryear {%
{Belkacem}%
, {Samadi}%
, {Mosser}%
, {Goupil}%
\BCBL {}\ \BBA {} {Ludwig}%
}{%
{Belkacem}%
\ \protect \BOthers {.}}{%
{\protect \APACyear {2013}}%
}]{%
2013ASPC..479...61B}
\APACinsertmetastar {%
2013ASPC..479...61B}%
\begin{APACrefauthors}%
{Belkacem}, K.%
, {Samadi}, R.%
, {Mosser}, B.%
, {Goupil}, M\BHBI J.%
\BCBL {}\ \BBA {} {Ludwig}, H\BHBI G.%
\end{APACrefauthors}%
\unskip\
\newblock
\APACrefYearMonthDay{2013}{{\APACmonth{12}}}{},
\newblock
{\BBOQ}\APACrefatitle {{On the Seismic Scaling Relations {$\Delta$}{$\nu$} {--}
  {$\rho$} and {$\nu$}$_{max}$ {--} {$\nu$}$_{c}$}} {{On the Seismic Scaling
  Relations {$\Delta$}{$\nu$} {--} {$\rho$} and {$\nu$}$_{max}$ {--}
  {$\nu$}$_{c}$}}.{\BBCQ}
\newblock
\BIn{} H.~{Shibahashi}\ \BBA {} A\BPBI E.~{Lynas-Gray}\ (\BEDS), \APACrefbtitle
  {Progress in Physics of the Sun and Stars: A New Era in Helio- and
  Asteroseismology} {Progress in Physics of the Sun and Stars: A New Era in
  Helio- and Asteroseismology}\ \BVOL~479, \BPG~61.
\PrintBackRefs{\CurrentBib}

\bibitem [\protect \citeauthoryear {%
{Borucki}%
\ \protect \BOthers {.}}{%
{Borucki}%
\ \protect \BOthers {.}}{%
{\protect \APACyear {2010}}%
}]{%
2010AAS...21510101B}
\APACinsertmetastar {%
2010AAS...21510101B}%
\begin{APACrefauthors}%
{Borucki}, W\BPBI J.%
, {Koch}, D.%
, {Basri}, G.%
\ et al.\end{APACrefauthors}%
\unskip\
\newblock
\APACrefYearMonthDay{2010}{{\APACmonth{01}}}{},
\newblock
{\BBOQ}\APACrefatitle {{Kepler Planet Detection Mission: Introduction and First
  Results}} {{Kepler Planet Detection Mission: Introduction and First
  Results}}.{\BBCQ}
\newblock
\BIn{} \APACrefbtitle {American Astronomical Society Meeting Abstracts 215}
  {American Astronomical Society Meeting Abstracts 215}\ \BVOL~42, \BPG~101.01.
\PrintBackRefs{\CurrentBib}

\bibitem [\protect \citeauthoryear {%
{Brown}%
, {Gilliland}%
, {Noyes}%
\BCBL {}\ \BBA {} {Ramsey}%
}{%
{Brown}%
\ \protect \BOthers {.}}{%
{\protect \APACyear {1991}}%
}]{%
1991ApJ...368..599B}
\APACinsertmetastar {%
1991ApJ...368..599B}%
\begin{APACrefauthors}%
{Brown}, T\BPBI M.%
, {Gilliland}, R\BPBI L.%
, {Noyes}, R\BPBI W.%
\BCBL {}\ \BBA {} {Ramsey}, L\BPBI W.%
\end{APACrefauthors}%
\unskip\
\newblock
\APACrefYearMonthDay{1991}{{\APACmonth{02}}}{},
\newblock
\unskip
\newblock
\APACjournalVolNumPages{\apj}{368}{}{599-609}.
\newblock
\begin{APACrefDOI} \doi{10.1086/169725} \end{APACrefDOI}
\PrintBackRefs{\CurrentBib}

\bibitem [\protect \citeauthoryear {%
{Brown}%
, {Latham}%
, {Everett}%
\BCBL {}\ \BBA {} {Esquerdo}%
}{%
{Brown}%
\ \protect \BOthers {.}}{%
{\protect \APACyear {2011}}%
}]{%
2011AJ....142..112B}
\APACinsertmetastar {%
2011AJ....142..112B}%
\begin{APACrefauthors}%
{Brown}, T\BPBI M.%
, {Latham}, D\BPBI W.%
, {Everett}, M\BPBI E.%
\BCBL {}\ \BBA {} {Esquerdo}, G\BPBI A.%
\end{APACrefauthors}%
\unskip\
\newblock
\APACrefYearMonthDay{2011}{{\APACmonth{10}}}{},
\newblock
\unskip
\newblock
\APACjournalVolNumPages{\aj}{142}{}{112}.
\newblock
\begin{APACrefDOI} \doi{10.1088/0004-6256/142/4/112} \end{APACrefDOI}
\PrintBackRefs{\CurrentBib}

\bibitem [\protect \citeauthoryear {%
{Casagrande}%
\ \protect \BOthers {.}}{%
{Casagrande}%
\ \protect \BOthers {.}}{%
{\protect \APACyear {2014}}%
}]{%
2014ApJ...787..110C}
\APACinsertmetastar {%
2014ApJ...787..110C}%
\begin{APACrefauthors}%
{Casagrande}, L.%
, {Silva Aguirre}, V.%
, {Stello}, D.%
\ et al.\end{APACrefauthors}%
\unskip\
\newblock
\APACrefYearMonthDay{2014}{{\APACmonth{06}}}{},
\newblock
\unskip
\newblock
\APACjournalVolNumPages{\apj}{787}{}{110}.
\newblock
\begin{APACrefDOI} \doi{10.1088/0004-637X/787/2/110} \end{APACrefDOI}
\PrintBackRefs{\CurrentBib}

\bibitem [\protect \citeauthoryear {%
{Chaplin}%
\ \protect \BOthers {.}}{%
{Chaplin}%
\ \protect \BOthers {.}}{%
{\protect \APACyear {2014}}%
}]{%
2014MNRAS.445..946C}
\APACinsertmetastar {%
2014MNRAS.445..946C}%
\begin{APACrefauthors}%
{Chaplin}, W\BPBI J.%
, {Elsworth}, Y.%
, {Davies}, G\BPBI R.%
, {Campante}, T\BPBI L.%
, {Handberg}, R.%
, {Miglio}, A.%
\BCBL {}\ \BBA {} {Basu}, S.%
\end{APACrefauthors}%
\unskip\
\newblock
\APACrefYearMonthDay{2014}{{\APACmonth{11}}}{},
\newblock
\unskip
\newblock
\APACjournalVolNumPages{\mnras}{445}{}{946-954}.
\newblock
\begin{APACrefDOI} \doi{10.1093/mnras/stu1811} \end{APACrefDOI}
\PrintBackRefs{\CurrentBib}

\bibitem [\protect \citeauthoryear {%
{Chaplin}%
\ \BBA {} {Miglio}%
}{%
{Chaplin}%
\ \BBA {} {Miglio}%
}{%
{\protect \APACyear {2013}}%
}]{%
2013ARA&A..51..353C}
\APACinsertmetastar {%
2013ARA&A..51..353C}%
\begin{APACrefauthors}%
{Chaplin}, W\BPBI J.%
\BCBT {}\ \BBA {} {Miglio}, A.%
\end{APACrefauthors}%
\unskip\
\newblock
\APACrefYearMonthDay{2013}{{\APACmonth{08}}}{},
\newblock
\unskip
\newblock
\APACjournalVolNumPages{\araa}{51}{}{353-392}.
\newblock
\begin{APACrefDOI} \doi{10.1146/annurev-astro-082812-140938} \end{APACrefDOI}
\PrintBackRefs{\CurrentBib}

\bibitem [\protect \citeauthoryear {%
{CoRoT Team}%
}{%
{CoRoT Team}%
}{%
{\protect \APACyear {2016}}%
}]{%
2016cole.book.....C}
\APACinsertmetastar {%
2016cole.book.....C}%
\begin{APACrefauthors}%
{CoRoT Team}.%
\end{APACrefauthors}%
\unskip\
\newblock
\APACrefYear{2016},
\newblock
\APACrefbtitle {{The CoRoT Legacy Book: The adventure of the ultra high
  precision photometry from space, by the CoRoT Team}} {{The CoRoT Legacy Book:
  The adventure of the ultra high precision photometry from space, by the CoRoT
  Team}}.
\newblock
\APACaddressPublisher{}{EDP Sciences}.
\newblock
\begin{APACrefDOI} \doi{10.1051/978-2-7598-1876-1} \end{APACrefDOI}
\PrintBackRefs{\CurrentBib}

\bibitem [\protect \citeauthoryear {%
{de Assis Peralta}%
, {Samadi}%
\BCBL {}\ \BBA {} {Michel}%
}{%
{de Assis Peralta}%
\ \protect \BOthers {.}}{%
{\protect \APACyear {2017}}%
}]{%
2017EPJWC.16001012D}
\APACinsertmetastar {%
2017EPJWC.16001012D}%
\begin{APACrefauthors}%
{de Assis Peralta}, R.%
, {Samadi}, R.%
\BCBL {}\ \BBA {} {Michel}, {\'E}.%
\end{APACrefauthors}%
\unskip\
\newblock
\APACrefYearMonthDay{2017}{{\APACmonth{10}}}{},
\newblock
{\BBOQ}\APACrefatitle {{Extraction of seismic indices and stellar granulation
  parameters for CoRoT and Kepler red giants using the MLEUP method. Main
  results and perspectives}} {{Extraction of seismic indices and stellar
  granulation parameters for CoRoT and Kepler red giants using the MLEUP
  method. Main results and perspectives}}.{\BBCQ}
\newblock
\BIn{} \APACrefbtitle {European Physical Journal Web of Conferences} {European
  Physical Journal Web of Conferences}\ \BVOL~160, \BPG~01012.
\newblock
\begin{APACrefDOI} \doi{10.1051/epjconf/201716001012} \end{APACrefDOI}
\PrintBackRefs{\CurrentBib}

\bibitem [\protect \citeauthoryear {%
{De Ridder}%
\ \protect \BOthers {.}}{%
{De Ridder}%
\ \protect \BOthers {.}}{%
{\protect \APACyear {2009}}%
}]{%
2009Natur.459..398D}
\APACinsertmetastar {%
2009Natur.459..398D}%
\begin{APACrefauthors}%
{De Ridder}, J.%
, {Barban}, C.%
, {Baudin}, F.%
\ et al.\end{APACrefauthors}%
\unskip\
\newblock
\APACrefYearMonthDay{2009}{{\APACmonth{05}}}{},
\newblock
\unskip
\newblock
\APACjournalVolNumPages{\nat}{459}{}{398-400}.
\newblock
\begin{APACrefDOI} \doi{10.1038/nature08022} \end{APACrefDOI}
\PrintBackRefs{\CurrentBib}

\bibitem [\protect \citeauthoryear {%
{Garc{\'{\i}}a}%
\ \protect \BOthers {.}}{%
{Garc{\'{\i}}a}%
\ \protect \BOthers {.}}{%
{\protect \APACyear {2010}}%
}]{%
2010Sci...329.1032G}
\APACinsertmetastar {%
2010Sci...329.1032G}%
\begin{APACrefauthors}%
{Garc{\'{\i}}a}, R\BPBI A.%
, {Mathur}, S.%
, {Salabert}, D.%
, {Ballot}, J.%
, {R{\'e}gulo}, C.%
, {Metcalfe}, T\BPBI S.%
\BCBL {}\ \BBA {} {Baglin}, A.%
\end{APACrefauthors}%
\unskip\
\newblock
\APACrefYearMonthDay{2010}{{\APACmonth{08}}}{},
\newblock
\unskip
\newblock
\APACjournalVolNumPages{Science}{329}{}{1032}.
\newblock
\begin{APACrefDOI} \doi{10.1126/science.1191064} \end{APACrefDOI}
\PrintBackRefs{\CurrentBib}

\bibitem [\protect \citeauthoryear {%
{Gilliland}%
\ \protect \BOthers {.}}{%
{Gilliland}%
\ \protect \BOthers {.}}{%
{\protect \APACyear {2010}}%
}]{%
2010PASP..122..131G}
\APACinsertmetastar {%
2010PASP..122..131G}%
\begin{APACrefauthors}%
{Gilliland}, R\BPBI L.%
, {Brown}, T\BPBI M.%
, {Christensen-Dalsgaard}, J.%
\ et al.\end{APACrefauthors}%
\unskip\
\newblock
\APACrefYearMonthDay{2010}{{\APACmonth{02}}}{},
\newblock
\unskip
\newblock
\APACjournalVolNumPages{\pasp}{122}{}{131-143}.
\newblock
\begin{APACrefDOI} \doi{10.1086/650399} \end{APACrefDOI}
\PrintBackRefs{\CurrentBib}

\bibitem [\protect \citeauthoryear {%
{Girardi}%
}{%
{Girardi}%
}{%
{\protect \APACyear {1999}}%
}]{%
1999MNRAS.308..818G}
\APACinsertmetastar {%
1999MNRAS.308..818G}%
\begin{APACrefauthors}%
{Girardi}, L.%
\end{APACrefauthors}%
\unskip\
\newblock
\APACrefYearMonthDay{1999}{{\APACmonth{09}}}{},
\newblock
\unskip
\newblock
\APACjournalVolNumPages{\mnras}{308}{}{818-832}.
\newblock
\begin{APACrefDOI} \doi{10.1046/j.1365-8711.1999.02746.x} \end{APACrefDOI}
\PrintBackRefs{\CurrentBib}

\bibitem [\protect \citeauthoryear {%
{Grosjean}%
\ \protect \BOthers {.}}{%
{Grosjean}%
\ \protect \BOthers {.}}{%
{\protect \APACyear {2014}}%
}]{%
2014A&A...572A..11G}
\APACinsertmetastar {%
2014A&A...572A..11G}%
\begin{APACrefauthors}%
{Grosjean}, M.%
, {Dupret}, M\BHBI A.%
, {Belkacem}, K.%
, {Montalban}, J.%
, {Samadi}, R.%
\BCBL {}\ \BBA {} {Mosser}, B.%
\end{APACrefauthors}%
\unskip\
\newblock
\APACrefYearMonthDay{2014}{{\APACmonth{12}}}{},
\newblock
\unskip
\newblock
\APACjournalVolNumPages{\aap}{572}{}{A11}.
\newblock
\begin{APACrefDOI} \doi{10.1051/0004-6361/201423827} \end{APACrefDOI}
\PrintBackRefs{\CurrentBib}

\bibitem [\protect \citeauthoryear {%
{Harvey}%
}{%
{Harvey}%
}{%
{\protect \APACyear {1985}}%
}]{%
1985ESASP.235..199H}
\APACinsertmetastar {%
1985ESASP.235..199H}%
\begin{APACrefauthors}%
{Harvey}, J.%
\end{APACrefauthors}%
\unskip\
\newblock
\APACrefYearMonthDay{1985}{{\APACmonth{06}}}{},
\newblock
{\BBOQ}\APACrefatitle {{High-resolution helioseismology}} {{High-resolution
  helioseismology}}.{\BBCQ}
\newblock
\BIn{} E.~{Rolfe}\ \BBA {} B.~{Battrick}\ (\BEDS), \APACrefbtitle {Future
  Missions in Solar, Heliospheric \& Space Plasma Physics} {Future Missions in
  Solar, Heliospheric \& Space Plasma Physics}\ \BVOL~235, \BPG~199-208.
\PrintBackRefs{\CurrentBib}

\bibitem [\protect \citeauthoryear {%
{Hekker}%
\ \protect \BOthers {.}}{%
{Hekker}%
\ \protect \BOthers {.}}{%
{\protect \APACyear {2010}}%
}]{%
2010MNRAS.402.2049H}
\APACinsertmetastar {%
2010MNRAS.402.2049H}%
\begin{APACrefauthors}%
{Hekker}, S.%
, {Broomhall}, A\BHBI M.%
, {Chaplin}, W\BPBI J.%
\ et al.\end{APACrefauthors}%
\unskip\
\newblock
\APACrefYearMonthDay{2010}{{\APACmonth{03}}}{},
\newblock
\unskip
\newblock
\APACjournalVolNumPages{\mnras}{402}{}{2049-2059}.
\newblock
\begin{APACrefDOI} \doi{10.1111/j.1365-2966.2009.16030.x} \end{APACrefDOI}
\PrintBackRefs{\CurrentBib}

\bibitem [\protect \citeauthoryear {%
{Hekker}%
\ \protect \BOthers {.}}{%
{Hekker}%
\ \protect \BOthers {.}}{%
{\protect \APACyear {2011}}%
}]{%
2011A&A...525A.131H}
\APACinsertmetastar {%
2011A&A...525A.131H}%
\begin{APACrefauthors}%
{Hekker}, S.%
, {Elsworth}, Y.%
, {De Ridder}, J.%
\ et al.\end{APACrefauthors}%
\unskip\
\newblock
\APACrefYearMonthDay{2011}{{\APACmonth{01}}}{},
\newblock
\unskip
\newblock
\APACjournalVolNumPages{\aap}{525}{}{A131}.
\newblock
\begin{APACrefDOI} \doi{10.1051/0004-6361/201015185} \end{APACrefDOI}
\PrintBackRefs{\CurrentBib}

\bibitem [\protect \citeauthoryear {%
{Hekker}%
\ \protect \BOthers {.}}{%
{Hekker}%
\ \protect \BOthers {.}}{%
{\protect \APACyear {2012}}%
}]{%
2012A&A...544A..90H}
\APACinsertmetastar {%
2012A&A...544A..90H}%
\begin{APACrefauthors}%
{Hekker}, S.%
, {Elsworth}, Y.%
, {Mosser}, B.%
\ et al.\end{APACrefauthors}%
\unskip\
\newblock
\APACrefYearMonthDay{2012}{{\APACmonth{08}}}{},
\newblock
\unskip
\newblock
\APACjournalVolNumPages{\aap}{544}{}{A90}.
\newblock
\begin{APACrefDOI} \doi{10.1051/0004-6361/201219328} \end{APACrefDOI}
\PrintBackRefs{\CurrentBib}

\bibitem [\protect \citeauthoryear {%
{Hekker}%
\ \protect \BOthers {.}}{%
{Hekker}%
\ \protect \BOthers {.}}{%
{\protect \APACyear {2009}}%
}]{%
2009A&A...506..465H}
\APACinsertmetastar {%
2009A&A...506..465H}%
\begin{APACrefauthors}%
{Hekker}, S.%
, {Kallinger}, T.%
, {Baudin}, F.%
\ et al.\end{APACrefauthors}%
\unskip\
\newblock
\APACrefYearMonthDay{2009}{{\APACmonth{10}}}{},
\newblock
\unskip
\newblock
\APACjournalVolNumPages{\aap}{506}{}{465-469}.
\newblock
\begin{APACrefDOI} \doi{10.1051/0004-6361/200911858} \end{APACrefDOI}
\PrintBackRefs{\CurrentBib}

\bibitem [\protect \citeauthoryear {%
{Huber}%
\ \protect \BOthers {.}}{%
{Huber}%
\ \protect \BOthers {.}}{%
{\protect \APACyear {2010}}%
}]{%
2010ApJ...723.1607H}
\APACinsertmetastar {%
2010ApJ...723.1607H}%
\begin{APACrefauthors}%
{Huber}, D.%
, {Bedding}, T\BPBI R.%
, {Stello}, D.%
\ et al.\end{APACrefauthors}%
\unskip\
\newblock
\APACrefYearMonthDay{2010}{{\APACmonth{11}}}{},
\newblock
\unskip
\newblock
\APACjournalVolNumPages{\apj}{723}{}{1607-1617}.
\newblock
\begin{APACrefDOI} \doi{10.1088/0004-637X/723/2/1607} \end{APACrefDOI}
\PrintBackRefs{\CurrentBib}

\bibitem [\protect \citeauthoryear {%
{Huber}%
\ \protect \BOthers {.}}{%
{Huber}%
\ \protect \BOthers {.}}{%
{\protect \APACyear {2014}}%
}]{%
2014ApJS..211....2H}
\APACinsertmetastar {%
2014ApJS..211....2H}%
\begin{APACrefauthors}%
{Huber}, D.%
, {Silva Aguirre}, V.%
, {Matthews}, J\BPBI M.%
\ et al.\end{APACrefauthors}%
\unskip\
\newblock
\APACrefYearMonthDay{2014}{{\APACmonth{03}}}{},
\newblock
\unskip
\newblock
\APACjournalVolNumPages{\apjs}{211}{}{2}.
\newblock
\begin{APACrefDOI} \doi{10.1088/0067-0049/211/1/2} \end{APACrefDOI}
\PrintBackRefs{\CurrentBib}

\bibitem [\protect \citeauthoryear {%
{Huber}%
\ \protect \BOthers {.}}{%
{Huber}%
\ \protect \BOthers {.}}{%
{\protect \APACyear {2009}}%
}]{%
2009CoAst.160...74H}
\APACinsertmetastar {%
2009CoAst.160...74H}%
\begin{APACrefauthors}%
{Huber}, D.%
, {Stello}, D.%
, {Bedding}, T\BPBI R.%
, {Chaplin}, W\BPBI J.%
, {Arentoft}, T.%
, {Quirion}, P\BHBI O.%
\BCBL {}\ \BBA {} {Kjeldsen}, H.%
\end{APACrefauthors}%
\unskip\
\newblock
\APACrefYearMonthDay{2009}{{\APACmonth{10}}}{},
\newblock
\unskip
\newblock
\APACjournalVolNumPages{Communications in Asteroseismology}{160}{}{74}.
\PrintBackRefs{\CurrentBib}

\bibitem [\protect \citeauthoryear {%
{Kallinger}%
\ \protect \BOthers {.}}{%
{Kallinger}%
\ \protect \BOthers {.}}{%
{\protect \APACyear {2014}}%
}]{%
2014A&A...570A..41K}
\APACinsertmetastar {%
2014A&A...570A..41K}%
\begin{APACrefauthors}%
{Kallinger}, T.%
, {De Ridder}, J.%
, {Hekker}, S.%
\ et al.\end{APACrefauthors}%
\unskip\
\newblock
\APACrefYearMonthDay{2014}{{\APACmonth{10}}}{},
\newblock
\unskip
\newblock
\APACjournalVolNumPages{\aap}{570}{}{A41}.
\newblock
\begin{APACrefDOI} \doi{10.1051/0004-6361/201424313} \end{APACrefDOI}
\PrintBackRefs{\CurrentBib}

\bibitem [\protect \citeauthoryear {%
{Kallinger}%
, {Mosser}%
\BCBL {}\ \protect \BOthers {.}}{%
{Kallinger}%
, {Mosser}%
\BCBL {}\ \protect \BOthers {.}}{%
{\protect \APACyear {2010}}%
}]{%
2010A&A...522A...1K}
\APACinsertmetastar {%
2010A&A...522A...1K}%
\begin{APACrefauthors}%
{Kallinger}, T.%
, {Mosser}, B.%
, {Hekker}, S.%
\ et al.\end{APACrefauthors}%
\unskip\
\newblock
\APACrefYearMonthDay{2010}{{\APACmonth{11}}}{},
\newblock
\unskip
\newblock
\APACjournalVolNumPages{\aap}{522}{}{A1}.
\newblock
\begin{APACrefDOI} \doi{10.1051/0004-6361/201015263} \end{APACrefDOI}
\PrintBackRefs{\CurrentBib}

\bibitem [\protect \citeauthoryear {%
{Kallinger}%
, {Weiss}%
\BCBL {}\ \protect \BOthers {.}}{%
{Kallinger}%
, {Weiss}%
\BCBL {}\ \protect \BOthers {.}}{%
{\protect \APACyear {2010}}%
}]{%
2010A&A...509A..77K}
\APACinsertmetastar {%
2010A&A...509A..77K}%
\begin{APACrefauthors}%
{Kallinger}, T.%
, {Weiss}, W\BPBI W.%
, {Barban}, C.%
\ et al.\end{APACrefauthors}%
\unskip\
\newblock
\APACrefYearMonthDay{2010}{{\APACmonth{01}}}{},
\newblock
\unskip
\newblock
\APACjournalVolNumPages{\aap}{509}{}{A77}.
\newblock
\begin{APACrefDOI} \doi{10.1051/0004-6361/200811437} \end{APACrefDOI}
\PrintBackRefs{\CurrentBib}

\bibitem [\protect \citeauthoryear {%
{Karoff}%
}{%
{Karoff}%
}{%
{\protect \APACyear {2012}}%
}]{%
2012MNRAS.421.3170K}
\APACinsertmetastar {%
2012MNRAS.421.3170K}%
\begin{APACrefauthors}%
{Karoff}, C.%
\end{APACrefauthors}%
\unskip\
\newblock
\APACrefYearMonthDay{2012}{{\APACmonth{04}}}{},
\newblock
\unskip
\newblock
\APACjournalVolNumPages{\mnras}{421}{}{3170-3179}.
\newblock
\begin{APACrefDOI} \doi{10.1111/j.1365-2966.2012.20542.x} \end{APACrefDOI}
\PrintBackRefs{\CurrentBib}

\bibitem [\protect \citeauthoryear {%
{Karoff}%
\ \protect \BOthers {.}}{%
{Karoff}%
\ \protect \BOthers {.}}{%
{\protect \APACyear {2013}}%
}]{%
2013ApJ...767...34K}
\APACinsertmetastar {%
2013ApJ...767...34K}%
\begin{APACrefauthors}%
{Karoff}, C.%
, {Campante}, T\BPBI L.%
, {Ballot}, J.%
\ et al.\end{APACrefauthors}%
\unskip\
\newblock
\APACrefYearMonthDay{2013}{{\APACmonth{04}}}{},
\newblock
\unskip
\newblock
\APACjournalVolNumPages{\apj}{767}{}{34}.
\newblock
\begin{APACrefDOI} \doi{10.1088/0004-637X/767/1/34} \end{APACrefDOI}
\PrintBackRefs{\CurrentBib}

\bibitem [\protect \citeauthoryear {%
{Kendall}%
\ \BBA {} {Stuart}%
}{%
{Kendall}%
\ \BBA {} {Stuart}%
}{%
{\protect \APACyear {1967}}%
}]{%
Kendall1967}
\APACinsertmetastar {%
Kendall1967}%
\begin{APACrefauthors}%
{Kendall}, M.%
\BCBT {}\ \BBA {} {Stuart}, A.%
\end{APACrefauthors}%
\unskip\
\newblock
\APACrefYear{1967},
\newblock
\APACrefbtitle {The advanced theory of Astrophysics: Inference and
  relationship} {The advanced theory of Astrophysics: Inference and
  relationship}\ (\BVOL~II).
\PrintBackRefs{\CurrentBib}

\bibitem [\protect \citeauthoryear {%
{Kippenhahn}%
\ \BBA {} {Weigert}%
}{%
{Kippenhahn}%
\ \BBA {} {Weigert}%
}{%
{\protect \APACyear {1994}}%
}]{%
1994sse..book.....K}
\APACinsertmetastar {%
1994sse..book.....K}%
\begin{APACrefauthors}%
{Kippenhahn}, R.%
\BCBT {}\ \BBA {} {Weigert}, A.%
\end{APACrefauthors}%
\unskip\
\newblock
\APACrefYear{1994},
\newblock
\APACrefbtitle {{Stellar Structure and Evolution}} {{Stellar Structure and
  Evolution}}.
\PrintBackRefs{\CurrentBib}

\bibitem [\protect \citeauthoryear {%
{Kjeldsen}%
\ \BBA {} {Bedding}%
}{%
{Kjeldsen}%
\ \BBA {} {Bedding}%
}{%
{\protect \APACyear {1995}}%
}]{%
1995A&A...293...87K}
\APACinsertmetastar {%
1995A&A...293...87K}%
\begin{APACrefauthors}%
{Kjeldsen}, H.%
\BCBT {}\ \BBA {} {Bedding}, T\BPBI R.%
\end{APACrefauthors}%
\unskip\
\newblock
\APACrefYearMonthDay{1995}{{\APACmonth{01}}}{},
\newblock
\unskip
\newblock
\APACjournalVolNumPages{\aap}{293}{}{87-106}.
\PrintBackRefs{\CurrentBib}

\bibitem [\protect \citeauthoryear {%
{Kjeldsen}%
\ \BBA {} {Bedding}%
}{%
{Kjeldsen}%
\ \BBA {} {Bedding}%
}{%
{\protect \APACyear {2011}}%
}]{%
2011A&A...529L...8K}
\APACinsertmetastar {%
2011A&A...529L...8K}%
\begin{APACrefauthors}%
{Kjeldsen}, H.%
\BCBT {}\ \BBA {} {Bedding}, T\BPBI R.%
\end{APACrefauthors}%
\unskip\
\newblock
\APACrefYearMonthDay{2011}{{\APACmonth{05}}}{},
\newblock
\unskip
\newblock
\APACjournalVolNumPages{\aap}{529}{}{L8}.
\newblock
\begin{APACrefDOI} \doi{10.1051/0004-6361/201116789} \end{APACrefDOI}
\PrintBackRefs{\CurrentBib}

\bibitem [\protect \citeauthoryear {%
{Lagarde}%
\ \protect \BOthers {.}}{%
{Lagarde}%
\ \protect \BOthers {.}}{%
{\protect \APACyear {2012}}%
}]{%
2012A&A...543A.108L}
\APACinsertmetastar {%
2012A&A...543A.108L}%
\begin{APACrefauthors}%
{Lagarde}, N.%
, {Decressin}, T.%
, {Charbonnel}, C.%
, {Eggenberger}, P.%
, {Ekstr{\"o}m}, S.%
\BCBL {}\ \BBA {} {Palacios}, A.%
\end{APACrefauthors}%
\unskip\
\newblock
\APACrefYearMonthDay{2012}{{\APACmonth{07}}}{},
\newblock
\unskip
\newblock
\APACjournalVolNumPages{\aap}{543}{}{A108}.
\newblock
\begin{APACrefDOI} \doi{10.1051/0004-6361/201118331} \end{APACrefDOI}
\PrintBackRefs{\CurrentBib}

\bibitem [\protect \citeauthoryear {%
{Ludwig}%
}{%
{Ludwig}%
}{%
{\protect \APACyear {2006}}%
}]{%
2006A&A...445..661L}
\APACinsertmetastar {%
2006A&A...445..661L}%
\begin{APACrefauthors}%
{Ludwig}, H\BHBI G.%
\end{APACrefauthors}%
\unskip\
\newblock
\APACrefYearMonthDay{2006}{{\APACmonth{01}}}{},
\newblock
\unskip
\newblock
\APACjournalVolNumPages{\aap}{445}{}{661-671}.
\newblock
\begin{APACrefDOI} \doi{10.1051/0004-6361:20042102} \end{APACrefDOI}
\PrintBackRefs{\CurrentBib}

\bibitem [\protect \citeauthoryear {%
{Luo}%
\ \protect \BOthers {.}}{%
{Luo}%
\ \protect \BOthers {.}}{%
{\protect \APACyear {2016}}%
}]{%
2016yCat.5149....0L}
\APACinsertmetastar {%
2016yCat.5149....0L}%
\begin{APACrefauthors}%
{Luo}, A\BHBI L.%
, {Zhao}, Y\BHBI H.%
, {Zhao}, G.%
\ et al.\end{APACrefauthors}%
\unskip\
\newblock
\APACrefYearMonthDay{2016}{{\APACmonth{11}}}{},
\newblock
\unskip
\newblock
\APACjournalVolNumPages{VizieR Online Data Catalog}{5149}{}{}.
\PrintBackRefs{\CurrentBib}

\bibitem [\protect \citeauthoryear {%
{Mathur}%
\ \protect \BOthers {.}}{%
{Mathur}%
\ \protect \BOthers {.}}{%
{\protect \APACyear {2014}}%
}]{%
2014A&A...562A.124M}
\APACinsertmetastar {%
2014A&A...562A.124M}%
\begin{APACrefauthors}%
{Mathur}, S.%
, {Garc{\'{\i}}a}, R\BPBI A.%
, {Ballot}, J.%
\ et al.\end{APACrefauthors}%
\unskip\
\newblock
\APACrefYearMonthDay{2014}{{\APACmonth{02}}}{},
\newblock
\unskip
\newblock
\APACjournalVolNumPages{\aap}{562}{}{A124}.
\newblock
\begin{APACrefDOI} \doi{10.1051/0004-6361/201322707} \end{APACrefDOI}
\PrintBackRefs{\CurrentBib}

\bibitem [\protect \citeauthoryear {%
{Mathur}%
\ \protect \BOthers {.}}{%
{Mathur}%
\ \protect \BOthers {.}}{%
{\protect \APACyear {2016}}%
}]{%
2016ApJ...827...50M}
\APACinsertmetastar {%
2016ApJ...827...50M}%
\begin{APACrefauthors}%
{Mathur}, S.%
, {Garc{\'{\i}}a}, R\BPBI A.%
, {Huber}, D.%
\ et al.\end{APACrefauthors}%
\unskip\
\newblock
\APACrefYearMonthDay{2016}{{\APACmonth{08}}}{},
\newblock
\unskip
\newblock
\APACjournalVolNumPages{\apj}{827}{}{50}.
\newblock
\begin{APACrefDOI} \doi{10.3847/0004-637X/827/1/50} \end{APACrefDOI}
\PrintBackRefs{\CurrentBib}

\bibitem [\protect \citeauthoryear {%
{Mathur}%
\ \protect \BOthers {.}}{%
{Mathur}%
\ \protect \BOthers {.}}{%
{\protect \APACyear {2010}}%
}]{%
2010A&A...511A..46M}
\APACinsertmetastar {%
2010A&A...511A..46M}%
\begin{APACrefauthors}%
{Mathur}, S.%
, {Garc{\'{\i}}a}, R\BPBI A.%
, {R{\'e}gulo}, C.%
\ et al.\end{APACrefauthors}%
\unskip\
\newblock
\APACrefYearMonthDay{2010}{{\APACmonth{02}}}{},
\newblock
\unskip
\newblock
\APACjournalVolNumPages{\aap}{511}{}{A46}.
\newblock
\begin{APACrefDOI} \doi{10.1051/0004-6361/200913266} \end{APACrefDOI}
\PrintBackRefs{\CurrentBib}

\bibitem [\protect \citeauthoryear {%
{Mathur}%
\ \protect \BOthers {.}}{%
{Mathur}%
\ \protect \BOthers {.}}{%
{\protect \APACyear {2011}}%
}]{%
2011ApJ...741..119M}
\APACinsertmetastar {%
2011ApJ...741..119M}%
\begin{APACrefauthors}%
{Mathur}, S.%
, {Hekker}, S.%
, {Trampedach}, R.%
\ et al.\end{APACrefauthors}%
\unskip\
\newblock
\APACrefYearMonthDay{2011}{{\APACmonth{11}}}{},
\newblock
\unskip
\newblock
\APACjournalVolNumPages{\apj}{741}{}{119}.
\newblock
\begin{APACrefDOI} \doi{10.1088/0004-637X/741/2/119} \end{APACrefDOI}
\PrintBackRefs{\CurrentBib}

\bibitem [\protect \citeauthoryear {%
{Mathur}%
\ \protect \BOthers {.}}{%
{Mathur}%
\ \protect \BOthers {.}}{%
{\protect \APACyear {2017}}%
}]{%
2017ApJS..229...30M}
\APACinsertmetastar {%
2017ApJS..229...30M}%
\begin{APACrefauthors}%
{Mathur}, S.%
, {Huber}, D.%
, {Batalha}, N\BPBI M.%
\ et al.\end{APACrefauthors}%
\unskip\
\newblock
\APACrefYearMonthDay{2017}{{\APACmonth{04}}}{},
\newblock
\unskip
\newblock
\APACjournalVolNumPages{\apjs}{229}{}{30}.
\newblock
\begin{APACrefDOI} \doi{10.3847/1538-4365/229/2/30} \end{APACrefDOI}
\PrintBackRefs{\CurrentBib}

\bibitem [\protect \citeauthoryear {%
{Michel}%
}{%
{Michel}%
}{%
{\protect \APACyear {1993}}%
}]{%
1993DSSN....6...19M}
\APACinsertmetastar {%
1993DSSN....6...19M}%
\begin{APACrefauthors}%
{Michel}, E.%
\end{APACrefauthors}%
\unskip\
\newblock
\APACrefYearMonthDay{1993}{{\APACmonth{05}}}{},
\newblock
\unskip
\newblock
\APACjournalVolNumPages{Delta Scuti Star Newsletter}{6}{}{19-23}.
\PrintBackRefs{\CurrentBib}

\bibitem [\protect \citeauthoryear {%
{Michel}%
\ \protect \BOthers {.}}{%
{Michel}%
\ \protect \BOthers {.}}{%
{\protect \APACyear {2008}}%
}]{%
2008Sci...322..558M}
\APACinsertmetastar {%
2008Sci...322..558M}%
\begin{APACrefauthors}%
{Michel}, E.%
, {Baglin}, A.%
, {Auvergne}, M.%
\ et al.\end{APACrefauthors}%
\unskip\
\newblock
\APACrefYearMonthDay{2008}{{\APACmonth{10}}}{},
\newblock
\unskip
\newblock
\APACjournalVolNumPages{Science}{322}{}{558}.
\newblock
\begin{APACrefDOI} \doi{10.1126/science.1163004} \end{APACrefDOI}
\PrintBackRefs{\CurrentBib}

\bibitem [\protect \citeauthoryear {%
{Miglio}%
\ \protect \BOthers {.}}{%
{Miglio}%
\ \protect \BOthers {.}}{%
{\protect \APACyear {2012}}%
}]{%
2012MNRAS.419.2077M}
\APACinsertmetastar {%
2012MNRAS.419.2077M}%
\begin{APACrefauthors}%
{Miglio}, A.%
, {Brogaard}, K.%
, {Stello}, D.%
\ et al.\end{APACrefauthors}%
\unskip\
\newblock
\APACrefYearMonthDay{2012}{{\APACmonth{01}}}{},
\newblock
\unskip
\newblock
\APACjournalVolNumPages{\mnras}{419}{}{2077-2088}.
\newblock
\begin{APACrefDOI} \doi{10.1111/j.1365-2966.2011.19859.x} \end{APACrefDOI}
\PrintBackRefs{\CurrentBib}

\bibitem [\protect \citeauthoryear {%
{Miglio}%
\ \protect \BOthers {.}}{%
{Miglio}%
\ \protect \BOthers {.}}{%
{\protect \APACyear {2009}}%
}]{%
2009A&A...503L..21M}
\APACinsertmetastar {%
2009A&A...503L..21M}%
\begin{APACrefauthors}%
{Miglio}, A.%
, {Montalb{\'a}n}, J.%
, {Baudin}, F.%
\ et al.\end{APACrefauthors}%
\unskip\
\newblock
\APACrefYearMonthDay{2009}{{\APACmonth{09}}}{},
\newblock
\unskip
\newblock
\APACjournalVolNumPages{\aap}{503}{}{L21-L24}.
\newblock
\begin{APACrefDOI} \doi{10.1051/0004-6361/200912822} \end{APACrefDOI}
\PrintBackRefs{\CurrentBib}

\bibitem [\protect \citeauthoryear {%
{Mosser}%
\ \BBA {} {Appourchaux}%
}{%
{Mosser}%
\ \BBA {} {Appourchaux}%
}{%
{\protect \APACyear {2009}}%
}]{%
2009A&A...508..877M}
\APACinsertmetastar {%
2009A&A...508..877M}%
\begin{APACrefauthors}%
{Mosser}, B.%
\BCBT {}\ \BBA {} {Appourchaux}, T.%
\end{APACrefauthors}%
\unskip\
\newblock
\APACrefYearMonthDay{2009}{{\APACmonth{12}}}{},
\newblock
\unskip
\newblock
\APACjournalVolNumPages{\aap}{508}{}{877-887}.
\newblock
\begin{APACrefDOI} \doi{10.1051/0004-6361/200912944} \end{APACrefDOI}
\PrintBackRefs{\CurrentBib}

\bibitem [\protect \citeauthoryear {%
{Mosser}%
, {Barban}%
\BCBL {}\ \protect \BOthers {.}}{%
{Mosser}%
, {Barban}%
\BCBL {}\ \protect \BOthers {.}}{%
{\protect \APACyear {2011}}%
}]{%
2011A&A...532A..86M}
\APACinsertmetastar {%
2011A&A...532A..86M}%
\begin{APACrefauthors}%
{Mosser}, B.%
, {Barban}, C.%
, {Montalb{\'a}n}, J.%
\ et al.\end{APACrefauthors}%
\unskip\
\newblock
\APACrefYearMonthDay{2011}{{\APACmonth{08}}}{},
\newblock
\unskip
\newblock
\APACjournalVolNumPages{\aap}{532}{}{A86}.
\newblock
\begin{APACrefDOI} \doi{10.1051/0004-6361/201116825} \end{APACrefDOI}
\PrintBackRefs{\CurrentBib}

\bibitem [\protect \citeauthoryear {%
{Mosser}%
, {Belkacem}%
\BCBL {}\ \protect \BOthers {.}}{%
{Mosser}%
, {Belkacem}%
\BCBL {}\ \protect \BOthers {.}}{%
{\protect \APACyear {2011}}%
}]{%
2011A&A...525L...9M}
\APACinsertmetastar {%
2011A&A...525L...9M}%
\begin{APACrefauthors}%
{Mosser}, B.%
, {Belkacem}, K.%
, {Goupil}, M.%
\ et al.\end{APACrefauthors}%
\unskip\
\newblock
\APACrefYearMonthDay{2011}{{\APACmonth{01}}}{},
\newblock
\unskip
\newblock
\APACjournalVolNumPages{\aap}{525}{}{L9}.
\newblock
\begin{APACrefDOI} \doi{10.1051/0004-6361/201015440} \end{APACrefDOI}
\PrintBackRefs{\CurrentBib}

\bibitem [\protect \citeauthoryear {%
{Mosser}%
\ \protect \BOthers {.}}{%
{Mosser}%
\ \protect \BOthers {.}}{%
{\protect \APACyear {2010}}%
}]{%
2010A&A...517A..22M}
\APACinsertmetastar {%
2010A&A...517A..22M}%
\begin{APACrefauthors}%
{Mosser}, B.%
, {Belkacem}, K.%
, {Goupil}, M\BHBI J.%
\ et al.\end{APACrefauthors}%
\unskip\
\newblock
\APACrefYearMonthDay{2010}{{\APACmonth{07}}}{},
\newblock
\unskip
\newblock
\APACjournalVolNumPages{\aap}{517}{}{A22}.
\newblock
\begin{APACrefDOI} \doi{10.1051/0004-6361/201014036} \end{APACrefDOI}
\PrintBackRefs{\CurrentBib}

\bibitem [\protect \citeauthoryear {%
{Mosser}%
, {Dziembowski}%
\BCBL {}\ \protect \BOthers {.}}{%
{Mosser}%
, {Dziembowski}%
\BCBL {}\ \protect \BOthers {.}}{%
{\protect \APACyear {2013}}%
}]{%
2013A&A...559A.137M}
\APACinsertmetastar {%
2013A&A...559A.137M}%
\begin{APACrefauthors}%
{Mosser}, B.%
, {Dziembowski}, W\BPBI A.%
, {Belkacem}, K.%
\ et al.\end{APACrefauthors}%
\unskip\
\newblock
\APACrefYearMonthDay{2013}{{\APACmonth{11}}}{},
\newblock
\unskip
\newblock
\APACjournalVolNumPages{\aap}{559}{}{A137}.
\newblock
\begin{APACrefDOI} \doi{10.1051/0004-6361/201322243} \end{APACrefDOI}
\PrintBackRefs{\CurrentBib}

\bibitem [\protect \citeauthoryear {%
{Mosser}%
, {Elsworth}%
\BCBL {}\ \protect \BOthers {.}}{%
{Mosser}%
, {Elsworth}%
\BCBL {}\ \protect \BOthers {.}}{%
{\protect \APACyear {2012}}%
}]{%
2012A&A...537A..30M}
\APACinsertmetastar {%
2012A&A...537A..30M}%
\begin{APACrefauthors}%
{Mosser}, B.%
, {Elsworth}, Y.%
, {Hekker}, S.%
\ et al.\end{APACrefauthors}%
\unskip\
\newblock
\APACrefYearMonthDay{2012}{{\APACmonth{01}}}{},
\newblock
\unskip
\newblock
\APACjournalVolNumPages{\aap}{537}{}{A30}.
\newblock
\begin{APACrefDOI} \doi{10.1051/0004-6361/201117352} \end{APACrefDOI}
\PrintBackRefs{\CurrentBib}

\bibitem [\protect \citeauthoryear {%
{Mosser}%
, {Goupil}%
\BCBL {}\ \protect \BOthers {.}}{%
{Mosser}%
, {Goupil}%
\BCBL {}\ \protect \BOthers {.}}{%
{\protect \APACyear {2012}}%
}]{%
2012A&A...540A.143M}
\APACinsertmetastar {%
2012A&A...540A.143M}%
\begin{APACrefauthors}%
{Mosser}, B.%
, {Goupil}, M\BPBI J.%
, {Belkacem}, K.%
\ et al.\end{APACrefauthors}%
\unskip\
\newblock
\APACrefYearMonthDay{2012}{{\APACmonth{04}}}{},
\newblock
\unskip
\newblock
\APACjournalVolNumPages{\aap}{540}{}{A143}.
\newblock
\begin{APACrefDOI} \doi{10.1051/0004-6361/201118519} \end{APACrefDOI}
\PrintBackRefs{\CurrentBib}

\bibitem [\protect \citeauthoryear {%
{Mosser}%
, {Michel}%
\BCBL {}\ \protect \BOthers {.}}{%
{Mosser}%
, {Michel}%
\BCBL {}\ \protect \BOthers {.}}{%
{\protect \APACyear {2013}}%
}]{%
2013A&A...550A.126M}
\APACinsertmetastar {%
2013A&A...550A.126M}%
\begin{APACrefauthors}%
{Mosser}, B.%
, {Michel}, E.%
, {Belkacem}, K.%
\ et al.\end{APACrefauthors}%
\unskip\
\newblock
\APACrefYearMonthDay{2013}{{\APACmonth{02}}}{},
\newblock
\unskip
\newblock
\APACjournalVolNumPages{\aap}{550}{}{A126}.
\newblock
\begin{APACrefDOI} \doi{10.1051/0004-6361/201220435} \end{APACrefDOI}
\PrintBackRefs{\CurrentBib}

\bibitem [\protect \citeauthoryear {%
{Mosser}%
, {Samadi}%
\BCBL {}\ \BBA {} {Belkacem}%
}{%
{Mosser}%
, {Samadi}%
\BCBL {}\ \BBA {} {Belkacem}%
}{%
{\protect \APACyear {2013}}%
}]{%
2013sf2a.conf...25M}
\APACinsertmetastar {%
2013sf2a.conf...25M}%
\begin{APACrefauthors}%
{Mosser}, B.%
, {Samadi}, R.%
\BCBL {}\ \BBA {} {Belkacem}, K.%
\end{APACrefauthors}%
\unskip\
\newblock
\APACrefYearMonthDay{2013}{{\APACmonth{11}}}{},
\newblock
{\BBOQ}\APACrefatitle {{Red giants seismology}} {{Red giants
  seismology}}.{\BBCQ}
\newblock
\BIn{} L.~{Cambresy}, F.~{Martins}, E.~{Nuss}\BCBL {}\ \BOthers {.}\ (\BEDS),
  \APACrefbtitle {SF2A-2013: Proceedings of the Annual meeting of the French
  Society of Astronomy and Astrophysics} {SF2A-2013: Proceedings of the Annual
  meeting of the French Society of Astronomy and Astrophysics}\ \BPG~25-36.
\PrintBackRefs{\CurrentBib}

\bibitem [\protect \citeauthoryear {%
{Ollivier}%
\ \protect \BOthers {.}}{%
{Ollivier}%
\ \protect \BOthers {.}}{%
{\protect \APACyear {2016}}%
}]{%
2016cole.book...41O}
\APACinsertmetastar {%
2016cole.book...41O}%
\begin{APACrefauthors}%
{Ollivier}, M.%
, {Deru}, A.%
, {Chaintreuil}, S.%
\ et al.\end{APACrefauthors}%
\unskip\
\newblock
\APACrefYearMonthDay{2016}{}{},
\newblock
{\BBOQ}\APACrefatitle {{II.2 Description of processes and corrections from
  observation to delivery}} {{II.2 Description of processes and corrections
  from observation to delivery}}.{\BBCQ}
\newblock
\BIn{} {CoRoT Team}\ (\BED), \APACrefbtitle {The CoRoT Legacy Book: The
  Adventure of the Ultra High Precision Photometry from Space} {The CoRoT
  Legacy Book: The Adventure of the Ultra High Precision Photometry from
  Space}\ \BPG~41.
\newblock
\begin{APACrefDOI} \doi{10.1051/978-2-7598-1876-1.c022} \end{APACrefDOI}
\PrintBackRefs{\CurrentBib}

\bibitem [\protect \citeauthoryear {%
{Pinsonneault}%
\ \protect \BOthers {.}}{%
{Pinsonneault}%
\ \protect \BOthers {.}}{%
{\protect \APACyear {2014}}%
}]{%
2014ApJS..215...19P}
\APACinsertmetastar {%
2014ApJS..215...19P}%
\begin{APACrefauthors}%
{Pinsonneault}, M\BPBI H.%
, {Elsworth}, Y.%
, {Epstein}, C.%
\ et al.\end{APACrefauthors}%
\unskip\
\newblock
\APACrefYearMonthDay{2014}{{\APACmonth{12}}}{},
\newblock
\unskip
\newblock
\APACjournalVolNumPages{\apjs}{215}{}{19}.
\newblock
\begin{APACrefDOI} \doi{10.1088/0067-0049/215/2/19} \end{APACrefDOI}
\PrintBackRefs{\CurrentBib}

\bibitem [\protect \citeauthoryear {%
{Powell}%
}{%
{Powell}%
}{%
{\protect \APACyear {1964}}%
}]{%
Powell1964}
\APACinsertmetastar {%
Powell1964}%
\begin{APACrefauthors}%
{Powell}, M.%
\end{APACrefauthors}%
\unskip\
\newblock
\APACrefYearMonthDay{1964}{}{},
\newblock
\unskip
\newblock
\APACjournalVolNumPages{The Computer Journal}{7}{}{(2):155-162}.
\PrintBackRefs{\CurrentBib}

\bibitem [\protect \citeauthoryear {%
{Press}%
, {Teukolsky}%
, {Vetterling}%
\BCBL {}\ \BBA {} {Flannery}%
}{%
{Press}%
\ \protect \BOthers {.}}{%
{\protect \APACyear {2007}}%
}]{%
numerical_recipes}
\APACinsertmetastar {%
numerical_recipes}%
\begin{APACrefauthors}%
{Press}, W.%
, {Teukolsky}, S.%
, {Vetterling}, W.%
\BCBL {}\ \BBA {} {Flannery}, B.%
\end{APACrefauthors}%
\unskip\
\newblock
\APACrefYear{2007},
\newblock
\APACrefbtitle {{Numerical Recipes (Third Edition)}} {{Numerical Recipes (Third
  Edition)}}\ (\BVOL~506; I\BPBI .~Cambridge University~Press, \BED{}).
\PrintBackRefs{\CurrentBib}

\bibitem [\protect \citeauthoryear {%
{Rauer}%
\ \protect \BOthers {.}}{%
{Rauer}%
\ \protect \BOthers {.}}{%
{\protect \APACyear {2014}}%
}]{%
2014ExA....38..249R}
\APACinsertmetastar {%
2014ExA....38..249R}%
\begin{APACrefauthors}%
{Rauer}, H.%
, {Catala}, C.%
, {Aerts}, C.%
\ et al.\end{APACrefauthors}%
\unskip\
\newblock
\APACrefYearMonthDay{2014}{{\APACmonth{11}}}{},
\newblock
\unskip
\newblock
\APACjournalVolNumPages{Experimental Astronomy}{38}{}{249-330}.
\newblock
\begin{APACrefDOI} \doi{10.1007/s10686-014-9383-4} \end{APACrefDOI}
\PrintBackRefs{\CurrentBib}

\bibitem [\protect \citeauthoryear {%
{Ricker}%
\ \protect \BOthers {.}}{%
{Ricker}%
\ \protect \BOthers {.}}{%
{\protect \APACyear {2015}}%
}]{%
2015JATIS...1a4003R}
\APACinsertmetastar {%
2015JATIS...1a4003R}%
\begin{APACrefauthors}%
{Ricker}, G\BPBI R.%
, {Winn}, J\BPBI N.%
, {Vanderspek}, R.%
\ et al.\end{APACrefauthors}%
\unskip\
\newblock
\APACrefYearMonthDay{2015}{{\APACmonth{01}}}{},
\newblock
\unskip
\newblock
\APACjournalVolNumPages{Journal of Astronomical Telescopes, Instruments, and
  Systems}{1}{1}{014003}.
\newblock
\begin{APACrefDOI} \doi{10.1117/1.JATIS.1.1.014003} \end{APACrefDOI}
\PrintBackRefs{\CurrentBib}

\bibitem [\protect \citeauthoryear {%
{Ruiz-Dern}%
, {Babusiaux}%
, {Arenou}%
, {Turon}%
\BCBL {}\ \BBA {} {Lallement}%
}{%
{Ruiz-Dern}%
\ \protect \BOthers {.}}{%
{\protect \APACyear {2018}}%
}]{%
2018A&A...609A.116R}
\APACinsertmetastar {%
2018A&A...609A.116R}%
\begin{APACrefauthors}%
{Ruiz-Dern}, L.%
, {Babusiaux}, C.%
, {Arenou}, F.%
, {Turon}, C.%
\BCBL {}\ \BBA {} {Lallement}, R.%
\end{APACrefauthors}%
\unskip\
\newblock
\APACrefYearMonthDay{2018}{{\APACmonth{01}}}{},
\newblock
\unskip
\newblock
\APACjournalVolNumPages{\aap}{609}{}{A116}.
\newblock
\begin{APACrefDOI} \doi{10.1051/0004-6361/201731572} \end{APACrefDOI}
\PrintBackRefs{\CurrentBib}

\bibitem [\protect \citeauthoryear {%
{Samadi}%
, {Belkacem}%
\BCBL {}\ \BBA {} {Ludwig}%
}{%
{Samadi}%
, {Belkacem}%
\BCBL {}\ \BBA {} {Ludwig}%
}{%
{\protect \APACyear {2013}}%
}]{%
2013A26A...559A..39S}
\APACinsertmetastar {%
2013A26A...559A..39S}%
\begin{APACrefauthors}%
{Samadi}, R.%
, {Belkacem}, K.%
\BCBL {}\ \BBA {} {Ludwig}, H\BHBI G.%
\end{APACrefauthors}%
\unskip\
\newblock
\APACrefYearMonthDay{2013}{{\APACmonth{11}}}{},
\newblock
\unskip
\newblock
\APACjournalVolNumPages{\aap}{559}{}{A39}.
\newblock
\begin{APACrefDOI} \doi{10.1051/0004-6361/201220816} \end{APACrefDOI}
\PrintBackRefs{\CurrentBib}

\bibitem [\protect \citeauthoryear {%
{Samadi}%
, {Belkacem}%
, {Ludwig}%
, {Caffau}%
\BCBL {}\ \protect \BOthers {.}}{%
{Samadi}%
, {Belkacem}%
, {Ludwig}%
, {Caffau}%
\BCBL {}\ \protect \BOthers {.}}{%
{\protect \APACyear {2013}}%
}]{%
2013A&A...559A..40S}
\APACinsertmetastar {%
2013A&A...559A..40S}%
\begin{APACrefauthors}%
{Samadi}, R.%
, {Belkacem}, K.%
, {Ludwig}, H\BHBI G.%
\ et al.\end{APACrefauthors}%
\unskip\
\newblock
\APACrefYearMonthDay{2013}{{\APACmonth{11}}}{},
\newblock
\unskip
\newblock
\APACjournalVolNumPages{\aap}{559}{}{A40}.
\newblock
\begin{APACrefDOI} \doi{10.1051/0004-6361/201220817} \end{APACrefDOI}
\PrintBackRefs{\CurrentBib}

\bibitem [\protect \citeauthoryear {%
{SDSS Collaboration}%
\ \protect \BOthers {.}}{%
{SDSS Collaboration}%
\ \protect \BOthers {.}}{%
{\protect \APACyear {2016}}%
}]{%
2016arXiv160802013S}
\APACinsertmetastar {%
2016arXiv160802013S}%
\begin{APACrefauthors}%
{SDSS Collaboration}%
, {Albareti}, F\BPBI D.%
, {Allende Prieto}, C.%
\ et al.\end{APACrefauthors}%
\unskip\
\newblock
\APACrefYearMonthDay{2016}{{\APACmonth{08}}}{},
\newblock
\unskip
\newblock
\APACjournalVolNumPages{ArXiv e-prints}{}{}{}.
\PrintBackRefs{\CurrentBib}

\bibitem [\protect \citeauthoryear {%
{Stello}%
, {Chaplin}%
, {Basu}%
, {Elsworth}%
\BCBL {}\ \BBA {} {Bedding}%
}{%
{Stello}%
\ \protect \BOthers {.}}{%
{\protect \APACyear {2009}}%
}]{%
2009MNRAS.400L..80S}
\APACinsertmetastar {%
2009MNRAS.400L..80S}%
\begin{APACrefauthors}%
{Stello}, D.%
, {Chaplin}, W\BPBI J.%
, {Basu}, S.%
, {Elsworth}, Y.%
\BCBL {}\ \BBA {} {Bedding}, T\BPBI R.%
\end{APACrefauthors}%
\unskip\
\newblock
\APACrefYearMonthDay{2009}{{\APACmonth{11}}}{},
\newblock
\unskip
\newblock
\APACjournalVolNumPages{\mnras}{400}{}{L80-L84}.
\newblock
\begin{APACrefDOI} \doi{10.1111/j.1745-3933.2009.00767.x} \end{APACrefDOI}
\PrintBackRefs{\CurrentBib}

\bibitem [\protect \citeauthoryear {%
{Tassoul}%
}{%
{Tassoul}%
}{%
{\protect \APACyear {1980}}%
}]{%
1980ApJS...43..469T}
\APACinsertmetastar {%
1980ApJS...43..469T}%
\begin{APACrefauthors}%
{Tassoul}, M.%
\end{APACrefauthors}%
\unskip\
\newblock
\APACrefYearMonthDay{1980}{{\APACmonth{08}}}{},
\newblock
\unskip
\newblock
\APACjournalVolNumPages{\apjs}{43}{}{469-490}.
\newblock
\begin{APACrefDOI} \doi{10.1086/190678} \end{APACrefDOI}
\PrintBackRefs{\CurrentBib}

\bibitem [\protect \citeauthoryear {%
{Toutain}%
\ \BBA {} {Appourchaux}%
}{%
{Toutain}%
\ \BBA {} {Appourchaux}%
}{%
{\protect \APACyear {1994}}%
}]{%
1994A&A...289..649T}
\APACinsertmetastar {%
1994A&A...289..649T}%
\begin{APACrefauthors}%
{Toutain}, T.%
\BCBT {}\ \BBA {} {Appourchaux}, T.%
\end{APACrefauthors}%
\unskip\
\newblock
\APACrefYearMonthDay{1994}{{\APACmonth{09}}}{},
\newblock
\unskip
\newblock
\APACjournalVolNumPages{\aap}{289}{}{649-658}.
\PrintBackRefs{\CurrentBib}

\bibitem [\protect \citeauthoryear {%
{Ulrich}%
}{%
{Ulrich}%
}{%
{\protect \APACyear {1986}}%
}]{%
1986ApJ...306L..37U}
\APACinsertmetastar {%
1986ApJ...306L..37U}%
\begin{APACrefauthors}%
{Ulrich}, R\BPBI K.%
\end{APACrefauthors}%
\unskip\
\newblock
\APACrefYearMonthDay{1986}{{\APACmonth{07}}}{},
\newblock
\unskip
\newblock
\APACjournalVolNumPages{\apjl}{306}{}{L37-L40}.
\newblock
\begin{APACrefDOI} \doi{10.1086/184700} \end{APACrefDOI}
\PrintBackRefs{\CurrentBib}

\bibitem [\protect \citeauthoryear {%
{V{\'a}zquez Rami{\'o}}%
, {R{\'e}gulo}%
\BCBL {}\ \BBA {} {Roca Cort{\'e}s}%
}{%
{V{\'a}zquez Rami{\'o}}%
\ \protect \BOthers {.}}{%
{\protect \APACyear {2005}}%
}]{%
2005A&A...443L..11V}
\APACinsertmetastar {%
2005A&A...443L..11V}%
\begin{APACrefauthors}%
{V{\'a}zquez Rami{\'o}}, H.%
, {R{\'e}gulo}, C.%
\BCBL {}\ \BBA {} {Roca Cort{\'e}s}, T.%
\end{APACrefauthors}%
\unskip\
\newblock
\APACrefYearMonthDay{2005}{{\APACmonth{11}}}{},
\newblock
\unskip
\newblock
\APACjournalVolNumPages{\aap}{443}{}{L11-L14}.
\newblock
\begin{APACrefDOI} \doi{10.1051/0004-6361:200500191} \end{APACrefDOI}
\PrintBackRefs{\CurrentBib}

\bibitem [\protect \citeauthoryear {%
{Verner}%
\ \protect \BOthers {.}}{%
{Verner}%
\ \protect \BOthers {.}}{%
{\protect \APACyear {2011}}%
}]{%
2011MNRAS.415.3539V}
\APACinsertmetastar {%
2011MNRAS.415.3539V}%
\begin{APACrefauthors}%
{Verner}, G\BPBI A.%
, {Elsworth}, Y.%
, {Chaplin}, W\BPBI J.%
\ et al.\end{APACrefauthors}%
\unskip\
\newblock
\APACrefYearMonthDay{2011}{{\APACmonth{08}}}{},
\newblock
\unskip
\newblock
\APACjournalVolNumPages{\mnras}{415}{}{3539-3551}.
\newblock
\begin{APACrefDOI} \doi{10.1111/j.1365-2966.2011.18968.x} \end{APACrefDOI}
\PrintBackRefs{\CurrentBib}

\bibitem [\protect \citeauthoryear {%
{Vrard}%
, {Mosser}%
\BCBL {}\ \BBA {} {Samadi}%
}{%
{Vrard}%
\ \protect \BOthers {.}}{%
{\protect \APACyear {2016}}%
}]{%
2016A&A...588A..87V}
\APACinsertmetastar {%
2016A&A...588A..87V}%
\begin{APACrefauthors}%
{Vrard}, M.%
, {Mosser}, B.%
\BCBL {}\ \BBA {} {Samadi}, R.%
\end{APACrefauthors}%
\unskip\
\newblock
\APACrefYearMonthDay{2016}{{\APACmonth{04}}}{},
\newblock
\unskip
\newblock
\APACjournalVolNumPages{\aap}{588}{}{A87}.
\newblock
\begin{APACrefDOI} \doi{10.1051/0004-6361/201527259} \end{APACrefDOI}
\PrintBackRefs{\CurrentBib}

\bibitem [\protect \citeauthoryear {%
Wilks%
}{%
Wilks%
}{%
{\protect \APACyear {1938}}%
}]{%
wilks1938}
\APACinsertmetastar {%
wilks1938}%
\begin{APACrefauthors}%
Wilks, S\BPBI S.%
\end{APACrefauthors}%
\unskip\
\newblock
\APACrefYearMonthDay{1938}{03}{},
\newblock
\unskip
\newblock
\APACjournalVolNumPages{Ann. Math. Statist.}{9}{1}{60--62}.
\newblock
\begin{APACrefURL} \url{https://doi.org/10.1214/aoms/1177732360}
  \end{APACrefURL}
\newblock
\begin{APACrefDOI} \doi{10.1214/aoms/1177732360} \end{APACrefDOI}
\PrintBackRefs{\CurrentBib}

\bibitem [\protect \citeauthoryear {%
{Woodard}%
}{%
{Woodard}%
}{%
{\protect \APACyear {1984}}%
}]{%
1984PhDT........34W}
\APACinsertmetastar {%
1984PhDT........34W}%
\begin{APACrefauthors}%
{Woodard}, M\BPBI F.%
\end{APACrefauthors}%
\unskip\
\newblock
\APACrefYear{1984}.
\unskip\
\newblock
\APACrefbtitle {{Short-Period Oscillations in the Total Solar Irradiance.}}
  {{Short-Period Oscillations in the Total Solar Irradiance.}}\
  \APACtypeAddressSchool {\BUPhD}{}{}.
\unskip\
\newblock
\APACaddressSchool {}{UNIVERSITY OF CALIFORNIA, SAN DIEGO.}
\PrintBackRefs{\CurrentBib}

\end{thebibliography}



\end{document}